\documentclass[twocolumn,superscriptaddress,floatfix,preprintnumbers,prd,nofootinbib]{revtex4-2}
\usepackage{bm}
\usepackage{bbm}
\usepackage{amsmath}
\usepackage{etoolbox}
\usepackage{float}

\usepackage{needspace}

\newcommand{\eg}{{\emph{e.g.~}}}
\newcommand{\ie}{{\emph{i.e.~}}}

\usepackage[utf8]{inputenc}
\usepackage{verbatim}
\usepackage{amsmath,amsfonts,amssymb}
\usepackage[breaklinks,colorlinks]{hyperref}
\usepackage{float}
\usepackage{natbib}
\usepackage{tikz}
\usepackage{float}
\usepackage{graphicx}
\usepackage{adjustbox}
\usepackage{tabularx}

\usepackage{amsbsy}
\usepackage{orcidlink}
\hypersetup{
    colorlinks=true,
    citecolor=[rgb]{.1, .7, .6},
    linkcolor=[rgb]{.2, .55, .95},
    filecolor=magenta,
    urlcolor=[rgb]{.1, .7, .6},
}

\begin{document}

\preprint{CERN-TH-2025-147}
\preprint{DESY-25-110}

\title{Superradiance Constraints from GW231123}

\author{Andrea Caputo\orcidlink{0000-0003-3516-8332}}\email{andrea.caputo@cern.ch} 
\affiliation{Department of Theoretical Physics, CERN, Esplanade des Particules 1, P.O. Box 1211, Geneva 23, Switzerland}
\affiliation{Dipartimento di Fisica, ``Sapienza'' Universit\`a di Roma \& Sezione INFN Roma1, Piazzale Aldo Moro
5, 00185, Roma, Italy}

\author{Gabriele Franciolini\orcidlink{0000-0003-3516-8332}}\email{gabriele.franciolini@cern.ch} 
\affiliation{Department of Theoretical Physics, CERN, Esplanade des Particules 1, P.O. Box 1211, Geneva 23, Switzerland}

\author{Samuel J. Witte\orcidlink{0000-0003-4649-3085}}\email{Samuel.Witte@physics.ox.ac.uk}
\affiliation{Rudolf Peierls Centre for Theoretical Physics, University of Oxford, UK}
\affiliation{Deutsches Elektronen-Synchrotron DESY, Notkestra\ss{}e 85, 22607 Hamburg, Germany}
\affiliation{
II. Institute of Theoretical Physics, Universit\"{a}t Hamburg, 22761, Hamburg, Germany}

\begin{abstract}
Gravitational wave observations have recently revealed with high significance, and high precision, the existence of $\mathcal{O}(100) \, M_\odot$ rapidly rotating black holes, allowing  gravitational wave events to be used for the first time to probe unexplored axion parameter space using the phenomenon known as black hole superradiance. Here, we present new limits on axions using the binary black hole merger event GW231123, whose constituent black holes are among the fastest spinning observed with gravitational waves to date. We demonstrate that the most viable binary formation channels lead to conservative constraints on axion masses $\mu \sim [0.6-5] \times \, 10^{-13}$ eV and decay constants $f_\Phi \gtrsim 10^{14}$ GeV, extending existing superradiance constraints derived using x-ray observations to yet lower axion masses.
\end{abstract}

\maketitle

\section{Introduction}

The LIGO-Virgo-KAGRA (LVK) collaboration has recently announced the detection of the merger of two constituent BHs (BHs) with masses  $M_1 = 137^{+22}_{-17} \, M_\odot$  and $M_2 = 103^{+20}_{-52} \, M_\odot$ (at $90\%$ credible intervals)~\cite{LIGOScientific:2025rsn}. From an astrophysical perspective, this event is remarkable, as the inferred masses of both constituents lie in the so-called ``pair instability mass gap''~\cite{1964ApJS....9..201F,1967PhRvL..18..379B,Woosley:2021xba} (\ie the range of masses in which pair instabilities prevent the formation of BHs directly from stellar collapse) and represent the heaviest high-significance gravitational wave (GW) event observed to date \cite{KAGRA:2021vkt,Wadekar:2023gea}. This event is even more interesting from the perspective of a particle physicist, as there is a strong preference in the waveform that at least one, if not both, constituent BHs have extremely high dimensionless spin, $\tilde{a}_1 = 0.90^{+0.10}_{-0.19}$ and $\tilde{a}_2 = 0.80^{+0.20}_{-0.51}$, with the primary exhibiting the highest black hole spin ever confidently measured through GWs. This unprecedented measurement is obtained thanks to the confident inference of a large effective precessing spin \cite{Schmidt:2014iyl}, a feature which appears robust against waveform systematics \cite{LIGOScientific:2025rsn}.
This is significant as it represents the first time GW data can be used to probe unexplored parameter space of new light fundamental bosonic particles via the phenomenon known as superradiance.

\begin{figure}[ht!]
    \centering
    \includegraphics[width=0.95\linewidth]{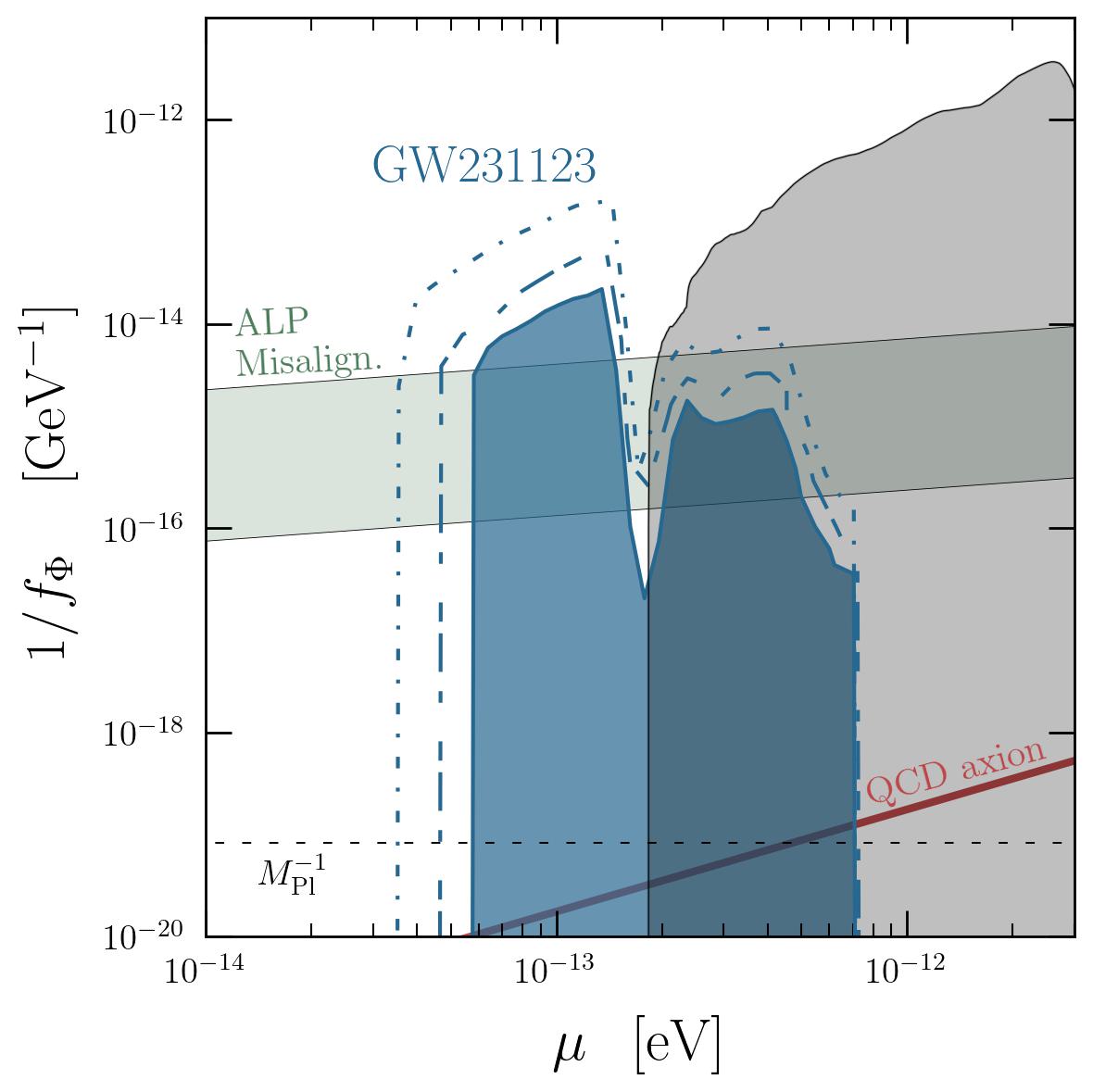}
    \caption{Constraint on the axion decay constant $f_\Phi$ and mass $\mu$ derived from GW231123 (blue); result is obtained using {\tt{NRSur}} waveform model analysis, using $N_{\rm max}=5$, and with $\tau_{\rm max} = 10^5$ (shaded), $10^6$ (dashed), and $10^7$ years (dot-dashed). Shown for comparison are superradiance constraints on solar-mass scale BHs~\cite{Witte:2024drg}, with spin inferences performed using x-ray data (black), the QCD axion line (red), and the parameter space where misalignment mechanism could produce the entirety of dark matter with  $\mathcal{O}(1)$ initial field values (green). The inverse Planck scale $M_{\rm Pl}$ is highlighted with a horizontal dashed line.
     }
    \label{fig:main}
\end{figure}

BH superradiance is a phenomenon in which low-energy bosonic fields experience an instability around Kerr BHs, causing them to rapidly transfer the rotational energy of the BH into high-occupation number quasi-bound states (see \eg\cite{Brito:2015oca} for a review). For a scalar field, this process, under optimal circumstances, can lead to an order one depletion of the BH spin on the timescale of hours~\cite{Detweiler:1980uk,Dolan:2007mj,East:2017ovw,Baryakhtar:2017ngi,Baumann:2019eav,Brito:2020lup,East:2023nsk}, suggesting an inherent incompatibility between rapidly rotating BHs and the existence of light bosons~\cite{Arvanitaki:2009fg,Arvanitaki:2010sy,cardoso2018constraining, Brito:2017wnc, Brito:2017zvb}.

 To date, superradiance constraints have been derived using BH spin measurements that are inferred from x-ray observations~\cite{Stott:2020gjj,Unal:2020jiy,Baryakhtar:2020gao,Hoof:2024quk,Witte:2024drg}; this approach, however, can be highly sensitive to the properties of matter near the innermost stable circular orbit (where the modeling is the least well-understood), the details of radiative transfer, and the thermodynamic properties of x-ray emitting regions -- see \eg~\cite{Witte:2024drg} for a detailed discussion of systematics. GW observations offer an alternative route to measuring high-spin BHs which is free from the modeling systematics that plague x-ray observations -- the only difficulty, until now, has been the lack of gravitationally-observed high-spin BHs.\footnote{There have been attempts to use radio observations of supermassive BHs~\cite{Unal:2020jiy}, however the large residual differences between analyses suggest uncertainties on the spin inference remain too large to be of use~\cite{nemmen2019spin,EventHorizonTelescope:2019pgp,tamburini2020measurement,cruz2022state}. It is worth highlighting that during the review of this manuscript, LVK released an analysis of GW241011~\cite{LIGOScientific:2025brd}, which yielded the most precise high-spin measurement to date; the mass of this event, however, is situated in the same range as BHs for which there exist x-ray spin measurements, and thus the derived constraints are sub-dominant to pre-existing limits.   }

The aim of this {\emph{Letter}} is to demonstrate the power of GW231123, and more generally the power of forthcoming GW observations, in using spin measurements to constrain the existence of light bosonic particles. We focus specifically on the extent to which the inferred spin distributions of the constituent BHs in GW231123 can be used to probe axions, as these particles constitute the most well-motivated bosons at low energies (we do, however, derive a constraint on non-interacting spin-1 fields in Supplemental Material, which includes Refs.~\cite{ligo_scientific_collaboration_2025_16004263,Cuceu:2025fzi,Spera:2017fyx,10.1093/mnras/207.3.585,Bromm:2003vv,Belczynski:2004gu,Tanikawa:2021qqi,Hijikawa:2021hrf,Belczynski:2016ieo,Liu:2020lmi,Zeldovich:1967lct,Hawking:1971ei,Carr:1974nx,Carr:1975qj,Bird:2016dcv,Clesse:2016vqa,Sasaki:2016jop,Eroshenko:2016hmn,Wang:2016ana,Clesse:2020ghq,Hall:2020daa,Franciolini:2022tfm,Escriva:2022bwe,Byrnes:2025tji,LISACosmologyWorkingGroup:2023njw,DeLuca:2020fpg,DeLuca:2020sae,Yuan:2025avq,Mirbabayi:2019uph,DeLuca:2019buf,Harada:2020pzb,DeLuca:2020bjf,DeLuca:2020qqa,Franciolini:2022iaa,Franciolini:2021xbq,Ali-Haimoud:2017rtz,Raidal:2018bbj,Vaskonen:2019jpv,Franciolini:2022ewd,Raidal:2024bmm,Hasinger:2020ptw,Arvanitaki:2014wva,Baumann:2018vus,Baumann:2022pkl,Tong:2022bbl,Fan:2023jjj,Tomaselli:2023ysb,Zhu:2024bqs,Takahashi:2024fyq,Boskovic:2024fga,Tomaselli:2024bdd,Tomaselli:2025jfo,raftery2001statistics}). Using the inferred properties of GW231123, we deduce the range of characteristic timescales over which superradiance could have spun down the BHs, within realistic assumptions motivated by the hierarchical formation scenario (see e.g.~\cite{Gerosa:2021mno} for a review). We then use the inferred posteriors on the spin of each constituent BH, along with motivated priors on their natal values, to simulate possible spin-down histories in the presence of an axion.  The main result of this paper is the limit shown in Fig.~\ref{fig:main}, which demonstrates that GW231123 has excluded a new, and well-motivated, region of axion parameter space that is effectively inaccessible to conventional searches.\footnote{GW evidence for
highly spinning BHs has also been presented in~\cite{Hannam:2021pit,LIGOScientific:2021usb,KAGRA:2021vkt,Wadekar:2023gea,Nitz:2020oeq,Williams:2024tna}. In particular, GW190517 was also characterized by high spins. However, the component masses of that binary only probe regions of axion parameter space excluded by other constraints. Moreover, GW190517 had a modest network signal-to-noise ratio ($\text{SNR} \sim 10$), below the former detection threshold of 12 (although it does satisfy the new ranking statistic requirement of the false alarm rate $\text{FAR}< 0.25\text{yr}^{-1}$).} Needless to say, the constraints derived here rely on the inferred spins of the constituent BHs in GW231123; while the inference of high spins appears to be robust to variations in waveform modeling, the spin posteriors are still sensitive systematic uncertainties that can induce small shifts into the inferred limits (we note, however, that the sensitivity is primarily affected by the small-spin region of the posterior -- see Supplemental Material~\ref{secApp:otherresults}). As waveform modeling improves in the future, so too will the robustness of these constraints; in addition, the approach and methodology used in the analysis here will undoubtedly prove invaluable as new detections of high-spin BHs emerge.

\section{Black Hole Superradiance}
Let us start with a brief review of BH superradiance. Here, we restrict our attention to scalar fields -- generalizations to the case of  higher spin fields can be found \eg in~\cite{Rosa:2011my,Witek:2012tr,Pani:2012vp,Pani:2012bp,Endlich:2016jgc,Baryakhtar:2017ngi,East:2017ovw,East:2017mrj,East:2018glu,Dolan:2018dqv,Brito:2020lup,brito2020black,Brito:2015oca} (in Supplemental Material ~\ref{app:spin1}, we also estimate constraints on non-interacting spin-1 fields). We focus specifically on the case of axions $\Phi$, \ie light pseudoscalars which have an intrinsic discrete shift symmetry $\Phi \rightarrow \Phi + 2\pi f_\Phi$ that protects the smallness of the mass, where $f_\Phi$ is the axion decay constant. The existence of the shift symmetry implies contributions to the axion potential are of the form $V(\Phi) = \mu^2 f_\Phi^2 \left[ 1 - \cos\left(\frac{\Phi}{f_\Phi}\right) \right] + \cdots$ (with higher order, shift-symmetry preserving terms, potentially being present), which for small field values reduces to $V(\Phi) \simeq \frac{1}{2} \, \mu^2 \,  \Phi^2 - \frac{1}{4!} \lambda \Phi^4 + \cdots$, with $\mu$ the axion mass, and $\lambda \equiv (\mu / f_\Phi)^2$.  Thus, a natural expectation of axions is the existence of quartic self-interactions; despite a very small coupling constant, this interaction plays an important role in the evolution of superradiant systems~\cite{Arvanitaki:2010sy,Yoshino:2012kn,Yoshino:2015nsa,Omiya:2022mwv,Gruzinov:2016hcq,Baryakhtar:2020gao,Omiya:2024xlz,Witte:2024drg}. 

The evolution of superradiant states can be determined by solving the equation of motion in a Kerr background. At leading order, one can neglect self-interactions and solve for the complex energy spectrum of quasi-bound states near the black hole; this energy spectrum is discrete, with eigenvalues $\omega = \omega_r + i \Gamma$ labeled by a set of quantum numbers $\left|n\ell m\right>$, see \eg~\cite{Detweiler:1980uk,Rosa:2009ei,Baumann:2019eav}; these quantum numbers are analogous to the principal $n$, orbital $\ell$, and magnetic $m$ quantum numbers of the hydrogen atom, obeying $n \geq  \ell + 1$, $\ell \geq 0$, and $|m| \leq \ell$. The fastest growing state is the $\left| 211 \right>$, which for a $\mathcal{O}(100) \, M_\odot$ BH has an $e-$fold  growth timescale that can be as short as $\tau_{211} \equiv (2\Gamma_{211})^{-1} \sim \mathcal{O}({\rm hours})$. 
For a free particle, the growth of the state saturates when the angular frequency of the horizon $\Omega_H$ reaches a value $\Omega_H \equiv \tilde{a} / (2 r_+) \sim \mu / m$\footnote{We adopt natural units throughout this work.} , where $\tilde{a} \equiv a / (G M)$ is the dimensionless spin and $r_+ = GM (1 + \sqrt{1 - \tilde{a}^2})$ is the outer horizon. This requires roughly $\mathcal{O}(180)$ e-folds of growth, implying the spin of a rapidly rotating BH can be modified on timescales of days. In the calculations below, we compute the eigen-frequencies of each state and at each point during spin-down using the continued fraction method \cite{leaver1985analytic,konoplya2006stability,cardoso2005superradiant,Dolan:2007mj}.

Self-interactions serve to coupled the different eigenstates of the system, transferring energy efficiently between them \footnote{Generally speaking, other interactions could play a role in quenching superradiant growth, see \eg~\cite{Fukuda:2019ewf,Rosa:2017ury,Ikeda:2018nhb,Mathur:2020aqv,Blas:2020kaa,Blas:2020nbs,Baryakhtar:2020gao,Caputo:2021efm,Siemonsen:2022ivj,Spieksma:2023vwl,Ferreira:2024ktd}, however the large occupation numbers $N$ of quasi-bound states stimulate the $2\leftrightarrow 2$ scattering process, causing the rate of energy dissipation to scale as $\propto N^3$ ensuring it dominates the energy dissipation.}. Rather than having decoupled growth of quasi-bound states, as in the case of the free particle solution, self-interactions induced coupled evolution of states in which the growth of any state can be quenched, saturating to a fixed occupation number, long before altering the spin of the BH~\cite{Gruzinov:2016hcq,Baryakhtar:2020gao,Omiya:2024xlz,Witte:2024drg}. The evolution of the superradiant system can be determined by following the energy injection and dissipation, which is described by the set of equations:
\begin{eqnarray}
    \dot{\epsilon}_{n\ell m} &=& 2\Gamma_{n\ell m} \, \epsilon_{n\ell m} +   \sum_{\mathcal{R}_{i,j,k}} \,  \zeta_{i,j} \, \gamma_{i,j}^{k, {\rm L}} \, \epsilon_i \, \epsilon_j \, \epsilon_k \label{eqns:main_e} \\
    \dot{\tilde{a}} &=&  -\sum_{n\ell m} 2 \, m \, \Gamma_{n\ell m} \, \epsilon_{n \ell m} \label{eqns:main_a}\\
    \frac{\dot{M}}{\mu G M^2}& =& - \sum_{n \ell m} 2 \Gamma_{n\ell m} \, \epsilon_{n \ell m} + \sum_{(i,j,k)} \gamma_{i,j}^{k, {\rm BH}} \, \epsilon_i \, \epsilon_j \, \epsilon_k \, . \label{eqns:main_M}
\end{eqnarray}
where $\epsilon_{n\ell m} \equiv N_{n\ell m} / (G M^2)$ is the normalized occupation number $N_{n\ell m}$ of a state, $\zeta_{i,j}$ is a degeneracy counting factor equal to 1 for $i \neq j$ and 2 for $i=j$. Here, we have introduced rate coefficients $\gamma_{i,j}^{k, {\rm L}}$, where the indices $(i,j,k)$ denote quasi-bound states (\ie $i = \left| n\ell m\right>$, $j=  \left| n^\prime\ell^\prime m^\prime\right>$, etc), and `L' represents energy loss, either to infinity $\infty$ (occurring when $\omega_i + \omega_j - \omega_{k} > \mu$), or into bound states which dissipate their energy into the BH (occurring when $\omega_i + \omega_j - \omega_{k} \leq  \mu$). The collective list of non-zero scattering rates relevant for $n \leq 5$ is presented in Table 1 and 2 of~\cite{Witte:2024drg}. Note that the axion mass implicitly enters in both the eigenvalues $\Gamma_{n\ell m}$ as well as the scattering rates $\gamma_{i,j}^{k, L}$, while the decay constant enters only through the scattering rates.

In general, one should include all states that obtain non-negligible occupation numbers on the timescales of interest; in practice, however, this is currently unfeasible, with the latest analyses only including states with $n \leq 5$~\cite{Witte:2024drg}. Here, we perform our fiducial analysis with all states $n\leq 5$, and illustrate the impact of taking instead $n\leq 3$ in the Supplemental Material. All rate coefficients are computed using the procedure of~\cite{Witte:2024drg}.

In solving Eqns.~\ref{eqns:main_e}-\ref{eqns:main_M}, there is an important subtlety which arises from the fact that we observe the spin and mass close to the merger, and these values may differ sizably from those at birth. For this reason, one typically evolves the superradiant system on timescales $t \lesssim \tau_{\rm max}$, where $\tau_{\rm max}$ is chosen to be sufficiently short that astrophysical processes cannot appreciably alter the mass or spin of the system; for isolated BHs, this is often taken to be the Salpeter timescale, $\tau_{\rm Salp} \sim 4.5 \times 10^7$ yr, however we argue in the following Section that for GW231123 the relevant timescale may be shorter, and thus we apply adopt a more conservative timescale in our analyses. In general, one could attempt to model the evolution on longer timescales, including \eg accretion, multiple mergers, etc.~\cite{Brito:2014wla}, but this requires complicated astrophysical modeling which is beyond the scope of this work.

\section{Determination of Merger Timescales}\label{sec:timescale}

As mentioned in the preceding section, testing superradiance typically requires defining the characteristic timescale over which the properties of the black hole can be assumed to remain constant. For isolated black holes, this is often taken to be the Salpeter timescale. For second or third generational hierarchical mergers, the relevant timescale is instead determined by the time between successive mergers, $\tau_{\rm merger}$, which can be significantly shorter than $\tau_{\rm Sal}$. Since the  large masses and spins of the BHs involved in the GW231123 event strongly suggest that one or both of the binary components may be the product of a previous binary BH merger \cite{LIGOScientific:2025rsn,Stegmann:2025cja,Li:2025fnf}, a scenario known as a hierarchical merger~\cite{Rodriguez:2019huv, Antonini:2016gqe, Mapelli:2021syv, Antonini:2022kly, Chattopadhyay:2023yyk, Mahapatra:2022orr, Yang:2019cbr, ArcaSedda:2020hbe, Vaccaro:2023gas, Gilbaum:2024xxq, Rodriguez:2020viw, AraujoAlvarez:2024xxx, Mapelli:2021syv, Mapelli:2020xeq,FragioneSilk} (see e.g.~\cite{Gerosa:2021mno} for a review), we endeavor to determine the minimal value of $\tau_{\rm merger}$ consistent with the properties of GW231123. In App.~\ref{secApp:Exotic} we further discuss more exotic scenarios, showing the constraint in Fig.~\ref{fig:main} remains valid.

Assuming a hierarchical origin, one can infer important information about the general properties of the local environment. For instance, the presence of high-spin BHs in GW231123 is consistent with significant anisotropic GW emission, which implies large recoil velocities for the remnants of first-generation mergers which constitute the progenitors of GW231123---typically on the order of several hundred kilometers per second~\cite{Alvarez:2024dpd}. If this ``birth kick'' exceeds the escape velocity of the host environment, further mergers would be prevented. This suggests that GW231123 likely originated in a deep potential well, \eg in a nuclear star cluster (NSC)~\cite{Mapelli:2020xeq, Mapelli:2021syv, Antonini:2018auk, PortegiesZwart:2002, Lupi:2014vza, Rodriguez:2017pec, Samsing:2020qqd} or an active galactic nucleus (AGN)~\cite{Tagawa:2019osr, Grobner:2020drr, Bartos:2016dgn, Leigh:2017zjd, McKernan:2017tvu, Secunda:2018zfg, 2015MNRAS.448..754H, OLeary:2008myb, McKernan_2018, McKernan:2012, McKernan:2014, McKernan:2020}, where escape velocities can reach $v_{\rm esc} \sim 100$--$500~\mathrm{km/s}$ or higher. In contrast, typical escape velocities from young stellar clusters and globular clusters are significantly lower, $v_{\rm esc} \lesssim 50~\mathrm{km/s}$~\cite{Weatherford_2023, Fragione:2018vty, Oh_2016}, making it unlikely that the remnants of earlier mergers could have been retained in such environments.

In general, one can write the rate density of BH-BH mergers as
\begin{equation}
\mathcal{R} = n_{\rm env} N_{\rm BH}(M_{\rm BH})
\Gamma_{\rm merger},
\label{eq:RateGeneral}
\end{equation}
where $n_{\rm env}$ is the average number density in the Universe of the environment under consideration, 
$N_{\rm BH}$ is the number of BHs within such an environment (where we included explicitly the dependence on the BH mass), and $\Gamma_{\rm merger}$ is the merger rate per BH. Such a rate can be decomposed into two contributions, $\Gamma_{\rm merger} \sim \frac{1}{\tau_{\rm pair}+\tau_{\rm coal}}$, where $\tau_{\rm pair}$ is the pairing timescale for two BHs, and $\tau_{\rm coal}$ is the time to coalescence once the binary has formed. 
In what follows, we will assume the former dominates the merger timescales and that the coalescing time is negligible. This is particularly justified for GW231123 which is expected to form from dynamical processes involving large velocity BHs. In fact, one can verify that for the parameters of interest, namely $v \sim v_{\rm recoil} \gtrsim 100\,\text{km/s}$, $\tau_{\rm pair} \gg \tau_{\rm coal}$, where  $v$ denotes the relative velocity of a BH pair and $v_{\rm recoil}$ represents the ``birth-kick'' recoil velocity of the remnants from first-generation mergers. Such a hierarchy is naturally expected, as both the binary formation rate (scaling as $\tau_{\rm pair} \propto v^{11/7}$ for dynamical capture~\cite{Mouri_2002, OLeary:2008myb}, and $\tau_{\rm pair}\propto v^{9}$ for three-body formation~\cite{Goodman1993, pina2023}) 
and the coalescing time (scaling as 
$\tau_{\rm coal} \propto v^{-36/7}$ for dynamical capture, and $\tau_{\rm coal}\propto v^{-8}$ for three-body formation, assuming Peter's formula \cite{Peters:1963ux,Peters:1964zz}) scale strongly with $v$. 
A similar argument applies to other channels, such as e.g. binary-single interactions. We will proceed under this assumption, which allows us to treat superradiance growth around both individual isolated BHs (see Supplemental Material for a discussion on the impact of the binary),  before they pair in the binary and merge soon after.

The rate density in Eq.~\ref{eq:RateGeneral} should then be compared with the one inferred by the LVK collaboration for GW231123, $\mathcal{R}_{\rm obs} = 0.08^{+0.19}_{-0.07}~\mathrm{Gpc}^{-3}~\mathrm{yr}^{-1}$. From such a comparison, one can infer the characteristic pairing time,
\begin{equation}
\label{eq:mergerGeneral}
\tau_{\rm pair} \simeq \tau_{\rm merger} \equiv \Gamma_{\rm merger}^{-1} = \frac{n_{\rm env} N_{\rm BH}(M_{\rm BH})}{\mathcal{R}_{\rm obs}}.
\end{equation}
This estimate is fully agnostic with respect to the physical mechanisms responsible for binary formation and coalescence. In order to provide realistic numbers, we will evaluate Eq.~\ref{eq:mergerGeneral} considering at least one of the BHs to be a remnant, and treating separately the possibility that this event occurred in an AGN or NSC.

\subsection{Active galactic nuclei}

In AGN, gas-rich accretion disks surround supermassive BHs, and stellar-mass BHs can become embedded and experience migration due to torques from the surrounding gas~\cite{McKernan_2018}. A particularly relevant feature of AGN disks is the presence of \emph{migration traps}---locations where the net torque acting on an object vanishes, halting its inward or outward drift~\cite{Bellovary:2015ifg}. These traps naturally concentrate compact objects, enhancing the probability of dynamical interactions and subsequent mergers, making AGN disks ideal for multiple merger generations. As such, they offer a promising, natural scenario for explaining the high-spin, hierarchical nature of events like GW231123.

In this environment, the merger timescale in Eq.~\ref{eq:mergerGeneral} is
\begin{align}
\tau_{\rm merger}^{\rm AGN} & = \text{Myr} \left(\frac{n_{\rm env}^{\rm AGN}}{10^{-5}\,\text{Mpc}^{-3}}\right) \left(\frac{N_{\rm BH}}{10}\right)
\nonumber \\
& 
\times\left(\frac{0.1 \, \mathrm{Gpc}^{-3}~\mathrm{yr}^{-1}}{\mathcal{R}_{\rm obs}}\right),
\end{align}
where for the AGN number density at low redshift (\(z \lesssim 0.5\)), we adopted a typical, yet conservative, value inferred from multi-wavelength luminosity functions (see, e.g., Fig.~7 in Ref.~\cite{2005ApJ...624..630S}). This number density only includes contributions from AGN with luminosities within the range where migration traps are expected to form and persist~\cite{Gilbaum:2024xxq}. The number of massive, second-generation BHs in the trap is instead estimated following Ref.~\cite{Secunda:2020mhd} (see their Fig.~9), where the authors simulate first-generation BH mergers embedded in the AGN disk and track their migration, binary pairing, and mergers. We stress that if the number of BHs is larger than expected, for a fixed observed rate, this would imply a larger merger timescale and therefore one could probe a larger region of the axion parameter space.
We also note that the observed rate, $\mathcal{R}_{\rm obs}$, ranges from $0.01$ to $0.3\;\mathrm{Gpc}^{-3}\,\mathrm{yr}^{-1}$ at $90\%$ C.L., implying that $\tau_{\rm pair}$ could be smaller by a factor of a few, or alternatively up to ten times longer.  Overall, adopting
\begin{equation}
\tau_{\rm merger}^{\rm AGN}\sim10^{-1}\text{--}10\;\mathrm{Myr}
\end{equation}
should abundantly encompass all uncertainties.

\subsection{Nuclear star clusters}

NSCs~\cite{Neumayer:2020gno} represent an alternative dynamical environment in which compact-object binaries can form and merge multiple times. Compared to AGN disks, NSCs are gas-poor systems, but they can nonetheless sustain high merger rates due to their large stellar densities and deep gravitational potentials. The dominant binary formation mechanism in NSCs depends on whether or not a central supermassive BH (SMBH) is present. In NSCs that do host an SMBH, the short relaxation time facilitates the development of a steep density cusp of stellar-mass BHs around the central object, enhancing the probability of binary formation via gravitational-wave capture~\cite{OLeary:2008myb}. In contrast, NSCs without a SMBH behave more like high-mass globular clusters, where binaries form and harden primarily through three-body encounters and binary single interactions~\cite{Antonini:2016gqe}. In either case, we can write an expression for the merger rate analogous to that in Eq.~\ref{eq:RateGeneral}.

The number density of NSCs in the local universe, relevant for events like GW231123, whose redshift is $z=0.39^{+0.27}_{-0.24}$, depends on the fraction of galaxies that host NSCs and on the galaxy stellar mass function. The former peaks at approximately 90\% for galaxies with a stellar mass around $\sim 10^9 \, M_\odot$, and decreases toward both higher and lower masses. Nonetheless, nucleation fractions remain at the level of $\sim 10\%$ even for galaxies as small as $M_* \sim 5 \times 10^5 \, M_\odot$ and as massive as $M_* \sim 5 \times 10^{11} \, M_\odot$ (see Fig.~3 of Ref.~\cite{NSCfraction}). Each galaxy can host at most one NSC, located at its photometric and dynamical center (unlike the case for globular clusters). This allows us to estimate $n_{\rm env}^{\rm NSC} \sim n_{\rm gal} f_{\rm NSC}$. Using results from galaxy stellar mass function studies~\cite{2017MNRAS.470..283W, Weigel_2016,2012MNRAS.421..621B}, and adopting a conservative value of $f_{\rm NSC} \sim 10\%$, we obtain (see also~\cite{Antonini:2018auk}) $n_{\rm env}^{\rm NSC} \sim 10^{-3} \, \mathrm{Mpc}^{-3}$.

As in the case of AGN, determining $N_{\rm BH}(M_{\rm BH})$ is less straightforward, but we can expect it to be in the same ballpark. In fact, assuming standard stellar initial mass functions~\cite{2001MNRAS.322..231K}, one expects $0.1-1 \%$ of the NSC mass to be in the form of seed BHs that may form binaries and merge~\cite{Mapelli:2016vca, Fryer_2012}. Typical NSC have masses $\sim 10^5-10^8\, M_\odot$, which implies roughly $M_{\rm BH, tot} \sim 10^3-10^4 \,M_\odot$. Of these BHs, we can expect $10-50 \%$ of them to be in binaries~\cite{Morscher2015}. The majority of these BH will on average have masses $\sim 10-20 \, M_\odot$, leaving us with $\sim \mathcal{O}(10-100)$ BHs which will then undergo repeated mergers.
We can thus write
\begin{equation}
\tau_{\rm merger}^{\rm NSC} 
= 10\,\text{Myr}\, N_{\rm BH} 
\left(\frac{n_{\rm env}^{\rm NSC}}{10^{-3}\,\text{Mpc}^{-3}}\right) \left(\frac{0.1 \, \mathrm{Gpc}^{-3}~\mathrm{yr}^{-1}}{\mathcal{R}_{\rm obs}}\right) \, ,
\end{equation}
implying
\begin{equation}
\tau_{\rm merger}^{\rm NSC}\sim1\text{--}100\;\mathrm{Myr}
\end{equation}
should bracket all uncertainties.

\section{Analysis and Conclusions}

Here, we begin by providing a heuristic description of our statistical framework that allows one to derive constraints on the axion parameter space; a statistically rigorous derivation of the relevant quantities from the LVK posterior is also included for completeness.

 First, we fix the axion mass; this is useful as our posterior will otherwise be unbounded in three directions (and thus risks having a high sensitivity to the prior). We adopt a log-flat prior on $ f_\Phi$ over the range $[10^{11}, 10^{20}]\,\mathrm{GeV}$.  We focus on understanding the evolution of the BHs from a time $t = t_{\rm 0} - \tau_{\rm max}$, to the time of merger $t_{\rm 0}$, as  uncertainties in the merger history do not allow us to infer the properties of the BHs on longer timescales. In our analysis, we vary $\tau_{\rm max}$ to be consistent with the shortest characteristic timescales identified in the viable formation channels associated with NSCs and AGNs, with our fiducial limits taken to be the minimum of these, namely $10^5$ years. Since the change in mass of each BH due to superradiance is small relative to the mass uncertainty, one can account for the mass measurement by adopting mass priors using the LVK joint prior distribution $p(M_1, M_2)$. The initial spins at a time $t = t_{\rm 0} - \tau_{\rm max}$ are, of course, unknown -- while astrophysical processes cannot alter the spins on these timescales, superradiance can. Since superradiance only serves to extract spin, we adopt spin priors which contain non-zero weight only above the inferred values of the spins at merger. More specifically, given a sample $i$ of the $X=1,2$ BH (with $`1'$ corresponding to the heavier BH), we compute the conditional distribution $p(\tilde{a}_X| M_{1,i}, M_{2,i})$; we then determine the mean $\zeta_X$ and standard deviation $\beta_X$ in the conditional probability $p(\tilde{a}_X| M_{1,i}, M_{2,i})$ (which we infer from the LVK posterior), and adopt a prior that is half-flat, for $\tilde{a}_{X, i} > \zeta_X$, and half-Gaussian, for the initial spin $\tilde{a}_{X,i} \leq \zeta_X$. The half-Gaussian distribution ensures an appropriate sampling of the initial spin distribution in the event that superradiance does not spin down the black hole, while the half-flat component of the prior ensures that initial high-spin black holes which have been spun down via superradiance to $\tilde{a} = \left<\tilde{a}_X\right>$ are perfectly compatible with the observed data\footnote{Within the hierarchical scenario, the prior distribution of spin for the multi-generation mergers can be derived from specific assumptions on the progenitors and the environment, see e.g. \cite{Gerosa:2017kvu,Vaccaro:2023cwr,Borchers:2025sid,Stegmann:2025cja}. In most cases, the distribution is rather peaked around $\tilde a \sim 0.7$, as obtained for a remnant of mergers of spinless BHs. However, this is reasonably broad assuming spinning progenitors and/or higher-generation BHs, supporting our choice.   }. Eqns.~\ref{eqns:main_e}-\ref{eqns:main_M} are then evolved from $t = t_{\rm 0} - \tau_{\rm max}$ to $t = t_{\rm 0}$, to determine the spins that would have been observed by LVK $\tilde{a}_{1,2}$ at merger. In order to evaluate the compatibility of the predicted spins with those observed, we construct a likelihood using two-dimensional kernel density estimation (KDE) on the set of conditioned posterior samples $p(\tilde{a}_1, \tilde{a}_1| M_{1,i}, M_{2,i})$ from the LVK analysis. The interested reader can find a more detailed derivation of the statistical procedure in the Supplemental Material.

  We run between five and ten Monte Carlo chains, reconstruct the one dimensional marginalized posterior on $f_\Phi$ from the Monte Carlo samples, and extract the $90\%$ lower limit. This procedure is repeated over many axion masses, and the collection of these points produces the limit shown in Fig.~\ref{fig:main}.   Note the multi-bump feature is driven primarily by the contributions of the heavier (left) and lighter (right) BHs (while multi-bumped features do also arise from the presence of different superradiant states for single BHs, the $\left|322\right>$ state of the heavy BH is sub-dominant to the $\left|211\right>$ state of the light BH at higher axion masses).  For reference, we also show in Fig.~\ref{fig:main} superradiance constraints (gray) derived using x-ray observations~\cite{Witte:2024drg}, the values of $f_\Phi-\mu$ corresponding to the QCD axion (red), and the parameter space where the misalignment mechanism can most naturally generate the dark matter abundance (green). Note that the limit derived in Fig.~\ref{fig:main} contains sensitivity to {\emph{both}} black holes (see Supplemental Material). The various features seen in the limit arise from the interplay of having BHs of different masses, and looking at the spin-down induced by the $m=1$ and $m=2$ states.

This analysis demonstrates that GW observations of high-spin BHs can be used to set competitive limits on axions (and, more generally, light bosonic fields) in unexplored regions of the parameter space. The dominant uncertainty entering this analysis comes from the unknown timescale between successive mergers of the constituent BHs -- this is in contrast to approaches using x-ray observations, where the dominant systematic arguably comes from the modeling of the accreting material near the horizon. Using a data-driven approach, we have argued that this uncertainty is not prohibitive, and that GW231123 can be used to derive meaningful constraints on feebly coupled axions.

The upcoming high-sensitivity LVK O5 run \cite{KAGRA:2013rdx} promises to significantly enhance discovery potential for similar events. Thanks to an expected signal-to-noise ratio improvement of ${\cal O}(3)$~\cite{Barsotti:18}, the observable volume will expand by nearly ${\cal O}(20)$. The detection of additional events of this kind in future LVK runs would be accompanied by narrower uncertainties on event parameters and will allow for improved constraints on the underlying spin distribution, along with other source population properties, thereby strengthening the bounds discussed in this work.

\section{Acknowledgments}%

The authors would like to thank Giovanni Maria Tomaselli, Paolo Pani, Enrico Barausse, and Davide Gerosa for their useful comments.
AC is supported by an ERC STG grant (``AstroDarkLS'', grant No. 101117510). SJW acknowledges support from a Royal Society University Research Fellowship (URF-R1-231065). This work is also
supported by the Deutsche Forschungsgemeinschaft under Germany's Excellence Strategy EXC 2121 ``Quantum Universe" 390833306. This article/publication is based upon work from COST Action COSMIC WISPers CA21106, supported by COST (European Cooperation in Science and Technology). The data that support the findings of this article are openly available~\cite{WitteGit}.
\\
\\


\bibliography{biblio}

\begin{thebibliography}{178}%
\makeatletter
\providecommand \@ifxundefined [1]{%
 \@ifx{#1\undefined}
}%
\providecommand \@ifnum [1]{%
 \ifnum #1\expandafter \@firstoftwo
 \else \expandafter \@secondoftwo
 \fi
}%
\providecommand \@ifx [1]{%
 \ifx #1\expandafter \@firstoftwo
 \else \expandafter \@secondoftwo
 \fi
}%
\providecommand \natexlab [1]{#1}%
\providecommand \enquote  [1]{``#1''}%
\providecommand \bibnamefont  [1]{#1}%
\providecommand \bibfnamefont [1]{#1}%
\providecommand \citenamefont [1]{#1}%
\providecommand \href@noop [0]{\@secondoftwo}%
\providecommand \href [0]{\begingroup \@sanitize@url \@href}%
\providecommand \@href[1]{\@@startlink{#1}\@@href}%
\providecommand \@@href[1]{\endgroup#1\@@endlink}%
\providecommand \@sanitize@url [0]{\catcode `\\12\catcode `\$12\catcode
  `\&12\catcode `\#12\catcode `\^12\catcode `\_12\catcode `\%12\relax}%
\providecommand \@@startlink[1]{}%
\providecommand \@@endlink[0]{}%
\providecommand \url  [0]{\begingroup\@sanitize@url \@url }%
\providecommand \@url [1]{\endgroup\@href {#1}{\urlprefix }}%
\providecommand \urlprefix  [0]{URL }%
\providecommand \Eprint [0]{\href }%
\providecommand \doibase [0]{https://doi.org/}%
\providecommand \selectlanguage [0]{\@gobble}%
\providecommand \bibinfo  [0]{\@secondoftwo}%
\providecommand \bibfield  [0]{\@secondoftwo}%
\providecommand \translation [1]{[#1]}%
\providecommand \BibitemOpen [0]{}%
\providecommand \bibitemStop [0]{}%
\providecommand \bibitemNoStop [0]{.\EOS\space}%
\providecommand \EOS [0]{\spacefactor3000\relax}%
\providecommand \BibitemShut  [1]{\csname bibitem#1\endcsname}%
\let\auto@bib@innerbib\@empty
\bibitem [{LIG(2025)}]{LIGOScientific:2025rsn}%
  \BibitemOpen
  \bibfield  {title} {\bibinfo {title} {{GW231123: a Binary Black Hole Merger
  with Total Mass 190-265 $M_{\odot}$}},\ }\href@noop {} {\  (\bibinfo {year}
  {2025})},\ \Eprint {https://arxiv.org/abs/2507.08219} {arXiv:2507.08219
  [astro-ph.HE]} \BibitemShut {NoStop}%
\bibitem [{\citenamefont {{Fowler}}\ and\ \citenamefont
  {{Hoyle}}(1964)}]{1964ApJS....9..201F}%
  \BibitemOpen
  \bibfield  {author} {\bibinfo {author} {\bibfnamefont {W.~A.}\ \bibnamefont
  {{Fowler}}}\ and\ \bibinfo {author} {\bibfnamefont {F.}~\bibnamefont
  {{Hoyle}}},\ }\bibfield  {title} {\bibinfo {title} {{Neutrino Processes and
  Pair Formation in Massive Stars and Supernovae.}},\ }\href
  {https://doi.org/10.1086/190103} {\bibfield  {journal} {\bibinfo  {journal}
  {The Astrophysical Journal Supplement}\ }\textbf {\bibinfo {volume} {9}},\
  \bibinfo {pages} {201} (\bibinfo {year} {1964})}\BibitemShut {NoStop}%
\bibitem [{\citenamefont {{Barkat}}\ \emph {et~al.}(1967)\citenamefont
  {{Barkat}}, \citenamefont {{Rakavy}},\ and\ \citenamefont
  {{Sack}}}]{1967PhRvL..18..379B}%
  \BibitemOpen
  \bibfield  {author} {\bibinfo {author} {\bibfnamefont {Z.}~\bibnamefont
  {{Barkat}}}, \bibinfo {author} {\bibfnamefont {G.}~\bibnamefont {{Rakavy}}},\
  and\ \bibinfo {author} {\bibfnamefont {N.}~\bibnamefont {{Sack}}},\
  }\bibfield  {title} {\bibinfo {title} {{Dynamics of Supernova Explosion
  Resulting from Pair Formation}},\ }\href
  {https://doi.org/10.1103/PhysRevLett.18.379} {\bibfield  {journal} {\bibinfo
  {journal} {\prl}\ }\textbf {\bibinfo {volume} {18}},\ \bibinfo {pages} {379}
  (\bibinfo {year} {1967})}\BibitemShut {NoStop}%
\bibitem [{\citenamefont {Woosley}\ and\ \citenamefont
  {Heger}(2021)}]{Woosley:2021xba}%
  \BibitemOpen
  \bibfield  {author} {\bibinfo {author} {\bibfnamefont {S.~E.}\ \bibnamefont
  {Woosley}}\ and\ \bibinfo {author} {\bibfnamefont {A.}~\bibnamefont
  {Heger}},\ }\bibfield  {title} {\bibinfo {title} {{The Pair-Instability Mass
  Gap for Black Holes}},\ }\href {https://doi.org/10.3847/2041-8213/abf2c4}
  {\bibfield  {journal} {\bibinfo  {journal} {Astrophys. J. Lett.}\ }\textbf
  {\bibinfo {volume} {912}},\ \bibinfo {pages} {L31} (\bibinfo {year}
  {2021})},\ \Eprint {https://arxiv.org/abs/2103.07933} {arXiv:2103.07933
  [astro-ph.SR]} \BibitemShut {NoStop}%
\bibitem [{\citenamefont {Abbott}\ \emph {et~al.}(2023)\citenamefont {Abbott}
  \emph {et~al.}}]{KAGRA:2021vkt}%
  \BibitemOpen
  \bibfield  {author} {\bibinfo {author} {\bibfnamefont {R.}~\bibnamefont
  {Abbott}} \emph {et~al.} (\bibinfo {collaboration} {KAGRA, VIRGO, LIGO
  Scientific}),\ }\bibfield  {title} {\bibinfo {title} {{GWTC-3: Compact Binary
  Coalescences Observed by LIGO and Virgo during the Second Part of the Third
  Observing Run}},\ }\href {https://doi.org/10.1103/PhysRevX.13.041039}
  {\bibfield  {journal} {\bibinfo  {journal} {Phys. Rev. X}\ }\textbf {\bibinfo
  {volume} {13}},\ \bibinfo {pages} {041039} (\bibinfo {year} {2023})},\
  \Eprint {https://arxiv.org/abs/2111.03606} {arXiv:2111.03606 [gr-qc]}
  \BibitemShut {NoStop}%
\bibitem [{\citenamefont {Wadekar}\ \emph {et~al.}(2023)\citenamefont
  {Wadekar}, \citenamefont {Roulet}, \citenamefont {Venumadhav}, \citenamefont
  {Mehta}, \citenamefont {Zackay}, \citenamefont {Mushkin}, \citenamefont
  {Olsen},\ and\ \citenamefont {Zaldarriaga}}]{Wadekar:2023gea}%
  \BibitemOpen
  \bibfield  {author} {\bibinfo {author} {\bibfnamefont {D.}~\bibnamefont
  {Wadekar}}, \bibinfo {author} {\bibfnamefont {J.}~\bibnamefont {Roulet}},
  \bibinfo {author} {\bibfnamefont {T.}~\bibnamefont {Venumadhav}}, \bibinfo
  {author} {\bibfnamefont {A.~K.}\ \bibnamefont {Mehta}}, \bibinfo {author}
  {\bibfnamefont {B.}~\bibnamefont {Zackay}}, \bibinfo {author} {\bibfnamefont
  {J.}~\bibnamefont {Mushkin}}, \bibinfo {author} {\bibfnamefont
  {S.}~\bibnamefont {Olsen}},\ and\ \bibinfo {author} {\bibfnamefont
  {M.}~\bibnamefont {Zaldarriaga}},\ }\bibfield  {title} {\bibinfo {title}
  {{New black hole mergers in the LIGO-Virgo O3 data from a gravitational wave
  search including higher-order harmonics}},\ }\href@noop {} {\  (\bibinfo
  {year} {2023})},\ \Eprint {https://arxiv.org/abs/2312.06631}
  {arXiv:2312.06631 [gr-qc]} \BibitemShut {NoStop}%
\bibitem [{\citenamefont {Schmidt}\ \emph {et~al.}(2015)\citenamefont
  {Schmidt}, \citenamefont {Ohme},\ and\ \citenamefont
  {Hannam}}]{Schmidt:2014iyl}%
  \BibitemOpen
  \bibfield  {author} {\bibinfo {author} {\bibfnamefont {P.}~\bibnamefont
  {Schmidt}}, \bibinfo {author} {\bibfnamefont {F.}~\bibnamefont {Ohme}},\ and\
  \bibinfo {author} {\bibfnamefont {M.}~\bibnamefont {Hannam}},\ }\bibfield
  {title} {\bibinfo {title} {{Towards models of gravitational waveforms from
  generic binaries II: Modelling precession effects with a single effective
  precession parameter}},\ }\href {https://doi.org/10.1103/PhysRevD.91.024043}
  {\bibfield  {journal} {\bibinfo  {journal} {Phys. Rev. D}\ }\textbf {\bibinfo
  {volume} {91}},\ \bibinfo {pages} {024043} (\bibinfo {year} {2015})},\
  \Eprint {https://arxiv.org/abs/1408.1810} {arXiv:1408.1810 [gr-qc]}
  \BibitemShut {NoStop}%
\bibitem [{\citenamefont {Witte}\ and\ \citenamefont
  {Mummery}(2025)}]{Witte:2024drg}%
  \BibitemOpen
  \bibfield  {author} {\bibinfo {author} {\bibfnamefont {S.~J.}\ \bibnamefont
  {Witte}}\ and\ \bibinfo {author} {\bibfnamefont {A.}~\bibnamefont
  {Mummery}},\ }\bibfield  {title} {\bibinfo {title} {{Stepping up
  superradiance constraints on axions}},\ }\href
  {https://doi.org/10.1103/PhysRevD.111.083044} {\bibfield  {journal} {\bibinfo
   {journal} {Phys. Rev. D}\ }\textbf {\bibinfo {volume} {111}},\ \bibinfo
  {pages} {083044} (\bibinfo {year} {2025})},\ \Eprint
  {https://arxiv.org/abs/2412.03655} {arXiv:2412.03655 [hep-ph]} \BibitemShut
  {NoStop}%
\bibitem [{\citenamefont {Brito}\ \emph
  {et~al.}(2015{\natexlab{a}})\citenamefont {Brito}, \citenamefont {Cardoso},\
  and\ \citenamefont {Pani}}]{Brito:2015oca}%
  \BibitemOpen
  \bibfield  {author} {\bibinfo {author} {\bibfnamefont {R.}~\bibnamefont
  {Brito}}, \bibinfo {author} {\bibfnamefont {V.}~\bibnamefont {Cardoso}},\
  and\ \bibinfo {author} {\bibfnamefont {P.}~\bibnamefont {Pani}},\ }\bibfield
  {title} {\bibinfo {title} {{Superradiance}: {New Frontiers in Black Hole
  Physics}},\ }\href {https://doi.org/10.1007/978-3-319-19000-6} {\bibfield
  {journal} {\bibinfo  {journal} {Lect. Notes Phys.}\ }\textbf {\bibinfo
  {volume} {906}},\ \bibinfo {pages} {pp.1} (\bibinfo {year}
  {2015}{\natexlab{a}})},\ \Eprint {https://arxiv.org/abs/1501.06570}
  {arXiv:1501.06570 [gr-qc]} \BibitemShut {NoStop}%
\bibitem [{\citenamefont {Detweiler}(1980)}]{Detweiler:1980uk}%
  \BibitemOpen
  \bibfield  {author} {\bibinfo {author} {\bibfnamefont {S.~L.}\ \bibnamefont
  {Detweiler}},\ }\bibfield  {title} {\bibinfo {title} {{Klein-Gordon Equation
  and Rotating Black Holes}},\ }\href
  {https://doi.org/10.1103/PhysRevD.22.2323} {\bibfield  {journal} {\bibinfo
  {journal} {Phys. Rev. D}\ }\textbf {\bibinfo {volume} {22}},\ \bibinfo
  {pages} {2323} (\bibinfo {year} {1980})}\BibitemShut {NoStop}%
\bibitem [{\citenamefont {Dolan}(2007)}]{Dolan:2007mj}%
  \BibitemOpen
  \bibfield  {author} {\bibinfo {author} {\bibfnamefont {S.~R.}\ \bibnamefont
  {Dolan}},\ }\bibfield  {title} {\bibinfo {title} {{Instability of the massive
  Klein-Gordon field on the Kerr spacetime}},\ }\href
  {https://doi.org/10.1103/PhysRevD.76.084001} {\bibfield  {journal} {\bibinfo
  {journal} {Phys. Rev. D}\ }\textbf {\bibinfo {volume} {76}},\ \bibinfo
  {pages} {084001} (\bibinfo {year} {2007})},\ \Eprint
  {https://arxiv.org/abs/0705.2880} {arXiv:0705.2880 [gr-qc]} \BibitemShut
  {NoStop}%
\bibitem [{\citenamefont {East}\ and\ \citenamefont
  {Pretorius}(2017)}]{East:2017ovw}%
  \BibitemOpen
  \bibfield  {author} {\bibinfo {author} {\bibfnamefont {W.~E.}\ \bibnamefont
  {East}}\ and\ \bibinfo {author} {\bibfnamefont {F.}~\bibnamefont
  {Pretorius}},\ }\bibfield  {title} {\bibinfo {title} {{Superradiant
  Instability and Backreaction of Massive Vector Fields around Kerr Black
  Holes}},\ }\href {https://doi.org/10.1103/PhysRevLett.119.041101} {\bibfield
  {journal} {\bibinfo  {journal} {Phys. Rev. Lett.}\ }\textbf {\bibinfo
  {volume} {119}},\ \bibinfo {pages} {041101} (\bibinfo {year} {2017})},\
  \Eprint {https://arxiv.org/abs/1704.04791} {arXiv:1704.04791 [gr-qc]}
  \BibitemShut {NoStop}%
\bibitem [{\citenamefont {Baryakhtar}\ \emph {et~al.}(2017)\citenamefont
  {Baryakhtar}, \citenamefont {Lasenby},\ and\ \citenamefont
  {Teo}}]{Baryakhtar:2017ngi}%
  \BibitemOpen
  \bibfield  {author} {\bibinfo {author} {\bibfnamefont {M.}~\bibnamefont
  {Baryakhtar}}, \bibinfo {author} {\bibfnamefont {R.}~\bibnamefont
  {Lasenby}},\ and\ \bibinfo {author} {\bibfnamefont {M.}~\bibnamefont {Teo}},\
  }\bibfield  {title} {\bibinfo {title} {{Black Hole Superradiance Signatures
  of Ultralight Vectors}},\ }\href {https://doi.org/10.1103/PhysRevD.96.035019}
  {\bibfield  {journal} {\bibinfo  {journal} {Phys. Rev. D}\ }\textbf {\bibinfo
  {volume} {96}},\ \bibinfo {pages} {035019} (\bibinfo {year} {2017})},\
  \Eprint {https://arxiv.org/abs/1704.05081} {arXiv:1704.05081 [hep-ph]}
  \BibitemShut {NoStop}%
\bibitem [{\citenamefont {Baumann}\ \emph
  {et~al.}(2019{\natexlab{a}})\citenamefont {Baumann}, \citenamefont {Chia},
  \citenamefont {Stout},\ and\ \citenamefont {ter Haar}}]{Baumann:2019eav}%
  \BibitemOpen
  \bibfield  {author} {\bibinfo {author} {\bibfnamefont {D.}~\bibnamefont
  {Baumann}}, \bibinfo {author} {\bibfnamefont {H.~S.}\ \bibnamefont {Chia}},
  \bibinfo {author} {\bibfnamefont {J.}~\bibnamefont {Stout}},\ and\ \bibinfo
  {author} {\bibfnamefont {L.}~\bibnamefont {ter Haar}},\ }\bibfield  {title}
  {\bibinfo {title} {{The Spectra of Gravitational Atoms}},\ }\href
  {https://doi.org/10.1088/1475-7516/2019/12/006} {\bibfield  {journal}
  {\bibinfo  {journal} {JCAP}\ }\textbf {\bibinfo {volume} {12}},\ \bibinfo
  {pages} {006}},\ \Eprint {https://arxiv.org/abs/1908.10370} {arXiv:1908.10370
  [gr-qc]} \BibitemShut {NoStop}%
\bibitem [{\citenamefont {Brito}\ \emph
  {et~al.}(2020{\natexlab{a}})\citenamefont {Brito}, \citenamefont {Grillo},\
  and\ \citenamefont {Pani}}]{Brito:2020lup}%
  \BibitemOpen
  \bibfield  {author} {\bibinfo {author} {\bibfnamefont {R.}~\bibnamefont
  {Brito}}, \bibinfo {author} {\bibfnamefont {S.}~\bibnamefont {Grillo}},\ and\
  \bibinfo {author} {\bibfnamefont {P.}~\bibnamefont {Pani}},\ }\bibfield
  {title} {\bibinfo {title} {{Black Hole Superradiant Instability from
  Ultralight Spin-2 Fields}},\ }\href
  {https://doi.org/10.1103/PhysRevLett.124.211101} {\bibfield  {journal}
  {\bibinfo  {journal} {Phys. Rev. Lett.}\ }\textbf {\bibinfo {volume} {124}},\
  \bibinfo {pages} {211101} (\bibinfo {year} {2020}{\natexlab{a}})},\ \Eprint
  {https://arxiv.org/abs/2002.04055} {arXiv:2002.04055 [gr-qc]} \BibitemShut
  {NoStop}%
\bibitem [{\citenamefont {East}\ and\ \citenamefont
  {Siemonsen}(2023)}]{East:2023nsk}%
  \BibitemOpen
  \bibfield  {author} {\bibinfo {author} {\bibfnamefont {W.~E.}\ \bibnamefont
  {East}}\ and\ \bibinfo {author} {\bibfnamefont {N.}~\bibnamefont
  {Siemonsen}},\ }\bibfield  {title} {\bibinfo {title} {{Instability and
  backreaction of massive spin-2 fields around black holes}},\ }\href
  {https://doi.org/10.1103/PhysRevD.108.124048} {\bibfield  {journal} {\bibinfo
   {journal} {Phys. Rev. D}\ }\textbf {\bibinfo {volume} {108}},\ \bibinfo
  {pages} {124048} (\bibinfo {year} {2023})},\ \Eprint
  {https://arxiv.org/abs/2309.05096} {arXiv:2309.05096 [gr-qc]} \BibitemShut
  {NoStop}%
\bibitem [{\citenamefont {Arvanitaki}\ \emph {et~al.}(2010)\citenamefont
  {Arvanitaki}, \citenamefont {Dimopoulos}, \citenamefont {Dubovsky},
  \citenamefont {Kaloper},\ and\ \citenamefont
  {March-Russell}}]{Arvanitaki:2009fg}%
  \BibitemOpen
  \bibfield  {author} {\bibinfo {author} {\bibfnamefont {A.}~\bibnamefont
  {Arvanitaki}}, \bibinfo {author} {\bibfnamefont {S.}~\bibnamefont
  {Dimopoulos}}, \bibinfo {author} {\bibfnamefont {S.}~\bibnamefont
  {Dubovsky}}, \bibinfo {author} {\bibfnamefont {N.}~\bibnamefont {Kaloper}},\
  and\ \bibinfo {author} {\bibfnamefont {J.}~\bibnamefont {March-Russell}},\
  }\bibfield  {title} {\bibinfo {title} {{String Axiverse}},\ }\href
  {https://doi.org/10.1103/PhysRevD.81.123530} {\bibfield  {journal} {\bibinfo
  {journal} {Phys. Rev. D}\ }\textbf {\bibinfo {volume} {81}},\ \bibinfo
  {pages} {123530} (\bibinfo {year} {2010})},\ \Eprint
  {https://arxiv.org/abs/0905.4720} {arXiv:0905.4720 [hep-th]} \BibitemShut
  {NoStop}%
\bibitem [{\citenamefont {Arvanitaki}\ and\ \citenamefont
  {Dubovsky}(2011)}]{Arvanitaki:2010sy}%
  \BibitemOpen
  \bibfield  {author} {\bibinfo {author} {\bibfnamefont {A.}~\bibnamefont
  {Arvanitaki}}\ and\ \bibinfo {author} {\bibfnamefont {S.}~\bibnamefont
  {Dubovsky}},\ }\bibfield  {title} {\bibinfo {title} {{Exploring the String
  Axiverse with Precision Black Hole Physics}},\ }\href
  {https://doi.org/10.1103/PhysRevD.83.044026} {\bibfield  {journal} {\bibinfo
  {journal} {Phys. Rev. D}\ }\textbf {\bibinfo {volume} {83}},\ \bibinfo
  {pages} {044026} (\bibinfo {year} {2011})},\ \Eprint
  {https://arxiv.org/abs/1004.3558} {arXiv:1004.3558 [hep-th]} \BibitemShut
  {NoStop}%
\bibitem [{\citenamefont {Cardoso}\ \emph {et~al.}(2018)\citenamefont
  {Cardoso}, \citenamefont {Dias}, \citenamefont {Hartnett}, \citenamefont
  {Middleton}, \citenamefont {Pani},\ and\ \citenamefont
  {Santos}}]{cardoso2018constraining}%
  \BibitemOpen
  \bibfield  {author} {\bibinfo {author} {\bibfnamefont {V.}~\bibnamefont
  {Cardoso}}, \bibinfo {author} {\bibfnamefont {{\'O}.~J.}\ \bibnamefont
  {Dias}}, \bibinfo {author} {\bibfnamefont {G.~S.}\ \bibnamefont {Hartnett}},
  \bibinfo {author} {\bibfnamefont {M.}~\bibnamefont {Middleton}}, \bibinfo
  {author} {\bibfnamefont {P.}~\bibnamefont {Pani}},\ and\ \bibinfo {author}
  {\bibfnamefont {J.~E.}\ \bibnamefont {Santos}},\ }\bibfield  {title}
  {\bibinfo {title} {Constraining the mass of dark photons and axion-like
  particles through black-hole superradiance},\ }\href@noop {} {\bibfield
  {journal} {\bibinfo  {journal} {Journal of Cosmology and Astroparticle
  Physics}\ }\textbf {\bibinfo {volume} {2018}}\bibinfo  {number} { (03)},\
  \bibinfo {pages} {043}}\BibitemShut {NoStop}%
\bibitem [{\citenamefont {Brito}\ \emph
  {et~al.}(2017{\natexlab{a}})\citenamefont {Brito}, \citenamefont {Ghosh},
  \citenamefont {Barausse}, \citenamefont {Berti}, \citenamefont {Cardoso},
  \citenamefont {Dvorkin}, \citenamefont {Klein},\ and\ \citenamefont
  {Pani}}]{Brito:2017wnc}%
  \BibitemOpen
\bibfield  {number} {  }\bibfield  {author} {\bibinfo {author} {\bibfnamefont
  {R.}~\bibnamefont {Brito}}, \bibinfo {author} {\bibfnamefont
  {S.}~\bibnamefont {Ghosh}}, \bibinfo {author} {\bibfnamefont
  {E.}~\bibnamefont {Barausse}}, \bibinfo {author} {\bibfnamefont
  {E.}~\bibnamefont {Berti}}, \bibinfo {author} {\bibfnamefont
  {V.}~\bibnamefont {Cardoso}}, \bibinfo {author} {\bibfnamefont
  {I.}~\bibnamefont {Dvorkin}}, \bibinfo {author} {\bibfnamefont
  {A.}~\bibnamefont {Klein}},\ and\ \bibinfo {author} {\bibfnamefont
  {P.}~\bibnamefont {Pani}},\ }\bibfield  {title} {\bibinfo {title}
  {{Stochastic and resolvable gravitational waves from ultralight bosons}},\
  }\href {https://doi.org/10.1103/PhysRevLett.119.131101} {\bibfield  {journal}
  {\bibinfo  {journal} {Phys. Rev. Lett.}\ }\textbf {\bibinfo {volume} {119}},\
  \bibinfo {pages} {131101} (\bibinfo {year} {2017}{\natexlab{a}})},\ \Eprint
  {https://arxiv.org/abs/1706.05097} {arXiv:1706.05097 [gr-qc]} \BibitemShut
  {NoStop}%
\bibitem [{\citenamefont {Brito}\ \emph
  {et~al.}(2017{\natexlab{b}})\citenamefont {Brito}, \citenamefont {Ghosh},
  \citenamefont {Barausse}, \citenamefont {Berti}, \citenamefont {Cardoso},
  \citenamefont {Dvorkin}, \citenamefont {Klein},\ and\ \citenamefont
  {Pani}}]{Brito:2017zvb}%
  \BibitemOpen
  \bibfield  {author} {\bibinfo {author} {\bibfnamefont {R.}~\bibnamefont
  {Brito}}, \bibinfo {author} {\bibfnamefont {S.}~\bibnamefont {Ghosh}},
  \bibinfo {author} {\bibfnamefont {E.}~\bibnamefont {Barausse}}, \bibinfo
  {author} {\bibfnamefont {E.}~\bibnamefont {Berti}}, \bibinfo {author}
  {\bibfnamefont {V.}~\bibnamefont {Cardoso}}, \bibinfo {author} {\bibfnamefont
  {I.}~\bibnamefont {Dvorkin}}, \bibinfo {author} {\bibfnamefont
  {A.}~\bibnamefont {Klein}},\ and\ \bibinfo {author} {\bibfnamefont
  {P.}~\bibnamefont {Pani}},\ }\bibfield  {title} {\bibinfo {title}
  {{Gravitational wave searches for ultralight bosons with LIGO and LISA}},\
  }\href {https://doi.org/10.1103/PhysRevD.96.064050} {\bibfield  {journal}
  {\bibinfo  {journal} {Phys. Rev. D}\ }\textbf {\bibinfo {volume} {96}},\
  \bibinfo {pages} {064050} (\bibinfo {year} {2017}{\natexlab{b}})},\ \Eprint
  {https://arxiv.org/abs/1706.06311} {arXiv:1706.06311 [gr-qc]} \BibitemShut
  {NoStop}%
\bibitem [{\citenamefont {Stott}(2020)}]{Stott:2020gjj}%
  \BibitemOpen
  \bibfield  {author} {\bibinfo {author} {\bibfnamefont {M.~J.}\ \bibnamefont
  {Stott}},\ }\bibfield  {title} {\bibinfo {title} {{Ultralight Bosonic Field
  Mass Bounds from Astrophysical Black Hole Spin}},\ }\href@noop {} {\
  (\bibinfo {year} {2020})},\ \Eprint {https://arxiv.org/abs/2009.07206}
  {arXiv:2009.07206 [hep-ph]} \BibitemShut {NoStop}%
\bibitem [{\citenamefont {\"Unal}\ \emph {et~al.}(2021)\citenamefont {\"Unal},
  \citenamefont {Pacucci},\ and\ \citenamefont {Loeb}}]{Unal:2020jiy}%
  \BibitemOpen
  \bibfield  {author} {\bibinfo {author} {\bibfnamefont {C.}~\bibnamefont
  {\"Unal}}, \bibinfo {author} {\bibfnamefont {F.}~\bibnamefont {Pacucci}},\
  and\ \bibinfo {author} {\bibfnamefont {A.}~\bibnamefont {Loeb}},\ }\bibfield
  {title} {\bibinfo {title} {{Properties of ultralight bosons from heavy quasar
  spins via superradiance}},\ }\href
  {https://doi.org/10.1088/1475-7516/2021/05/007} {\bibfield  {journal}
  {\bibinfo  {journal} {JCAP}\ }\textbf {\bibinfo {volume} {05}},\ \bibinfo
  {pages} {007}},\ \Eprint {https://arxiv.org/abs/2012.12790} {arXiv:2012.12790
  [hep-ph]} \BibitemShut {NoStop}%
\bibitem [{\citenamefont {Baryakhtar}\ \emph {et~al.}(2021)\citenamefont
  {Baryakhtar}, \citenamefont {Galanis}, \citenamefont {Lasenby},\ and\
  \citenamefont {Simon}}]{Baryakhtar:2020gao}%
  \BibitemOpen
  \bibfield  {author} {\bibinfo {author} {\bibfnamefont {M.}~\bibnamefont
  {Baryakhtar}}, \bibinfo {author} {\bibfnamefont {M.}~\bibnamefont {Galanis}},
  \bibinfo {author} {\bibfnamefont {R.}~\bibnamefont {Lasenby}},\ and\ \bibinfo
  {author} {\bibfnamefont {O.}~\bibnamefont {Simon}},\ }\bibfield  {title}
  {\bibinfo {title} {{Black hole superradiance of self-interacting scalar
  fields}},\ }\href {https://doi.org/10.1103/PhysRevD.103.095019} {\bibfield
  {journal} {\bibinfo  {journal} {Phys. Rev. D}\ }\textbf {\bibinfo {volume}
  {103}},\ \bibinfo {pages} {095019} (\bibinfo {year} {2021})},\ \Eprint
  {https://arxiv.org/abs/2011.11646} {arXiv:2011.11646 [hep-ph]} \BibitemShut
  {NoStop}%
\bibitem [{\citenamefont {Hoof}\ \emph {et~al.}(2024)\citenamefont {Hoof},
  \citenamefont {Marsh}, \citenamefont {Sisk-Reyn\'es}, \citenamefont
  {Matthews},\ and\ \citenamefont {Reynolds}}]{Hoof:2024quk}%
  \BibitemOpen
  \bibfield  {author} {\bibinfo {author} {\bibfnamefont {S.}~\bibnamefont
  {Hoof}}, \bibinfo {author} {\bibfnamefont {D.~J.~E.}\ \bibnamefont {Marsh}},
  \bibinfo {author} {\bibfnamefont {J.}~\bibnamefont {Sisk-Reyn\'es}}, \bibinfo
  {author} {\bibfnamefont {J.~H.}\ \bibnamefont {Matthews}},\ and\ \bibinfo
  {author} {\bibfnamefont {C.}~\bibnamefont {Reynolds}},\ }\bibfield  {title}
  {\bibinfo {title} {{Getting More Out of Black Hole Superradiance: a
  Statistically Rigorous Approach to Ultralight Boson Constraints}},\
  }\href@noop {} {\  (\bibinfo {year} {2024})},\ \Eprint
  {https://arxiv.org/abs/2406.10337} {arXiv:2406.10337 [hep-ph]} \BibitemShut
  {NoStop}%
\bibitem [{\citenamefont {Nemmen}(2019)}]{nemmen2019spin}%
  \BibitemOpen
  \bibfield  {author} {\bibinfo {author} {\bibfnamefont {R.}~\bibnamefont
  {Nemmen}},\ }\bibfield  {title} {\bibinfo {title} {The spin of m87},\
  }\href@noop {} {\bibfield  {journal} {\bibinfo  {journal} {The Astrophysical
  Journal Letters}\ }\textbf {\bibinfo {volume} {880}},\ \bibinfo {pages} {L26}
  (\bibinfo {year} {2019})}\BibitemShut {NoStop}%
\bibitem [{\citenamefont {Akiyama}\ \emph {et~al.}(2019)\citenamefont {Akiyama}
  \emph {et~al.}}]{EventHorizonTelescope:2019pgp}%
  \BibitemOpen
  \bibfield  {author} {\bibinfo {author} {\bibfnamefont {K.}~\bibnamefont
  {Akiyama}} \emph {et~al.} (\bibinfo {collaboration} {Event Horizon
  Telescope}),\ }\bibfield  {title} {\bibinfo {title} {{First M87 Event Horizon
  Telescope Results. V. Physical Origin of the Asymmetric Ring}},\ }\href
  {https://doi.org/10.3847/2041-8213/ab0f43} {\bibfield  {journal} {\bibinfo
  {journal} {Astrophys. J. Lett.}\ }\textbf {\bibinfo {volume} {875}},\
  \bibinfo {pages} {L5} (\bibinfo {year} {2019})},\ \Eprint
  {https://arxiv.org/abs/1906.11242} {arXiv:1906.11242 [astro-ph.GA]}
  \BibitemShut {NoStop}%
\bibitem [{\citenamefont {Tamburini}\ \emph {et~al.}(2020)\citenamefont
  {Tamburini}, \citenamefont {Thid{\'e}},\ and\ \citenamefont
  {Della~Valle}}]{tamburini2020measurement}%
  \BibitemOpen
  \bibfield  {author} {\bibinfo {author} {\bibfnamefont {F.}~\bibnamefont
  {Tamburini}}, \bibinfo {author} {\bibfnamefont {B.}~\bibnamefont
  {Thid{\'e}}},\ and\ \bibinfo {author} {\bibfnamefont {M.}~\bibnamefont
  {Della~Valle}},\ }\bibfield  {title} {\bibinfo {title} {Measurement of the
  spin of the m87 black hole from its observed twisted light},\ }\href@noop {}
  {\bibfield  {journal} {\bibinfo  {journal} {Monthly Notices of the Royal
  Astronomical Society: Letters}\ }\textbf {\bibinfo {volume} {492}},\ \bibinfo
  {pages} {L22} (\bibinfo {year} {2020})}\BibitemShut {NoStop}%
\bibitem [{\citenamefont {Cruz-Osorio}\ \emph {et~al.}(2022)\citenamefont
  {Cruz-Osorio}, \citenamefont {Fromm}, \citenamefont {Mizuno}, \citenamefont
  {Nathanail}, \citenamefont {Younsi}, \citenamefont {Porth}, \citenamefont
  {Davelaar}, \citenamefont {Falcke}, \citenamefont {Kramer},\ and\
  \citenamefont {Rezzolla}}]{cruz2022state}%
  \BibitemOpen
  \bibfield  {author} {\bibinfo {author} {\bibfnamefont {A.}~\bibnamefont
  {Cruz-Osorio}}, \bibinfo {author} {\bibfnamefont {C.~M.}\ \bibnamefont
  {Fromm}}, \bibinfo {author} {\bibfnamefont {Y.}~\bibnamefont {Mizuno}},
  \bibinfo {author} {\bibfnamefont {A.}~\bibnamefont {Nathanail}}, \bibinfo
  {author} {\bibfnamefont {Z.}~\bibnamefont {Younsi}}, \bibinfo {author}
  {\bibfnamefont {O.}~\bibnamefont {Porth}}, \bibinfo {author} {\bibfnamefont
  {J.}~\bibnamefont {Davelaar}}, \bibinfo {author} {\bibfnamefont
  {H.}~\bibnamefont {Falcke}}, \bibinfo {author} {\bibfnamefont
  {M.}~\bibnamefont {Kramer}},\ and\ \bibinfo {author} {\bibfnamefont
  {L.}~\bibnamefont {Rezzolla}},\ }\bibfield  {title} {\bibinfo {title}
  {State-of-the-art energetic and morphological modelling of the launching site
  of the m87 jet},\ }\href@noop {} {\bibfield  {journal} {\bibinfo  {journal}
  {Nature Astronomy}\ }\textbf {\bibinfo {volume} {6}},\ \bibinfo {pages} {103}
  (\bibinfo {year} {2022})}\BibitemShut {NoStop}%
\bibitem [{\citenamefont {Abac}\ \emph {et~al.}(2025)\citenamefont {Abac} \emph
  {et~al.}}]{LIGOScientific:2025brd}%
  \BibitemOpen
  \bibfield  {author} {\bibinfo {author} {\bibfnamefont {A.~G.}\ \bibnamefont
  {Abac}} \emph {et~al.} (\bibinfo {collaboration} {LIGO Scientific, Virgo,
  KAGRA}),\ }\bibfield  {title} {\bibinfo {title} {{GW241011 and GW241110:
  Exploring Binary Formation and Fundamental Physics with Asymmetric, High-spin
  Black Hole Coalescences}},\ }\href {https://doi.org/10.3847/2041-8213/ae0d54}
  {\bibfield  {journal} {\bibinfo  {journal} {Astrophys. J. Lett.}\ }\textbf
  {\bibinfo {volume} {993}},\ \bibinfo {pages} {L21} (\bibinfo {year}
  {2025})},\ \Eprint {https://arxiv.org/abs/2510.26931} {arXiv:2510.26931
  [astro-ph.HE]} \BibitemShut {NoStop}%
\bibitem [{\citenamefont {Collaboration}\ \emph {et~al.}(2025)\citenamefont
  {Collaboration}, \citenamefont {Collaboration},\ and\ \citenamefont
  {Collaboration}}]{ligo_scientific_collaboration_2025_16004263}%
  \BibitemOpen
  \bibfield  {author} {\bibinfo {author} {\bibfnamefont {L.~S.}\ \bibnamefont
  {Collaboration}}, \bibinfo {author} {\bibfnamefont {V.}~\bibnamefont
  {Collaboration}},\ and\ \bibinfo {author} {\bibfnamefont {K.}~\bibnamefont
  {Collaboration}},\ }\bibfield  {title} {\bibinfo {title} {Gw231123: a binary
  black hole merger with total mass 190-265 msun},\ }\href
  {https://doi.org/10.5281/zenodo.16004263} {10.5281/zenodo.16004263} (\bibinfo
  {year} {2025})\BibitemShut {NoStop}%
\bibitem [{\citenamefont {Cuceu}\ \emph {et~al.}(2025)\citenamefont {Cuceu},
  \citenamefont {Bizouard}, \citenamefont {Christensen},\ and\ \citenamefont
  {Sakellariadou}}]{Cuceu:2025fzi}%
  \BibitemOpen
  \bibfield  {author} {\bibinfo {author} {\bibfnamefont {I.}~\bibnamefont
  {Cuceu}}, \bibinfo {author} {\bibfnamefont {M.~A.}\ \bibnamefont {Bizouard}},
  \bibinfo {author} {\bibfnamefont {N.}~\bibnamefont {Christensen}},\ and\
  \bibinfo {author} {\bibfnamefont {M.}~\bibnamefont {Sakellariadou}},\
  }\bibfield  {title} {\bibinfo {title} {{GW231123: Binary Black Hole Merger or
  Cosmic String?}},\ }\href@noop {} {\  (\bibinfo {year} {2025})},\ \Eprint
  {https://arxiv.org/abs/2507.20778} {arXiv:2507.20778 [gr-qc]} \BibitemShut
  {NoStop}%
\bibitem [{\citenamefont {Spera}\ and\ \citenamefont
  {Mapelli}(2017)}]{Spera:2017fyx}%
  \BibitemOpen
  \bibfield  {author} {\bibinfo {author} {\bibfnamefont {M.}~\bibnamefont
  {Spera}}\ and\ \bibinfo {author} {\bibfnamefont {M.}~\bibnamefont
  {Mapelli}},\ }\bibfield  {title} {\bibinfo {title} {{Very massive stars,
  pair-instability supernovae and intermediate-mass black holes with the SEVN
  code}},\ }\href {https://doi.org/10.1093/mnras/stx1576} {\bibfield  {journal}
  {\bibinfo  {journal} {Mon. Not. Roy. Astron. Soc.}\ }\textbf {\bibinfo
  {volume} {470}},\ \bibinfo {pages} {4739} (\bibinfo {year} {2017})},\ \Eprint
  {https://arxiv.org/abs/1706.06109} {arXiv:1706.06109 [astro-ph.SR]}
  \BibitemShut {NoStop}%
\bibitem [{\citenamefont {Bond}\ and\ \citenamefont
  {Carr}(1984)}]{10.1093/mnras/207.3.585}%
  \BibitemOpen
  \bibfield  {author} {\bibinfo {author} {\bibfnamefont {J.~R.}\ \bibnamefont
  {Bond}}\ and\ \bibinfo {author} {\bibfnamefont {B.~J.}\ \bibnamefont
  {Carr}},\ }\bibfield  {title} {\bibinfo {title} {Gravitational waves from a
  population of binary black holes},\ }\href
  {https://doi.org/10.1093/mnras/207.3.585} {\bibfield  {journal} {\bibinfo
  {journal} {Monthly Notices of the Royal Astronomical Society}\ }\textbf
  {\bibinfo {volume} {207}},\ \bibinfo {pages} {585} (\bibinfo {year}
  {1984})},\ \Eprint
  {https://arxiv.org/abs/https://academic.oup.com/mnras/article-pdf/207/3/585/18521578/mnras207-0585.pdf}
  {https://academic.oup.com/mnras/article-pdf/207/3/585/18521578/mnras207-0585.pdf}
  \BibitemShut {NoStop}%
\bibitem [{\citenamefont {Bromm}\ and\ \citenamefont
  {Larson}(2004)}]{Bromm:2003vv}%
  \BibitemOpen
  \bibfield  {author} {\bibinfo {author} {\bibfnamefont {V.}~\bibnamefont
  {Bromm}}\ and\ \bibinfo {author} {\bibfnamefont {R.~B.}\ \bibnamefont
  {Larson}},\ }\bibfield  {title} {\bibinfo {title} {{The First stars}},\
  }\href {https://doi.org/10.1146/annurev.astro.42.053102.134034} {\bibfield
  {journal} {\bibinfo  {journal} {Ann. Rev. Astron. Astrophys.}\ }\textbf
  {\bibinfo {volume} {42}},\ \bibinfo {pages} {79} (\bibinfo {year} {2004})},\
  \Eprint {https://arxiv.org/abs/astro-ph/0311019} {arXiv:astro-ph/0311019}
  \BibitemShut {NoStop}%
\bibitem [{\citenamefont {Belczynski}\ \emph {et~al.}(2004)\citenamefont
  {Belczynski}, \citenamefont {Bulik},\ and\ \citenamefont
  {Rudak}}]{Belczynski:2004gu}%
  \BibitemOpen
  \bibfield  {author} {\bibinfo {author} {\bibfnamefont {K.}~\bibnamefont
  {Belczynski}}, \bibinfo {author} {\bibfnamefont {T.}~\bibnamefont {Bulik}},\
  and\ \bibinfo {author} {\bibfnamefont {B.}~\bibnamefont {Rudak}},\ }\bibfield
   {title} {\bibinfo {title} {{First stellar binary black holes: strongest
  gravitational wave burst sources}},\ }\href {https://doi.org/10.1086/422172}
  {\bibfield  {journal} {\bibinfo  {journal} {Astrophys. J. Lett.}\ }\textbf
  {\bibinfo {volume} {608}},\ \bibinfo {pages} {L45} (\bibinfo {year}
  {2004})},\ \Eprint {https://arxiv.org/abs/astro-ph/0403361}
  {arXiv:astro-ph/0403361} \BibitemShut {NoStop}%
\bibitem [{\citenamefont {Tanikawa}\ \emph {et~al.}(2022)\citenamefont
  {Tanikawa}, \citenamefont {Yoshida}, \citenamefont {Kinugawa}, \citenamefont
  {Trani}, \citenamefont {Hosokawa}, \citenamefont {Susa},\ and\ \citenamefont
  {Omukai}}]{Tanikawa:2021qqi}%
  \BibitemOpen
  \bibfield  {author} {\bibinfo {author} {\bibfnamefont {A.}~\bibnamefont
  {Tanikawa}}, \bibinfo {author} {\bibfnamefont {T.}~\bibnamefont {Yoshida}},
  \bibinfo {author} {\bibfnamefont {T.}~\bibnamefont {Kinugawa}}, \bibinfo
  {author} {\bibfnamefont {A.~A.}\ \bibnamefont {Trani}}, \bibinfo {author}
  {\bibfnamefont {T.}~\bibnamefont {Hosokawa}}, \bibinfo {author}
  {\bibfnamefont {H.}~\bibnamefont {Susa}},\ and\ \bibinfo {author}
  {\bibfnamefont {K.}~\bibnamefont {Omukai}},\ }\bibfield  {title} {\bibinfo
  {title} {{Merger Rate Density of Binary Black Holes through Isolated
  Population I, II, III and Extremely Metal-poor Binary Star Evolution}},\
  }\href {https://doi.org/10.3847/1538-4357/ac4247} {\bibfield  {journal}
  {\bibinfo  {journal} {Astrophys. J.}\ }\textbf {\bibinfo {volume} {926}},\
  \bibinfo {pages} {83} (\bibinfo {year} {2022})},\ \Eprint
  {https://arxiv.org/abs/2110.10846} {arXiv:2110.10846 [astro-ph.HE]}
  \BibitemShut {NoStop}%
\bibitem [{\citenamefont {Hijikawa}\ \emph {et~al.}(2021)\citenamefont
  {Hijikawa}, \citenamefont {Tanikawa}, \citenamefont {Kinugawa}, \citenamefont
  {Yoshida},\ and\ \citenamefont {Umeda}}]{Hijikawa:2021hrf}%
  \BibitemOpen
  \bibfield  {author} {\bibinfo {author} {\bibfnamefont {K.}~\bibnamefont
  {Hijikawa}}, \bibinfo {author} {\bibfnamefont {A.}~\bibnamefont {Tanikawa}},
  \bibinfo {author} {\bibfnamefont {T.}~\bibnamefont {Kinugawa}}, \bibinfo
  {author} {\bibfnamefont {T.}~\bibnamefont {Yoshida}},\ and\ \bibinfo {author}
  {\bibfnamefont {H.}~\bibnamefont {Umeda}},\ }\bibfield  {title} {\bibinfo
  {title} {{On the population III binary black hole mergers beyond the
  pair-instability mass gap}},\ }\href {https://doi.org/10.1093/mnrasl/slab052}
  {\bibfield  {journal} {\bibinfo  {journal} {Mon. Not. Roy. Astron. Soc.}\
  }\textbf {\bibinfo {volume} {505}},\ \bibinfo {pages} {L69} (\bibinfo {year}
  {2021})},\ \Eprint {https://arxiv.org/abs/2104.13384} {arXiv:2104.13384
  [astro-ph.HE]} \BibitemShut {NoStop}%
\bibitem [{\citenamefont {Belczynski}\ \emph {et~al.}(2017)\citenamefont
  {Belczynski}, \citenamefont {Ryu}, \citenamefont {Perna}, \citenamefont
  {Berti}, \citenamefont {Tanaka},\ and\ \citenamefont
  {Bulik}}]{Belczynski:2016ieo}%
  \BibitemOpen
  \bibfield  {author} {\bibinfo {author} {\bibfnamefont {K.}~\bibnamefont
  {Belczynski}}, \bibinfo {author} {\bibfnamefont {T.}~\bibnamefont {Ryu}},
  \bibinfo {author} {\bibfnamefont {R.}~\bibnamefont {Perna}}, \bibinfo
  {author} {\bibfnamefont {E.}~\bibnamefont {Berti}}, \bibinfo {author}
  {\bibfnamefont {T.~L.}\ \bibnamefont {Tanaka}},\ and\ \bibinfo {author}
  {\bibfnamefont {T.}~\bibnamefont {Bulik}},\ }\bibfield  {title} {\bibinfo
  {title} {{On the likelihood of detecting gravitational waves from Population
  III compact object binaries}},\ }\href
  {https://doi.org/10.1093/mnras/stx1759} {\bibfield  {journal} {\bibinfo
  {journal} {Mon. Not. Roy. Astron. Soc.}\ }\textbf {\bibinfo {volume} {471}},\
  \bibinfo {pages} {4702} (\bibinfo {year} {2017})},\ \Eprint
  {https://arxiv.org/abs/1612.01524} {arXiv:1612.01524 [astro-ph.HE]}
  \BibitemShut {NoStop}%
\bibitem [{\citenamefont {Liu}\ and\ \citenamefont
  {Bromm}(2020)}]{Liu:2020lmi}%
  \BibitemOpen
  \bibfield  {author} {\bibinfo {author} {\bibfnamefont {B.}~\bibnamefont
  {Liu}}\ and\ \bibinfo {author} {\bibfnamefont {V.}~\bibnamefont {Bromm}},\
  }\bibfield  {title} {\bibinfo {title} {{The Population III origin of
  GW190521}},\ }\href {https://doi.org/10.3847/2041-8213/abc552} {\bibfield
  {journal} {\bibinfo  {journal} {Astrophys. J. Lett.}\ }\textbf {\bibinfo
  {volume} {903}},\ \bibinfo {pages} {L40} (\bibinfo {year} {2020})},\ \Eprint
  {https://arxiv.org/abs/2009.11447} {arXiv:2009.11447 [astro-ph.GA]}
  \BibitemShut {NoStop}%
\bibitem [{\citenamefont {Zel'dovich}\ and\ \citenamefont
  {Novikov}(1967)}]{Zeldovich:1967lct}%
  \BibitemOpen
  \bibfield  {author} {\bibinfo {author} {\bibfnamefont {Y.~B.}\ \bibnamefont
  {Zel'dovich}}\ and\ \bibinfo {author} {\bibfnamefont {I.~D.}\ \bibnamefont
  {Novikov}},\ }\bibfield  {title} {\bibinfo {title} {{The Hypothesis of Cores
  Retarded during Expansion and the Hot Cosmological Model}},\ }\href@noop {}
  {\bibfield  {journal} {\bibinfo  {journal} {Sov. Astron.}\ }\textbf {\bibinfo
  {volume} {10}},\ \bibinfo {pages} {602} (\bibinfo {year} {1967})}\BibitemShut
  {NoStop}%
\bibitem [{\citenamefont {Hawking}(1971)}]{Hawking:1971ei}%
  \BibitemOpen
  \bibfield  {author} {\bibinfo {author} {\bibfnamefont {S.}~\bibnamefont
  {Hawking}},\ }\bibfield  {title} {\bibinfo {title} {{Gravitationally
  collapsed objects of very low mass}},\ }\href
  {https://doi.org/10.1093/mnras/152.1.75} {\bibfield  {journal} {\bibinfo
  {journal} {Mon. Not. Roy. Astron. Soc.}\ }\textbf {\bibinfo {volume} {152}},\
  \bibinfo {pages} {75} (\bibinfo {year} {1971})}\BibitemShut {NoStop}%
\bibitem [{\citenamefont {Carr}\ and\ \citenamefont
  {Hawking}(1974)}]{Carr:1974nx}%
  \BibitemOpen
  \bibfield  {author} {\bibinfo {author} {\bibfnamefont {B.~J.}\ \bibnamefont
  {Carr}}\ and\ \bibinfo {author} {\bibfnamefont {S.~W.}\ \bibnamefont
  {Hawking}},\ }\bibfield  {title} {\bibinfo {title} {{Black holes in the early
  Universe}},\ }\href {https://doi.org/10.1093/mnras/168.2.399} {\bibfield
  {journal} {\bibinfo  {journal} {Mon. Not. Roy. Astron. Soc.}\ }\textbf
  {\bibinfo {volume} {168}},\ \bibinfo {pages} {399} (\bibinfo {year}
  {1974})}\BibitemShut {NoStop}%
\bibitem [{\citenamefont {Carr}(1975)}]{Carr:1975qj}%
  \BibitemOpen
  \bibfield  {author} {\bibinfo {author} {\bibfnamefont {B.~J.}\ \bibnamefont
  {Carr}},\ }\bibfield  {title} {\bibinfo {title} {{The Primordial black hole
  mass spectrum}},\ }\href {https://doi.org/10.1086/153853} {\bibfield
  {journal} {\bibinfo  {journal} {Astrophys. J.}\ }\textbf {\bibinfo {volume}
  {201}},\ \bibinfo {pages} {1} (\bibinfo {year} {1975})}\BibitemShut {NoStop}%
\bibitem [{\citenamefont {Bird}\ \emph {et~al.}(2016)\citenamefont {Bird},
  \citenamefont {Cholis}, \citenamefont {Mu{\~n}oz}, \citenamefont
  {Ali-Ha{\"\i}moud}, \citenamefont {Kamionkowski}, \citenamefont {Kovetz},
  \citenamefont {Raccanelli},\ and\ \citenamefont {Riess}}]{Bird:2016dcv}%
  \BibitemOpen
  \bibfield  {author} {\bibinfo {author} {\bibfnamefont {S.}~\bibnamefont
  {Bird}}, \bibinfo {author} {\bibfnamefont {I.}~\bibnamefont {Cholis}},
  \bibinfo {author} {\bibfnamefont {J.~B.}\ \bibnamefont {Mu{\~n}oz}}, \bibinfo
  {author} {\bibfnamefont {Y.}~\bibnamefont {Ali-Ha{\"\i}moud}}, \bibinfo
  {author} {\bibfnamefont {M.}~\bibnamefont {Kamionkowski}}, \bibinfo {author}
  {\bibfnamefont {E.~D.}\ \bibnamefont {Kovetz}}, \bibinfo {author}
  {\bibfnamefont {A.}~\bibnamefont {Raccanelli}},\ and\ \bibinfo {author}
  {\bibfnamefont {A.~G.}\ \bibnamefont {Riess}},\ }\bibfield  {title} {\bibinfo
  {title} {{Did LIGO detect dark matter?}},\ }\href
  {https://doi.org/10.1103/PhysRevLett.116.201301} {\bibfield  {journal}
  {\bibinfo  {journal} {Phys. Rev. Lett.}\ }\textbf {\bibinfo {volume} {116}},\
  \bibinfo {pages} {201301} (\bibinfo {year} {2016})},\ \Eprint
  {https://arxiv.org/abs/1603.00464} {arXiv:1603.00464 [astro-ph.CO]}
  \BibitemShut {NoStop}%
\bibitem [{\citenamefont {Clesse}\ and\ \citenamefont
  {Garc{\'\i}a-Bellido}(2017)}]{Clesse:2016vqa}%
  \BibitemOpen
  \bibfield  {author} {\bibinfo {author} {\bibfnamefont {S.}~\bibnamefont
  {Clesse}}\ and\ \bibinfo {author} {\bibfnamefont {J.}~\bibnamefont
  {Garc{\'\i}a-Bellido}},\ }\bibfield  {title} {\bibinfo {title} {{The
  clustering of massive Primordial Black Holes as Dark Matter: measuring their
  mass distribution with Advanced LIGO}},\ }\href
  {https://doi.org/10.1016/j.dark.2016.10.002} {\bibfield  {journal} {\bibinfo
  {journal} {Phys. Dark Univ.}\ }\textbf {\bibinfo {volume} {15}},\ \bibinfo
  {pages} {142} (\bibinfo {year} {2017})},\ \Eprint
  {https://arxiv.org/abs/1603.05234} {arXiv:1603.05234 [astro-ph.CO]}
  \BibitemShut {NoStop}%
\bibitem [{\citenamefont {Sasaki}\ \emph {et~al.}(2016)\citenamefont {Sasaki},
  \citenamefont {Suyama}, \citenamefont {Tanaka},\ and\ \citenamefont
  {Yokoyama}}]{Sasaki:2016jop}%
  \BibitemOpen
  \bibfield  {author} {\bibinfo {author} {\bibfnamefont {M.}~\bibnamefont
  {Sasaki}}, \bibinfo {author} {\bibfnamefont {T.}~\bibnamefont {Suyama}},
  \bibinfo {author} {\bibfnamefont {T.}~\bibnamefont {Tanaka}},\ and\ \bibinfo
  {author} {\bibfnamefont {S.}~\bibnamefont {Yokoyama}},\ }\bibfield  {title}
  {\bibinfo {title} {{Primordial Black Hole Scenario for the Gravitational-Wave
  Event GW150914}},\ }\href {https://doi.org/10.1103/PhysRevLett.117.061101}
  {\bibfield  {journal} {\bibinfo  {journal} {Phys. Rev. Lett.}\ }\textbf
  {\bibinfo {volume} {117}},\ \bibinfo {pages} {061101} (\bibinfo {year}
  {2016})},\ \bibinfo {note} {[Erratum: Phys.Rev.Lett. 121, 059901 (2018)]},\
  \Eprint {https://arxiv.org/abs/1603.08338} {arXiv:1603.08338 [astro-ph.CO]}
  \BibitemShut {NoStop}%
\bibitem [{\citenamefont {Eroshenko}(2018)}]{Eroshenko:2016hmn}%
  \BibitemOpen
  \bibfield  {author} {\bibinfo {author} {\bibfnamefont {Y.~N.}\ \bibnamefont
  {Eroshenko}},\ }\bibfield  {title} {\bibinfo {title} {{Gravitational waves
  from primordial black holes collisions in binary systems}},\ }\href
  {https://doi.org/10.1088/1742-6596/1051/1/012010} {\bibfield  {journal}
  {\bibinfo  {journal} {J. Phys. Conf. Ser.}\ }\textbf {\bibinfo {volume}
  {1051}},\ \bibinfo {pages} {012010} (\bibinfo {year} {2018})},\ \Eprint
  {https://arxiv.org/abs/1604.04932} {arXiv:1604.04932 [astro-ph.CO]}
  \BibitemShut {NoStop}%
\bibitem [{\citenamefont {Wang}\ \emph {et~al.}(2018)\citenamefont {Wang},
  \citenamefont {Wang}, \citenamefont {Huang},\ and\ \citenamefont
  {Li}}]{Wang:2016ana}%
  \BibitemOpen
  \bibfield  {author} {\bibinfo {author} {\bibfnamefont {S.}~\bibnamefont
  {Wang}}, \bibinfo {author} {\bibfnamefont {Y.-F.}\ \bibnamefont {Wang}},
  \bibinfo {author} {\bibfnamefont {Q.-G.}\ \bibnamefont {Huang}},\ and\
  \bibinfo {author} {\bibfnamefont {T.~G.~F.}\ \bibnamefont {Li}},\ }\bibfield
  {title} {\bibinfo {title} {{Constraints on the Primordial Black Hole
  Abundance from the First Advanced LIGO Observation Run Using the Stochastic
  Gravitational-Wave Background}},\ }\href
  {https://doi.org/10.1103/PhysRevLett.120.191102} {\bibfield  {journal}
  {\bibinfo  {journal} {Phys. Rev. Lett.}\ }\textbf {\bibinfo {volume} {120}},\
  \bibinfo {pages} {191102} (\bibinfo {year} {2018})},\ \Eprint
  {https://arxiv.org/abs/1610.08725} {arXiv:1610.08725 [astro-ph.CO]}
  \BibitemShut {NoStop}%
\bibitem [{\citenamefont {Clesse}\ and\ \citenamefont
  {Garcia-Bellido}(2022)}]{Clesse:2020ghq}%
  \BibitemOpen
  \bibfield  {author} {\bibinfo {author} {\bibfnamefont {S.}~\bibnamefont
  {Clesse}}\ and\ \bibinfo {author} {\bibfnamefont {J.}~\bibnamefont
  {Garcia-Bellido}},\ }\bibfield  {title} {\bibinfo {title} {{GW190425,
  GW190521 and GW190814: Three candidate mergers of primordial black holes from
  the QCD epoch}},\ }\href {https://doi.org/10.1016/j.dark.2022.101111}
  {\bibfield  {journal} {\bibinfo  {journal} {Phys. Dark Univ.}\ }\textbf
  {\bibinfo {volume} {38}},\ \bibinfo {pages} {101111} (\bibinfo {year}
  {2022})},\ \Eprint {https://arxiv.org/abs/2007.06481} {arXiv:2007.06481
  [astro-ph.CO]} \BibitemShut {NoStop}%
\bibitem [{\citenamefont {Hall}\ \emph {et~al.}(2020)\citenamefont {Hall},
  \citenamefont {Gow},\ and\ \citenamefont {Byrnes}}]{Hall:2020daa}%
  \BibitemOpen
  \bibfield  {author} {\bibinfo {author} {\bibfnamefont {A.}~\bibnamefont
  {Hall}}, \bibinfo {author} {\bibfnamefont {A.~D.}\ \bibnamefont {Gow}},\ and\
  \bibinfo {author} {\bibfnamefont {C.~T.}\ \bibnamefont {Byrnes}},\ }\bibfield
   {title} {\bibinfo {title} {{Bayesian analysis of LIGO-Virgo mergers:
  Primordial vs. astrophysical black hole populations}},\ }\href
  {https://doi.org/10.1103/PhysRevD.102.123524} {\bibfield  {journal} {\bibinfo
   {journal} {Phys. Rev. D}\ }\textbf {\bibinfo {volume} {102}},\ \bibinfo
  {pages} {123524} (\bibinfo {year} {2020})},\ \Eprint
  {https://arxiv.org/abs/2008.13704} {arXiv:2008.13704 [astro-ph.CO]}
  \BibitemShut {NoStop}%
\bibitem [{\citenamefont {Franciolini}\ \emph
  {et~al.}(2022{\natexlab{a}})\citenamefont {Franciolini}, \citenamefont
  {Musco}, \citenamefont {Pani},\ and\ \citenamefont
  {Urbano}}]{Franciolini:2022tfm}%
  \BibitemOpen
  \bibfield  {author} {\bibinfo {author} {\bibfnamefont {G.}~\bibnamefont
  {Franciolini}}, \bibinfo {author} {\bibfnamefont {I.}~\bibnamefont {Musco}},
  \bibinfo {author} {\bibfnamefont {P.}~\bibnamefont {Pani}},\ and\ \bibinfo
  {author} {\bibfnamefont {A.}~\bibnamefont {Urbano}},\ }\bibfield  {title}
  {\bibinfo {title} {{From inflation to black hole mergers and back again:
  Gravitational-wave data-driven constraints on inflationary scenarios with a
  first-principle model of primordial black holes across the QCD epoch}},\
  }\href {https://doi.org/10.1103/PhysRevD.106.123526} {\bibfield  {journal}
  {\bibinfo  {journal} {Phys. Rev. D}\ }\textbf {\bibinfo {volume} {106}},\
  \bibinfo {pages} {123526} (\bibinfo {year} {2022}{\natexlab{a}})},\ \Eprint
  {https://arxiv.org/abs/2209.05959} {arXiv:2209.05959 [astro-ph.CO]}
  \BibitemShut {NoStop}%
\bibitem [{\citenamefont {Escriv{\`a}}\ \emph {et~al.}(2023)\citenamefont
  {Escriv{\`a}}, \citenamefont {Bagui},\ and\ \citenamefont
  {Clesse}}]{Escriva:2022bwe}%
  \BibitemOpen
  \bibfield  {author} {\bibinfo {author} {\bibfnamefont {A.}~\bibnamefont
  {Escriv{\`a}}}, \bibinfo {author} {\bibfnamefont {E.}~\bibnamefont {Bagui}},\
  and\ \bibinfo {author} {\bibfnamefont {S.}~\bibnamefont {Clesse}},\
  }\bibfield  {title} {\bibinfo {title} {{Simulations of PBH formation at the
  QCD epoch and comparison with the GWTC-3 catalog}},\ }\href
  {https://doi.org/10.1088/1475-7516/2023/05/004} {\bibfield  {journal}
  {\bibinfo  {journal} {JCAP}\ }\textbf {\bibinfo {volume} {05}},\ \bibinfo
  {pages} {004}},\ \Eprint {https://arxiv.org/abs/2209.06196} {arXiv:2209.06196
  [astro-ph.CO]} \BibitemShut {NoStop}%
\bibitem [{\citenamefont {Byrnes}\ \emph {et~al.}(2025)\citenamefont {Byrnes},
  \citenamefont {Franciolini}, \citenamefont {Harada}, \citenamefont {Pani},\
  and\ \citenamefont {Sasaki}}]{Byrnes:2025tji}%
  \BibitemOpen
  \bibinfo {editor} {\bibfnamefont {C.}~\bibnamefont {Byrnes}}, \bibinfo
  {editor} {\bibfnamefont {G.}~\bibnamefont {Franciolini}}, \bibinfo {editor}
  {\bibfnamefont {T.}~\bibnamefont {Harada}}, \bibinfo {editor} {\bibfnamefont
  {P.}~\bibnamefont {Pani}},\ and\ \bibinfo {editor} {\bibfnamefont
  {M.}~\bibnamefont {Sasaki}},\ eds.,\ \href
  {https://doi.org/10.1007/978-981-97-8887-3} {\emph {\bibinfo {title}
  {{Primordial Black Holes}}}},\ Springer Series in Astrophysics and Cosmology\
  (\bibinfo  {publisher} {Springer},\ \bibinfo {year} {2025})\BibitemShut
  {NoStop}%
\bibitem [{\citenamefont {Bagui}\ \emph {et~al.}(2025)\citenamefont {Bagui}
  \emph {et~al.}}]{LISACosmologyWorkingGroup:2023njw}%
  \BibitemOpen
  \bibfield  {author} {\bibinfo {author} {\bibfnamefont {E.}~\bibnamefont
  {Bagui}} \emph {et~al.} (\bibinfo {collaboration} {LISA Cosmology Working
  Group}),\ }\bibfield  {title} {\bibinfo {title} {{Primordial black holes and
  their gravitational-wave signatures}},\ }\href
  {https://doi.org/10.1007/s41114-024-00053-w} {\bibfield  {journal} {\bibinfo
  {journal} {Living Rev. Rel.}\ }\textbf {\bibinfo {volume} {28}},\ \bibinfo
  {pages} {1} (\bibinfo {year} {2025})},\ \Eprint
  {https://arxiv.org/abs/2310.19857} {arXiv:2310.19857 [astro-ph.CO]}
  \BibitemShut {NoStop}%
\bibitem [{\citenamefont {De~Luca}\ \emph
  {et~al.}(2020{\natexlab{a}})\citenamefont {De~Luca}, \citenamefont
  {Franciolini}, \citenamefont {Pani},\ and\ \citenamefont
  {Riotto}}]{DeLuca:2020fpg}%
  \BibitemOpen
  \bibfield  {author} {\bibinfo {author} {\bibfnamefont {V.}~\bibnamefont
  {De~Luca}}, \bibinfo {author} {\bibfnamefont {G.}~\bibnamefont
  {Franciolini}}, \bibinfo {author} {\bibfnamefont {P.}~\bibnamefont {Pani}},\
  and\ \bibinfo {author} {\bibfnamefont {A.}~\bibnamefont {Riotto}},\
  }\bibfield  {title} {\bibinfo {title} {{Constraints on Primordial Black
  Holes: the Importance of Accretion}},\ }\href
  {https://doi.org/10.1103/PhysRevD.102.043505} {\bibfield  {journal} {\bibinfo
   {journal} {Phys. Rev. D}\ }\textbf {\bibinfo {volume} {102}},\ \bibinfo
  {pages} {043505} (\bibinfo {year} {2020}{\natexlab{a}})},\ \Eprint
  {https://arxiv.org/abs/2003.12589} {arXiv:2003.12589 [astro-ph.CO]}
  \BibitemShut {NoStop}%
\bibitem [{\citenamefont {De~Luca}\ \emph {et~al.}(2021)\citenamefont
  {De~Luca}, \citenamefont {Desjacques}, \citenamefont {Franciolini},
  \citenamefont {Pani},\ and\ \citenamefont {Riotto}}]{DeLuca:2020sae}%
  \BibitemOpen
  \bibfield  {author} {\bibinfo {author} {\bibfnamefont {V.}~\bibnamefont
  {De~Luca}}, \bibinfo {author} {\bibfnamefont {V.}~\bibnamefont {Desjacques}},
  \bibinfo {author} {\bibfnamefont {G.}~\bibnamefont {Franciolini}}, \bibinfo
  {author} {\bibfnamefont {P.}~\bibnamefont {Pani}},\ and\ \bibinfo {author}
  {\bibfnamefont {A.}~\bibnamefont {Riotto}},\ }\bibfield  {title} {\bibinfo
  {title} {{GW190521 Mass Gap Event and the Primordial Black Hole Scenario}},\
  }\href {https://doi.org/10.1103/PhysRevLett.126.051101} {\bibfield  {journal}
  {\bibinfo  {journal} {Phys. Rev. Lett.}\ }\textbf {\bibinfo {volume} {126}},\
  \bibinfo {pages} {051101} (\bibinfo {year} {2021})},\ \Eprint
  {https://arxiv.org/abs/2009.01728} {arXiv:2009.01728 [astro-ph.CO]}
  \BibitemShut {NoStop}%
\bibitem [{\citenamefont {Yuan}\ \emph {et~al.}(2025)\citenamefont {Yuan},
  \citenamefont {Chen},\ and\ \citenamefont {Liu}}]{Yuan:2025avq}%
  \BibitemOpen
  \bibfield  {author} {\bibinfo {author} {\bibfnamefont {C.}~\bibnamefont
  {Yuan}}, \bibinfo {author} {\bibfnamefont {Z.-C.}\ \bibnamefont {Chen}},\
  and\ \bibinfo {author} {\bibfnamefont {L.}~\bibnamefont {Liu}},\ }\bibfield
  {title} {\bibinfo {title} {{GW231123 Mass Gap Event and the Primordial Black
  Hole Scenario}},\ }\href@noop {} {\  (\bibinfo {year} {2025})},\ \Eprint
  {https://arxiv.org/abs/2507.15701} {arXiv:2507.15701 [astro-ph.CO]}
  \BibitemShut {NoStop}%
\bibitem [{\citenamefont {Mirbabayi}\ \emph {et~al.}(2020)\citenamefont
  {Mirbabayi}, \citenamefont {Gruzinov},\ and\ \citenamefont
  {Nore{\~n}a}}]{Mirbabayi:2019uph}%
  \BibitemOpen
  \bibfield  {author} {\bibinfo {author} {\bibfnamefont {M.}~\bibnamefont
  {Mirbabayi}}, \bibinfo {author} {\bibfnamefont {A.}~\bibnamefont
  {Gruzinov}},\ and\ \bibinfo {author} {\bibfnamefont {J.}~\bibnamefont
  {Nore{\~n}a}},\ }\bibfield  {title} {\bibinfo {title} {{Spin of Primordial
  Black Holes}},\ }\href {https://doi.org/10.1088/1475-7516/2020/03/017}
  {\bibfield  {journal} {\bibinfo  {journal} {JCAP}\ }\textbf {\bibinfo
  {volume} {03}},\ \bibinfo {pages} {017}},\ \Eprint
  {https://arxiv.org/abs/1901.05963} {arXiv:1901.05963 [astro-ph.CO]}
  \BibitemShut {NoStop}%
\bibitem [{\citenamefont {De~Luca}\ \emph {et~al.}(2019)\citenamefont
  {De~Luca}, \citenamefont {Desjacques}, \citenamefont {Franciolini},
  \citenamefont {Malhotra},\ and\ \citenamefont {Riotto}}]{DeLuca:2019buf}%
  \BibitemOpen
  \bibfield  {author} {\bibinfo {author} {\bibfnamefont {V.}~\bibnamefont
  {De~Luca}}, \bibinfo {author} {\bibfnamefont {V.}~\bibnamefont {Desjacques}},
  \bibinfo {author} {\bibfnamefont {G.}~\bibnamefont {Franciolini}}, \bibinfo
  {author} {\bibfnamefont {A.}~\bibnamefont {Malhotra}},\ and\ \bibinfo
  {author} {\bibfnamefont {A.}~\bibnamefont {Riotto}},\ }\bibfield  {title}
  {\bibinfo {title} {{The initial spin probability distribution of primordial
  black holes}},\ }\href {https://doi.org/10.1088/1475-7516/2019/05/018}
  {\bibfield  {journal} {\bibinfo  {journal} {JCAP}\ }\textbf {\bibinfo
  {volume} {05}},\ \bibinfo {pages} {018}},\ \Eprint
  {https://arxiv.org/abs/1903.01179} {arXiv:1903.01179 [astro-ph.CO]}
  \BibitemShut {NoStop}%
\bibitem [{\citenamefont {Harada}\ \emph {et~al.}(2021)\citenamefont {Harada},
  \citenamefont {Yoo}, \citenamefont {Kohri}, \citenamefont {Koga},\ and\
  \citenamefont {Monobe}}]{Harada:2020pzb}%
  \BibitemOpen
  \bibfield  {author} {\bibinfo {author} {\bibfnamefont {T.}~\bibnamefont
  {Harada}}, \bibinfo {author} {\bibfnamefont {C.-M.}\ \bibnamefont {Yoo}},
  \bibinfo {author} {\bibfnamefont {K.}~\bibnamefont {Kohri}}, \bibinfo
  {author} {\bibfnamefont {Y.}~\bibnamefont {Koga}},\ and\ \bibinfo {author}
  {\bibfnamefont {T.}~\bibnamefont {Monobe}},\ }\bibfield  {title} {\bibinfo
  {title} {{Spins of primordial black holes formed in the radiation-dominated
  phase of the universe: first-order effect}},\ }\href
  {https://doi.org/10.3847/1538-4357/abd9b9} {\bibfield  {journal} {\bibinfo
  {journal} {Astrophys. J.}\ }\textbf {\bibinfo {volume} {908}},\ \bibinfo
  {pages} {140} (\bibinfo {year} {2021})},\ \Eprint
  {https://arxiv.org/abs/2011.00710} {arXiv:2011.00710 [astro-ph.CO]}
  \BibitemShut {NoStop}%
\bibitem [{\citenamefont {De~Luca}\ \emph
  {et~al.}(2020{\natexlab{b}})\citenamefont {De~Luca}, \citenamefont
  {Franciolini}, \citenamefont {Pani},\ and\ \citenamefont
  {Riotto}}]{DeLuca:2020bjf}%
  \BibitemOpen
  \bibfield  {author} {\bibinfo {author} {\bibfnamefont {V.}~\bibnamefont
  {De~Luca}}, \bibinfo {author} {\bibfnamefont {G.}~\bibnamefont
  {Franciolini}}, \bibinfo {author} {\bibfnamefont {P.}~\bibnamefont {Pani}},\
  and\ \bibinfo {author} {\bibfnamefont {A.}~\bibnamefont {Riotto}},\
  }\bibfield  {title} {\bibinfo {title} {{The evolution of primordial black
  holes and their final observable spins}},\ }\href
  {https://doi.org/10.1088/1475-7516/2020/04/052} {\bibfield  {journal}
  {\bibinfo  {journal} {JCAP}\ }\textbf {\bibinfo {volume} {04}},\ \bibinfo
  {pages} {052}},\ \Eprint {https://arxiv.org/abs/2003.02778} {arXiv:2003.02778
  [astro-ph.CO]} \BibitemShut {NoStop}%
\bibitem [{\citenamefont {De~Luca}\ \emph
  {et~al.}(2020{\natexlab{c}})\citenamefont {De~Luca}, \citenamefont
  {Franciolini}, \citenamefont {Pani},\ and\ \citenamefont
  {Riotto}}]{DeLuca:2020qqa}%
  \BibitemOpen
  \bibfield  {author} {\bibinfo {author} {\bibfnamefont {V.}~\bibnamefont
  {De~Luca}}, \bibinfo {author} {\bibfnamefont {G.}~\bibnamefont
  {Franciolini}}, \bibinfo {author} {\bibfnamefont {P.}~\bibnamefont {Pani}},\
  and\ \bibinfo {author} {\bibfnamefont {A.}~\bibnamefont {Riotto}},\
  }\bibfield  {title} {\bibinfo {title} {{Primordial Black Holes Confront
  LIGO/Virgo data: Current situation}},\ }\href
  {https://doi.org/10.1088/1475-7516/2020/06/044} {\bibfield  {journal}
  {\bibinfo  {journal} {JCAP}\ }\textbf {\bibinfo {volume} {06}},\ \bibinfo
  {pages} {044}},\ \Eprint {https://arxiv.org/abs/2005.05641} {arXiv:2005.05641
  [astro-ph.CO]} \BibitemShut {NoStop}%
\bibitem [{\citenamefont {Franciolini}\ and\ \citenamefont
  {Pani}(2022)}]{Franciolini:2022iaa}%
  \BibitemOpen
  \bibfield  {author} {\bibinfo {author} {\bibfnamefont {G.}~\bibnamefont
  {Franciolini}}\ and\ \bibinfo {author} {\bibfnamefont {P.}~\bibnamefont
  {Pani}},\ }\bibfield  {title} {\bibinfo {title} {{Searching for mass-spin
  correlations in the population of gravitational-wave events: The GWTC-3 case
  study}},\ }\href {https://doi.org/10.1103/PhysRevD.105.123024} {\bibfield
  {journal} {\bibinfo  {journal} {Phys. Rev. D}\ }\textbf {\bibinfo {volume}
  {105}},\ \bibinfo {pages} {123024} (\bibinfo {year} {2022})},\ \Eprint
  {https://arxiv.org/abs/2201.13098} {arXiv:2201.13098 [astro-ph.HE]}
  \BibitemShut {NoStop}%
\bibitem [{\citenamefont {Franciolini}\ \emph
  {et~al.}(2022{\natexlab{b}})\citenamefont {Franciolini}, \citenamefont
  {Cotesta}, \citenamefont {Loutrel}, \citenamefont {Berti}, \citenamefont
  {Pani},\ and\ \citenamefont {Riotto}}]{Franciolini:2021xbq}%
  \BibitemOpen
  \bibfield  {author} {\bibinfo {author} {\bibfnamefont {G.}~\bibnamefont
  {Franciolini}}, \bibinfo {author} {\bibfnamefont {R.}~\bibnamefont
  {Cotesta}}, \bibinfo {author} {\bibfnamefont {N.}~\bibnamefont {Loutrel}},
  \bibinfo {author} {\bibfnamefont {E.}~\bibnamefont {Berti}}, \bibinfo
  {author} {\bibfnamefont {P.}~\bibnamefont {Pani}},\ and\ \bibinfo {author}
  {\bibfnamefont {A.}~\bibnamefont {Riotto}},\ }\bibfield  {title} {\bibinfo
  {title} {{How to assess the primordial origin of single gravitational-wave
  events with mass, spin, eccentricity, and deformability measurements}},\
  }\href {https://doi.org/10.1103/PhysRevD.105.063510} {\bibfield  {journal}
  {\bibinfo  {journal} {Phys. Rev. D}\ }\textbf {\bibinfo {volume} {105}},\
  \bibinfo {pages} {063510} (\bibinfo {year} {2022}{\natexlab{b}})},\ \Eprint
  {https://arxiv.org/abs/2112.10660} {arXiv:2112.10660 [astro-ph.CO]}
  \BibitemShut {NoStop}%
\bibitem [{\citenamefont {Ali-Ha{\"\i}moud}\ \emph {et~al.}(2017)\citenamefont
  {Ali-Ha{\"\i}moud}, \citenamefont {Kovetz},\ and\ \citenamefont
  {Kamionkowski}}]{Ali-Haimoud:2017rtz}%
  \BibitemOpen
  \bibfield  {author} {\bibinfo {author} {\bibfnamefont {Y.}~\bibnamefont
  {Ali-Ha{\"\i}moud}}, \bibinfo {author} {\bibfnamefont {E.~D.}\ \bibnamefont
  {Kovetz}},\ and\ \bibinfo {author} {\bibfnamefont {M.}~\bibnamefont
  {Kamionkowski}},\ }\bibfield  {title} {\bibinfo {title} {{Merger rate of
  primordial black-hole binaries}},\ }\href
  {https://doi.org/10.1103/PhysRevD.96.123523} {\bibfield  {journal} {\bibinfo
  {journal} {Phys. Rev. D}\ }\textbf {\bibinfo {volume} {96}},\ \bibinfo
  {pages} {123523} (\bibinfo {year} {2017})},\ \Eprint
  {https://arxiv.org/abs/1709.06576} {arXiv:1709.06576 [astro-ph.CO]}
  \BibitemShut {NoStop}%
\bibitem [{\citenamefont {Raidal}\ \emph {et~al.}(2019)\citenamefont {Raidal},
  \citenamefont {Spethmann}, \citenamefont {Vaskonen},\ and\ \citenamefont
  {Veerm{\"a}e}}]{Raidal:2018bbj}%
  \BibitemOpen
  \bibfield  {author} {\bibinfo {author} {\bibfnamefont {M.}~\bibnamefont
  {Raidal}}, \bibinfo {author} {\bibfnamefont {C.}~\bibnamefont {Spethmann}},
  \bibinfo {author} {\bibfnamefont {V.}~\bibnamefont {Vaskonen}},\ and\
  \bibinfo {author} {\bibfnamefont {H.}~\bibnamefont {Veerm{\"a}e}},\
  }\bibfield  {title} {\bibinfo {title} {{Formation and Evolution of Primordial
  Black Hole Binaries in the Early Universe}},\ }\href
  {https://doi.org/10.1088/1475-7516/2019/02/018} {\bibfield  {journal}
  {\bibinfo  {journal} {JCAP}\ }\textbf {\bibinfo {volume} {02}},\ \bibinfo
  {pages} {018}},\ \Eprint {https://arxiv.org/abs/1812.01930} {arXiv:1812.01930
  [astro-ph.CO]} \BibitemShut {NoStop}%
\bibitem [{\citenamefont {Vaskonen}\ and\ \citenamefont
  {Veerm{\"a}e}(2020)}]{Vaskonen:2019jpv}%
  \BibitemOpen
  \bibfield  {author} {\bibinfo {author} {\bibfnamefont {V.}~\bibnamefont
  {Vaskonen}}\ and\ \bibinfo {author} {\bibfnamefont {H.}~\bibnamefont
  {Veerm{\"a}e}},\ }\bibfield  {title} {\bibinfo {title} {{Lower bound on the
  primordial black hole merger rate}},\ }\href
  {https://doi.org/10.1103/PhysRevD.101.043015} {\bibfield  {journal} {\bibinfo
   {journal} {Phys. Rev. D}\ }\textbf {\bibinfo {volume} {101}},\ \bibinfo
  {pages} {043015} (\bibinfo {year} {2020})},\ \Eprint
  {https://arxiv.org/abs/1908.09752} {arXiv:1908.09752 [astro-ph.CO]}
  \BibitemShut {NoStop}%
\bibitem [{\citenamefont {Franciolini}\ \emph
  {et~al.}(2022{\natexlab{c}})\citenamefont {Franciolini}, \citenamefont
  {Kritos}, \citenamefont {Berti},\ and\ \citenamefont
  {Silk}}]{Franciolini:2022ewd}%
  \BibitemOpen
  \bibfield  {author} {\bibinfo {author} {\bibfnamefont {G.}~\bibnamefont
  {Franciolini}}, \bibinfo {author} {\bibfnamefont {K.}~\bibnamefont {Kritos}},
  \bibinfo {author} {\bibfnamefont {E.}~\bibnamefont {Berti}},\ and\ \bibinfo
  {author} {\bibfnamefont {J.}~\bibnamefont {Silk}},\ }\bibfield  {title}
  {\bibinfo {title} {{Primordial black hole mergers from three-body
  interactions}},\ }\href {https://doi.org/10.1103/PhysRevD.106.083529}
  {\bibfield  {journal} {\bibinfo  {journal} {Phys. Rev. D}\ }\textbf {\bibinfo
  {volume} {106}},\ \bibinfo {pages} {083529} (\bibinfo {year}
  {2022}{\natexlab{c}})},\ \Eprint {https://arxiv.org/abs/2205.15340}
  {arXiv:2205.15340 [astro-ph.CO]} \BibitemShut {NoStop}%
\bibitem [{\citenamefont {Raidal}\ \emph {et~al.}(2025)\citenamefont {Raidal},
  \citenamefont {Vaskonen},\ and\ \citenamefont
  {Veerm{\"a}e}}]{Raidal:2024bmm}%
  \BibitemOpen
  \bibfield  {author} {\bibinfo {author} {\bibfnamefont {M.}~\bibnamefont
  {Raidal}}, \bibinfo {author} {\bibfnamefont {V.}~\bibnamefont {Vaskonen}},\
  and\ \bibinfo {author} {\bibfnamefont {H.}~\bibnamefont {Veerm{\"a}e}},\
  }\bibinfo {title} {{Formation of~Primordial Black Hole Binaries and~Their
  Merger Rates}},\ in\ \href {https://doi.org/10.1007/978-981-97-8887-3_16}
  {\emph {\bibinfo {booktitle} {{Primordial Black Holes}}}},\ \bibinfo {editor}
  {edited by\ \bibinfo {editor} {\bibfnamefont {C.}~\bibnamefont {Byrnes}},
  \bibinfo {editor} {\bibfnamefont {G.}~\bibnamefont {Franciolini}}, \bibinfo
  {editor} {\bibfnamefont {T.}~\bibnamefont {Harada}}, \bibinfo {editor}
  {\bibfnamefont {P.}~\bibnamefont {Pani}},\ and\ \bibinfo {editor}
  {\bibfnamefont {M.}~\bibnamefont {Sasaki}}}\ (\bibinfo {year} {2025})\
  \Eprint {https://arxiv.org/abs/2404.08416} {arXiv:2404.08416 [astro-ph.CO]}
  \BibitemShut {NoStop}%
\bibitem [{\citenamefont {Hasinger}(2020)}]{Hasinger:2020ptw}%
  \BibitemOpen
  \bibfield  {author} {\bibinfo {author} {\bibfnamefont {G.}~\bibnamefont
  {Hasinger}},\ }\bibfield  {title} {\bibinfo {title} {{Illuminating the dark
  ages: Cosmic backgrounds from accretion onto primordial black hole dark
  matter}},\ }\href {https://doi.org/10.1088/1475-7516/2020/07/022} {\bibfield
  {journal} {\bibinfo  {journal} {JCAP}\ }\textbf {\bibinfo {volume} {07}},\
  \bibinfo {pages} {022}},\ \Eprint {https://arxiv.org/abs/2003.05150}
  {arXiv:2003.05150 [astro-ph.CO]} \BibitemShut {NoStop}%
\bibitem [{\citenamefont {Arvanitaki}\ \emph {et~al.}(2015)\citenamefont
  {Arvanitaki}, \citenamefont {Baryakhtar},\ and\ \citenamefont
  {Huang}}]{Arvanitaki:2014wva}%
  \BibitemOpen
  \bibfield  {author} {\bibinfo {author} {\bibfnamefont {A.}~\bibnamefont
  {Arvanitaki}}, \bibinfo {author} {\bibfnamefont {M.}~\bibnamefont
  {Baryakhtar}},\ and\ \bibinfo {author} {\bibfnamefont {X.}~\bibnamefont
  {Huang}},\ }\bibfield  {title} {\bibinfo {title} {{Discovering the QCD Axion
  with Black Holes and Gravitational Waves}},\ }\href
  {https://doi.org/10.1103/PhysRevD.91.084011} {\bibfield  {journal} {\bibinfo
  {journal} {Phys. Rev. D}\ }\textbf {\bibinfo {volume} {91}},\ \bibinfo
  {pages} {084011} (\bibinfo {year} {2015})},\ \Eprint
  {https://arxiv.org/abs/1411.2263} {arXiv:1411.2263 [hep-ph]} \BibitemShut
  {NoStop}%
\bibitem [{\citenamefont {Baumann}\ \emph
  {et~al.}(2019{\natexlab{b}})\citenamefont {Baumann}, \citenamefont {Chia},\
  and\ \citenamefont {Porto}}]{Baumann:2018vus}%
  \BibitemOpen
  \bibfield  {author} {\bibinfo {author} {\bibfnamefont {D.}~\bibnamefont
  {Baumann}}, \bibinfo {author} {\bibfnamefont {H.~S.}\ \bibnamefont {Chia}},\
  and\ \bibinfo {author} {\bibfnamefont {R.~A.}\ \bibnamefont {Porto}},\
  }\bibfield  {title} {\bibinfo {title} {{Probing Ultralight Bosons with Binary
  Black Holes}},\ }\href {https://doi.org/10.1103/PhysRevD.99.044001}
  {\bibfield  {journal} {\bibinfo  {journal} {Phys. Rev. D}\ }\textbf {\bibinfo
  {volume} {99}},\ \bibinfo {pages} {044001} (\bibinfo {year}
  {2019}{\natexlab{b}})},\ \Eprint {https://arxiv.org/abs/1804.03208}
  {arXiv:1804.03208 [gr-qc]} \BibitemShut {NoStop}%
\bibitem [{\citenamefont {Baumann}\ \emph {et~al.}(2022)\citenamefont
  {Baumann}, \citenamefont {Bertone}, \citenamefont {Stout},\ and\
  \citenamefont {Tomaselli}}]{Baumann:2022pkl}%
  \BibitemOpen
  \bibfield  {author} {\bibinfo {author} {\bibfnamefont {D.}~\bibnamefont
  {Baumann}}, \bibinfo {author} {\bibfnamefont {G.}~\bibnamefont {Bertone}},
  \bibinfo {author} {\bibfnamefont {J.}~\bibnamefont {Stout}},\ and\ \bibinfo
  {author} {\bibfnamefont {G.~M.}\ \bibnamefont {Tomaselli}},\ }\bibfield
  {title} {\bibinfo {title} {{Sharp Signals of Boson Clouds in Black Hole
  Binary Inspirals}},\ }\href {https://doi.org/10.1103/PhysRevLett.128.221102}
  {\bibfield  {journal} {\bibinfo  {journal} {Phys. Rev. Lett.}\ }\textbf
  {\bibinfo {volume} {128}},\ \bibinfo {pages} {221102} (\bibinfo {year}
  {2022})},\ \Eprint {https://arxiv.org/abs/2206.01212} {arXiv:2206.01212
  [gr-qc]} \BibitemShut {NoStop}%
\bibitem [{\citenamefont {Tong}\ \emph {et~al.}(2022)\citenamefont {Tong},
  \citenamefont {Wang},\ and\ \citenamefont {Zhu}}]{Tong:2022bbl}%
  \BibitemOpen
  \bibfield  {author} {\bibinfo {author} {\bibfnamefont {X.}~\bibnamefont
  {Tong}}, \bibinfo {author} {\bibfnamefont {Y.}~\bibnamefont {Wang}},\ and\
  \bibinfo {author} {\bibfnamefont {H.-Y.}\ \bibnamefont {Zhu}},\ }\bibfield
  {title} {\bibinfo {title} {{Termination of superradiance from a binary
  companion}},\ }\href {https://doi.org/10.1103/PhysRevD.106.043002} {\bibfield
   {journal} {\bibinfo  {journal} {Phys. Rev. D}\ }\textbf {\bibinfo {volume}
  {106}},\ \bibinfo {pages} {043002} (\bibinfo {year} {2022})},\ \Eprint
  {https://arxiv.org/abs/2205.10527} {arXiv:2205.10527 [gr-qc]} \BibitemShut
  {NoStop}%
\bibitem [{\citenamefont {Fan}\ \emph {et~al.}(2024)\citenamefont {Fan},
  \citenamefont {Tong}, \citenamefont {Wang},\ and\ \citenamefont
  {Zhu}}]{Fan:2023jjj}%
  \BibitemOpen
  \bibfield  {author} {\bibinfo {author} {\bibfnamefont {K.}~\bibnamefont
  {Fan}}, \bibinfo {author} {\bibfnamefont {X.}~\bibnamefont {Tong}}, \bibinfo
  {author} {\bibfnamefont {Y.}~\bibnamefont {Wang}},\ and\ \bibinfo {author}
  {\bibfnamefont {H.-Y.}\ \bibnamefont {Zhu}},\ }\bibfield  {title} {\bibinfo
  {title} {{Modulating binary dynamics via the termination of black hole
  superradiance}},\ }\href {https://doi.org/10.1103/PhysRevD.109.024059}
  {\bibfield  {journal} {\bibinfo  {journal} {Phys. Rev. D}\ }\textbf {\bibinfo
  {volume} {109}},\ \bibinfo {pages} {024059} (\bibinfo {year} {2024})},\
  \Eprint {https://arxiv.org/abs/2311.17013} {arXiv:2311.17013 [gr-qc]}
  \BibitemShut {NoStop}%
\bibitem [{\citenamefont {Tomaselli}\ \emph {et~al.}(2023)\citenamefont
  {Tomaselli}, \citenamefont {Spieksma},\ and\ \citenamefont
  {Bertone}}]{Tomaselli:2023ysb}%
  \BibitemOpen
  \bibfield  {author} {\bibinfo {author} {\bibfnamefont {G.~M.}\ \bibnamefont
  {Tomaselli}}, \bibinfo {author} {\bibfnamefont {T.~F.~M.}\ \bibnamefont
  {Spieksma}},\ and\ \bibinfo {author} {\bibfnamefont {G.}~\bibnamefont
  {Bertone}},\ }\bibfield  {title} {\bibinfo {title} {{Dynamical friction in
  gravitational atoms}},\ }\href
  {https://doi.org/10.1088/1475-7516/2023/07/070} {\bibfield  {journal}
  {\bibinfo  {journal} {JCAP}\ }\textbf {\bibinfo {volume} {07}},\ \bibinfo
  {pages} {070}},\ \Eprint {https://arxiv.org/abs/2305.15460} {arXiv:2305.15460
  [gr-qc]} \BibitemShut {NoStop}%
\bibitem [{\citenamefont {Zhu}\ \emph {et~al.}(2024)\citenamefont {Zhu},
  \citenamefont {Tong}, \citenamefont {Manzoni},\ and\ \citenamefont
  {Ma}}]{Zhu:2024bqs}%
  \BibitemOpen
  \bibfield  {author} {\bibinfo {author} {\bibfnamefont {H.-Y.}\ \bibnamefont
  {Zhu}}, \bibinfo {author} {\bibfnamefont {X.}~\bibnamefont {Tong}}, \bibinfo
  {author} {\bibfnamefont {G.}~\bibnamefont {Manzoni}},\ and\ \bibinfo {author}
  {\bibfnamefont {Y.}~\bibnamefont {Ma}},\ }\bibfield  {title} {\bibinfo
  {title} {{Survival of the Fittest: Testing Superradiance Termination with
  Simulated Binary Black Hole Statistics}},\ }\href@noop {} {\  (\bibinfo
  {year} {2024})},\ \Eprint {https://arxiv.org/abs/2409.14159}
  {arXiv:2409.14159 [gr-qc]} \BibitemShut {NoStop}%
\bibitem [{\citenamefont {Takahashi}\ \emph {et~al.}(2024)\citenamefont
  {Takahashi}, \citenamefont {Omiya},\ and\ \citenamefont
  {Tanaka}}]{Takahashi:2024fyq}%
  \BibitemOpen
  \bibfield  {author} {\bibinfo {author} {\bibfnamefont {T.}~\bibnamefont
  {Takahashi}}, \bibinfo {author} {\bibfnamefont {H.}~\bibnamefont {Omiya}},\
  and\ \bibinfo {author} {\bibfnamefont {T.}~\bibnamefont {Tanaka}},\
  }\bibfield  {title} {\bibinfo {title} {{Self-interacting axion clouds around
  rotating black holes in binary systems}},\ }\href@noop {} {\  (\bibinfo
  {year} {2024})},\ \Eprint {https://arxiv.org/abs/2408.08349}
  {arXiv:2408.08349 [gr-qc]} \BibitemShut {NoStop}%
\bibitem [{\citenamefont {Boskovic}\ \emph {et~al.}(2024)\citenamefont
  {Boskovic}, \citenamefont {Koschnitzke},\ and\ \citenamefont
  {Porto}}]{Boskovic:2024fga}%
  \BibitemOpen
  \bibfield  {author} {\bibinfo {author} {\bibfnamefont {M.}~\bibnamefont
  {Boskovic}}, \bibinfo {author} {\bibfnamefont {M.}~\bibnamefont
  {Koschnitzke}},\ and\ \bibinfo {author} {\bibfnamefont {R.~A.}\ \bibnamefont
  {Porto}},\ }\bibfield  {title} {\bibinfo {title} {{Signatures of ultralight
  bosons in the orbital eccentricity of binary black holes}},\ }\href@noop {}
  {\  (\bibinfo {year} {2024})}\BibitemShut {NoStop}%
\bibitem [{\citenamefont {Tomaselli}\ \emph {et~al.}(2024)\citenamefont
  {Tomaselli}, \citenamefont {Spieksma},\ and\ \citenamefont
  {Bertone}}]{Tomaselli:2024bdd}%
  \BibitemOpen
  \bibfield  {author} {\bibinfo {author} {\bibfnamefont {G.~M.}\ \bibnamefont
  {Tomaselli}}, \bibinfo {author} {\bibfnamefont {T.~F.~M.}\ \bibnamefont
  {Spieksma}},\ and\ \bibinfo {author} {\bibfnamefont {G.}~\bibnamefont
  {Bertone}},\ }\bibfield  {title} {\bibinfo {title} {{The resonant history of
  gravitational atoms in black hole binaries}},\ }\href@noop {} {\  (\bibinfo
  {year} {2024})},\ \Eprint {https://arxiv.org/abs/2403.03147}
  {arXiv:2403.03147 [gr-qc]} \BibitemShut {NoStop}%
\bibitem [{\citenamefont {Tomaselli}(2025)}]{Tomaselli:2025jfo}%
  \BibitemOpen
  \bibfield  {author} {\bibinfo {author} {\bibfnamefont {G.~M.}\ \bibnamefont
  {Tomaselli}},\ }\bibfield  {title} {\bibinfo {title} {{Smooth binary
  evolution from wide resonances in boson clouds}},\ }\href@noop {} {\
  (\bibinfo {year} {2025})},\ \Eprint {https://arxiv.org/abs/2507.15110}
  {arXiv:2507.15110 [gr-qc]} \BibitemShut {NoStop}%
\bibitem [{\citenamefont {Raftery}\ \emph {et~al.}(2001)\citenamefont
  {Raftery}, \citenamefont {Tanner},\ and\ \citenamefont
  {Wells}}]{raftery2001statistics}%
  \BibitemOpen
  \bibfield  {author} {\bibinfo {author} {\bibfnamefont {A.~E.}\ \bibnamefont
  {Raftery}}, \bibinfo {author} {\bibfnamefont {M.~A.}\ \bibnamefont
  {Tanner}},\ and\ \bibinfo {author} {\bibfnamefont {M.~T.}\ \bibnamefont
  {Wells}},\ }\href@noop {} {\emph {\bibinfo {title} {Statistics in the 21st
  Century}}}\ (\bibinfo  {publisher} {CRC Press},\ \bibinfo {year}
  {2001})\BibitemShut {NoStop}%
\bibitem [{\citenamefont {Gerosa}\ and\ \citenamefont
  {Fishbach}(2021)}]{Gerosa:2021mno}%
  \BibitemOpen
  \bibfield  {author} {\bibinfo {author} {\bibfnamefont {D.}~\bibnamefont
  {Gerosa}}\ and\ \bibinfo {author} {\bibfnamefont {M.}~\bibnamefont
  {Fishbach}},\ }\bibfield  {title} {\bibinfo {title} {{Hierarchical mergers of
  stellar-mass black holes and their gravitational-wave signatures}},\ }\href
  {https://doi.org/10.1038/s41550-021-01398-w} {\bibfield  {journal} {\bibinfo
  {journal} {Nature Astron.}\ }\textbf {\bibinfo {volume} {5}},\ \bibinfo
  {pages} {749} (\bibinfo {year} {2021})},\ \Eprint
  {https://arxiv.org/abs/2105.03439} {arXiv:2105.03439 [astro-ph.HE]}
  \BibitemShut {NoStop}%
\bibitem [{\citenamefont {Hannam}\ \emph {et~al.}(2022)\citenamefont {Hannam}
  \emph {et~al.}}]{Hannam:2021pit}%
  \BibitemOpen
  \bibfield  {author} {\bibinfo {author} {\bibfnamefont {M.}~\bibnamefont
  {Hannam}} \emph {et~al.},\ }\bibfield  {title} {\bibinfo {title}
  {{General-relativistic precession in a black-hole binary}},\ }\href
  {https://doi.org/10.1038/s41586-022-05212-z} {\bibfield  {journal} {\bibinfo
  {journal} {Nature}\ }\textbf {\bibinfo {volume} {610}},\ \bibinfo {pages}
  {652} (\bibinfo {year} {2022})},\ \Eprint {https://arxiv.org/abs/2112.11300}
  {arXiv:2112.11300 [gr-qc]} \BibitemShut {NoStop}%
\bibitem [{\citenamefont {Abbott}\ \emph {et~al.}(2024)\citenamefont {Abbott}
  \emph {et~al.}}]{LIGOScientific:2021usb}%
  \BibitemOpen
  \bibfield  {author} {\bibinfo {author} {\bibfnamefont {R.}~\bibnamefont
  {Abbott}} \emph {et~al.} (\bibinfo {collaboration} {LIGO Scientific,
  VIRGO}),\ }\bibfield  {title} {\bibinfo {title} {{GWTC-2.1: Deep extended
  catalog of compact binary coalescences observed by LIGO and Virgo during the
  first half of the third observing run}},\ }\href
  {https://doi.org/10.1103/PhysRevD.109.022001} {\bibfield  {journal} {\bibinfo
   {journal} {Phys. Rev. D}\ }\textbf {\bibinfo {volume} {109}},\ \bibinfo
  {pages} {022001} (\bibinfo {year} {2024})},\ \Eprint
  {https://arxiv.org/abs/2108.01045} {arXiv:2108.01045 [gr-qc]} \BibitemShut
  {NoStop}%
\bibitem [{\citenamefont {Nitz}\ \emph {et~al.}(2020)\citenamefont {Nitz},
  \citenamefont {Dent}, \citenamefont {Davies}, \citenamefont {Kumar},
  \citenamefont {Capano}, \citenamefont {Harry}, \citenamefont {Mozzon},
  \citenamefont {Nuttall}, \citenamefont {Lundgren},\ and\ \citenamefont
  {T{\'a}pai}}]{Nitz:2020oeq}%
  \BibitemOpen
  \bibfield  {author} {\bibinfo {author} {\bibfnamefont {A.~H.}\ \bibnamefont
  {Nitz}}, \bibinfo {author} {\bibfnamefont {T.}~\bibnamefont {Dent}}, \bibinfo
  {author} {\bibfnamefont {G.~S.}\ \bibnamefont {Davies}}, \bibinfo {author}
  {\bibfnamefont {S.}~\bibnamefont {Kumar}}, \bibinfo {author} {\bibfnamefont
  {C.~D.}\ \bibnamefont {Capano}}, \bibinfo {author} {\bibfnamefont
  {I.}~\bibnamefont {Harry}}, \bibinfo {author} {\bibfnamefont
  {S.}~\bibnamefont {Mozzon}}, \bibinfo {author} {\bibfnamefont
  {L.}~\bibnamefont {Nuttall}}, \bibinfo {author} {\bibfnamefont
  {A.}~\bibnamefont {Lundgren}},\ and\ \bibinfo {author} {\bibfnamefont
  {M.}~\bibnamefont {T{\'a}pai}},\ }\bibfield  {title} {\bibinfo {title}
  {{2-OGC: Open Gravitational-wave Catalog of binary mergers from analysis of
  public Advanced LIGO and Virgo data}},\ }\href
  {https://doi.org/10.3847/1538-4357/ab733f} {\bibfield  {journal} {\bibinfo
  {journal} {Astrophys. J.}\ }\textbf {\bibinfo {volume} {891}},\ \bibinfo
  {pages} {123} (\bibinfo {year} {2020})},\ \Eprint
  {https://arxiv.org/abs/1910.05331} {arXiv:1910.05331 [astro-ph.HE]}
  \BibitemShut {NoStop}%
\bibitem [{\citenamefont {Williams}(2025)}]{Williams:2024tna}%
  \BibitemOpen
  \bibfield  {author} {\bibinfo {author} {\bibfnamefont {D.}~\bibnamefont
  {Williams}},\ }\bibfield  {title} {\bibinfo {title} {{Beyond GWTC-3:
  analyzing and verifying new gravitational-wave events from community
  catalogues}},\ }\href {https://doi.org/10.1088/1361-6382/add233} {\bibfield
  {journal} {\bibinfo  {journal} {Class. Quant. Grav.}\ }\textbf {\bibinfo
  {volume} {42}},\ \bibinfo {pages} {105012} (\bibinfo {year} {2025})},\
  \Eprint {https://arxiv.org/abs/2401.08709} {arXiv:2401.08709 [astro-ph.HE]}
  \BibitemShut {NoStop}%
\bibitem [{\citenamefont {Rosa}\ and\ \citenamefont
  {Dolan}(2012)}]{Rosa:2011my}%
  \BibitemOpen
  \bibfield  {author} {\bibinfo {author} {\bibfnamefont {J.~G.}\ \bibnamefont
  {Rosa}}\ and\ \bibinfo {author} {\bibfnamefont {S.~R.}\ \bibnamefont
  {Dolan}},\ }\bibfield  {title} {\bibinfo {title} {{Massive vector fields on
  the Schwarzschild spacetime: quasi-normal modes and bound states}},\ }\href
  {https://doi.org/10.1103/PhysRevD.85.044043} {\bibfield  {journal} {\bibinfo
  {journal} {Phys. Rev. D}\ }\textbf {\bibinfo {volume} {85}},\ \bibinfo
  {pages} {044043} (\bibinfo {year} {2012})},\ \Eprint
  {https://arxiv.org/abs/1110.4494} {arXiv:1110.4494 [hep-th]} \BibitemShut
  {NoStop}%
\bibitem [{\citenamefont {Witek}\ \emph {et~al.}(2013)\citenamefont {Witek},
  \citenamefont {Cardoso}, \citenamefont {Ishibashi},\ and\ \citenamefont
  {Sperhake}}]{Witek:2012tr}%
  \BibitemOpen
  \bibfield  {author} {\bibinfo {author} {\bibfnamefont {H.}~\bibnamefont
  {Witek}}, \bibinfo {author} {\bibfnamefont {V.}~\bibnamefont {Cardoso}},
  \bibinfo {author} {\bibfnamefont {A.}~\bibnamefont {Ishibashi}},\ and\
  \bibinfo {author} {\bibfnamefont {U.}~\bibnamefont {Sperhake}},\ }\bibfield
  {title} {\bibinfo {title} {{Superradiant instabilities in astrophysical
  systems}},\ }\href {https://doi.org/10.1103/PhysRevD.87.043513} {\bibfield
  {journal} {\bibinfo  {journal} {Phys. Rev. D}\ }\textbf {\bibinfo {volume}
  {87}},\ \bibinfo {pages} {043513} (\bibinfo {year} {2013})},\ \Eprint
  {https://arxiv.org/abs/1212.0551} {arXiv:1212.0551 [gr-qc]} \BibitemShut
  {NoStop}%
\bibitem [{\citenamefont {Pani}\ \emph
  {et~al.}(2012{\natexlab{a}})\citenamefont {Pani}, \citenamefont {Cardoso},
  \citenamefont {Gualtieri}, \citenamefont {Berti},\ and\ \citenamefont
  {Ishibashi}}]{Pani:2012vp}%
  \BibitemOpen
  \bibfield  {author} {\bibinfo {author} {\bibfnamefont {P.}~\bibnamefont
  {Pani}}, \bibinfo {author} {\bibfnamefont {V.}~\bibnamefont {Cardoso}},
  \bibinfo {author} {\bibfnamefont {L.}~\bibnamefont {Gualtieri}}, \bibinfo
  {author} {\bibfnamefont {E.}~\bibnamefont {Berti}},\ and\ \bibinfo {author}
  {\bibfnamefont {A.}~\bibnamefont {Ishibashi}},\ }\bibfield  {title} {\bibinfo
  {title} {{Black hole bombs and photon mass bounds}},\ }\href
  {https://doi.org/10.1103/PhysRevLett.109.131102} {\bibfield  {journal}
  {\bibinfo  {journal} {Phys. Rev. Lett.}\ }\textbf {\bibinfo {volume} {109}},\
  \bibinfo {pages} {131102} (\bibinfo {year} {2012}{\natexlab{a}})},\ \Eprint
  {https://arxiv.org/abs/1209.0465} {arXiv:1209.0465 [gr-qc]} \BibitemShut
  {NoStop}%
\bibitem [{\citenamefont {Pani}\ \emph
  {et~al.}(2012{\natexlab{b}})\citenamefont {Pani}, \citenamefont {Cardoso},
  \citenamefont {Gualtieri}, \citenamefont {Berti},\ and\ \citenamefont
  {Ishibashi}}]{Pani:2012bp}%
  \BibitemOpen
  \bibfield  {author} {\bibinfo {author} {\bibfnamefont {P.}~\bibnamefont
  {Pani}}, \bibinfo {author} {\bibfnamefont {V.}~\bibnamefont {Cardoso}},
  \bibinfo {author} {\bibfnamefont {L.}~\bibnamefont {Gualtieri}}, \bibinfo
  {author} {\bibfnamefont {E.}~\bibnamefont {Berti}},\ and\ \bibinfo {author}
  {\bibfnamefont {A.}~\bibnamefont {Ishibashi}},\ }\bibfield  {title} {\bibinfo
  {title} {{Perturbations of slowly rotating black holes: massive vector fields
  in the Kerr metric}},\ }\href {https://doi.org/10.1103/PhysRevD.86.104017}
  {\bibfield  {journal} {\bibinfo  {journal} {Phys. Rev. D}\ }\textbf {\bibinfo
  {volume} {86}},\ \bibinfo {pages} {104017} (\bibinfo {year}
  {2012}{\natexlab{b}})},\ \Eprint {https://arxiv.org/abs/1209.0773}
  {arXiv:1209.0773 [gr-qc]} \BibitemShut {NoStop}%
\bibitem [{\citenamefont {Endlich}\ and\ \citenamefont
  {Penco}(2017)}]{Endlich:2016jgc}%
  \BibitemOpen
  \bibfield  {author} {\bibinfo {author} {\bibfnamefont {S.}~\bibnamefont
  {Endlich}}\ and\ \bibinfo {author} {\bibfnamefont {R.}~\bibnamefont
  {Penco}},\ }\bibfield  {title} {\bibinfo {title} {{A Modern Approach to
  Superradiance}},\ }\href {https://doi.org/10.1007/JHEP05(2017)052} {\bibfield
   {journal} {\bibinfo  {journal} {JHEP}\ }\textbf {\bibinfo {volume} {05}},\
  \bibinfo {pages} {052}},\ \Eprint {https://arxiv.org/abs/1609.06723}
  {arXiv:1609.06723 [hep-th]} \BibitemShut {NoStop}%
\bibitem [{\citenamefont {East}(2017)}]{East:2017mrj}%
  \BibitemOpen
  \bibfield  {author} {\bibinfo {author} {\bibfnamefont {W.~E.}\ \bibnamefont
  {East}},\ }\bibfield  {title} {\bibinfo {title} {{Superradiant instability of
  massive vector fields around spinning black holes in the relativistic
  regime}},\ }\href {https://doi.org/10.1103/PhysRevD.96.024004} {\bibfield
  {journal} {\bibinfo  {journal} {Phys. Rev. D}\ }\textbf {\bibinfo {volume}
  {96}},\ \bibinfo {pages} {024004} (\bibinfo {year} {2017})},\ \Eprint
  {https://arxiv.org/abs/1705.01544} {arXiv:1705.01544 [gr-qc]} \BibitemShut
  {NoStop}%
\bibitem [{\citenamefont {East}(2018)}]{East:2018glu}%
  \BibitemOpen
  \bibfield  {author} {\bibinfo {author} {\bibfnamefont {W.~E.}\ \bibnamefont
  {East}},\ }\bibfield  {title} {\bibinfo {title} {{Massive Boson Superradiant
  Instability of Black Holes: Nonlinear Growth, Saturation, and Gravitational
  Radiation}},\ }\href {https://doi.org/10.1103/PhysRevLett.121.131104}
  {\bibfield  {journal} {\bibinfo  {journal} {Phys. Rev. Lett.}\ }\textbf
  {\bibinfo {volume} {121}},\ \bibinfo {pages} {131104} (\bibinfo {year}
  {2018})},\ \Eprint {https://arxiv.org/abs/1807.00043} {arXiv:1807.00043
  [gr-qc]} \BibitemShut {NoStop}%
\bibitem [{\citenamefont {Dolan}(2018)}]{Dolan:2018dqv}%
  \BibitemOpen
  \bibfield  {author} {\bibinfo {author} {\bibfnamefont {S.~R.}\ \bibnamefont
  {Dolan}},\ }\bibfield  {title} {\bibinfo {title} {{Instability of the Proca
  field on Kerr spacetime}},\ }\href
  {https://doi.org/10.1103/PhysRevD.98.104006} {\bibfield  {journal} {\bibinfo
  {journal} {Phys. Rev. D}\ }\textbf {\bibinfo {volume} {98}},\ \bibinfo
  {pages} {104006} (\bibinfo {year} {2018})},\ \Eprint
  {https://arxiv.org/abs/1806.01604} {arXiv:1806.01604 [gr-qc]} \BibitemShut
  {NoStop}%
\bibitem [{\citenamefont {Brito}\ \emph
  {et~al.}(2020{\natexlab{b}})\citenamefont {Brito}, \citenamefont {Grillo},\
  and\ \citenamefont {Pani}}]{brito2020black}%
  \BibitemOpen
  \bibfield  {author} {\bibinfo {author} {\bibfnamefont {R.}~\bibnamefont
  {Brito}}, \bibinfo {author} {\bibfnamefont {S.}~\bibnamefont {Grillo}},\ and\
  \bibinfo {author} {\bibfnamefont {P.}~\bibnamefont {Pani}},\ }\bibfield
  {title} {\bibinfo {title} {Black hole superradiant instability from
  ultralight spin-2 fields},\ }\href@noop {} {\bibfield  {journal} {\bibinfo
  {journal} {Physical Review Letters}\ }\textbf {\bibinfo {volume} {124}},\
  \bibinfo {pages} {211101} (\bibinfo {year} {2020}{\natexlab{b}})}\BibitemShut
  {NoStop}%
\bibitem [{\citenamefont {Yoshino}\ and\ \citenamefont
  {Kodama}(2012)}]{Yoshino:2012kn}%
  \BibitemOpen
  \bibfield  {author} {\bibinfo {author} {\bibfnamefont {H.}~\bibnamefont
  {Yoshino}}\ and\ \bibinfo {author} {\bibfnamefont {H.}~\bibnamefont
  {Kodama}},\ }\bibfield  {title} {\bibinfo {title} {{Bosenova collapse of
  axion cloud around a rotating black hole}},\ }\href
  {https://doi.org/10.1143/PTP.128.153} {\bibfield  {journal} {\bibinfo
  {journal} {Prog. Theor. Phys.}\ }\textbf {\bibinfo {volume} {128}},\ \bibinfo
  {pages} {153} (\bibinfo {year} {2012})},\ \Eprint
  {https://arxiv.org/abs/1203.5070} {arXiv:1203.5070 [gr-qc]} \BibitemShut
  {NoStop}%
\bibitem [{\citenamefont {Yoshino}\ and\ \citenamefont
  {Kodama}(2015)}]{Yoshino:2015nsa}%
  \BibitemOpen
  \bibfield  {author} {\bibinfo {author} {\bibfnamefont {H.}~\bibnamefont
  {Yoshino}}\ and\ \bibinfo {author} {\bibfnamefont {H.}~\bibnamefont
  {Kodama}},\ }\bibfield  {title} {\bibinfo {title} {{The bosenova and
  axiverse}},\ }\href {https://doi.org/10.1088/0264-9381/32/21/214001}
  {\bibfield  {journal} {\bibinfo  {journal} {Class. Quant. Grav.}\ }\textbf
  {\bibinfo {volume} {32}},\ \bibinfo {pages} {214001} (\bibinfo {year}
  {2015})},\ \Eprint {https://arxiv.org/abs/1505.00714} {arXiv:1505.00714
  [gr-qc]} \BibitemShut {NoStop}%
\bibitem [{\citenamefont {Omiya}\ \emph {et~al.}(2022)\citenamefont {Omiya},
  \citenamefont {Takahashi},\ and\ \citenamefont {Tanaka}}]{Omiya:2022mwv}%
  \BibitemOpen
  \bibfield  {author} {\bibinfo {author} {\bibfnamefont {H.}~\bibnamefont
  {Omiya}}, \bibinfo {author} {\bibfnamefont {T.}~\bibnamefont {Takahashi}},\
  and\ \bibinfo {author} {\bibfnamefont {T.}~\bibnamefont {Tanaka}},\
  }\bibfield  {title} {\bibinfo {title} {{Adiabatic evolution of the
  self-interacting axion field around rotating black holes}},\ }\href
  {https://doi.org/10.1093/ptep/ptac058} {\bibfield  {journal} {\bibinfo
  {journal} {PTEP}\ }\textbf {\bibinfo {volume} {2022}},\ \bibinfo {pages}
  {043E03} (\bibinfo {year} {2022})},\ \Eprint
  {https://arxiv.org/abs/2201.04382} {arXiv:2201.04382 [gr-qc]} \BibitemShut
  {NoStop}%
\bibitem [{\citenamefont {Gruzinov}(2016)}]{Gruzinov:2016hcq}%
  \BibitemOpen
  \bibfield  {author} {\bibinfo {author} {\bibfnamefont {A.}~\bibnamefont
  {Gruzinov}},\ }\bibfield  {title} {\bibinfo {title} {{Black Hole Spindown by
  Light Bosons}},\ }\href@noop {} {\  (\bibinfo {year} {2016})},\ \Eprint
  {https://arxiv.org/abs/1604.06422} {arXiv:1604.06422 [astro-ph.HE]}
  \BibitemShut {NoStop}%
\bibitem [{\citenamefont {Omiya}\ \emph {et~al.}(2024)\citenamefont {Omiya},
  \citenamefont {Takahashi}, \citenamefont {Tanaka},\ and\ \citenamefont
  {Yoshino}}]{Omiya:2024xlz}%
  \BibitemOpen
  \bibfield  {author} {\bibinfo {author} {\bibfnamefont {H.}~\bibnamefont
  {Omiya}}, \bibinfo {author} {\bibfnamefont {T.}~\bibnamefont {Takahashi}},
  \bibinfo {author} {\bibfnamefont {T.}~\bibnamefont {Tanaka}},\ and\ \bibinfo
  {author} {\bibfnamefont {H.}~\bibnamefont {Yoshino}},\ }\bibfield  {title}
  {\bibinfo {title} {{Deci-Hz gravitational waves from the self-interacting
  axion cloud around a rotating stellar mass black hole}},\ }\href
  {https://doi.org/10.1103/PhysRevD.110.044002} {\bibfield  {journal} {\bibinfo
   {journal} {Phys. Rev. D}\ }\textbf {\bibinfo {volume} {110}},\ \bibinfo
  {pages} {044002} (\bibinfo {year} {2024})},\ \Eprint
  {https://arxiv.org/abs/2404.16265} {arXiv:2404.16265 [gr-qc]} \BibitemShut
  {NoStop}%
\bibitem [{\citenamefont {Rosa}(2010)}]{Rosa:2009ei}%
  \BibitemOpen
  \bibfield  {author} {\bibinfo {author} {\bibfnamefont {J.~G.}\ \bibnamefont
  {Rosa}},\ }\bibfield  {title} {\bibinfo {title} {{The Extremal black hole
  bomb}},\ }\href {https://doi.org/10.1007/JHEP06(2010)015} {\bibfield
  {journal} {\bibinfo  {journal} {JHEP}\ }\textbf {\bibinfo {volume} {06}},\
  \bibinfo {pages} {015}},\ \Eprint {https://arxiv.org/abs/0912.1780}
  {arXiv:0912.1780 [hep-th]} \BibitemShut {NoStop}%
\bibitem [{\citenamefont {Leaver}(1985)}]{leaver1985analytic}%
  \BibitemOpen
  \bibfield  {author} {\bibinfo {author} {\bibfnamefont {E.~W.}\ \bibnamefont
  {Leaver}},\ }\bibfield  {title} {\bibinfo {title} {An analytic representation
  for the quasi-normal modes of kerr black holes},\ }\href@noop {} {\bibfield
  {journal} {\bibinfo  {journal} {Proceedings of the Royal Society of London.
  A. Mathematical and Physical Sciences}\ }\textbf {\bibinfo {volume} {402}},\
  \bibinfo {pages} {285} (\bibinfo {year} {1985})}\BibitemShut {NoStop}%
\bibitem [{\citenamefont {Konoplya}\ and\ \citenamefont
  {Zhidenko}(2006)}]{konoplya2006stability}%
  \BibitemOpen
  \bibfield  {author} {\bibinfo {author} {\bibfnamefont {R.}~\bibnamefont
  {Konoplya}}\ and\ \bibinfo {author} {\bibfnamefont {A.}~\bibnamefont
  {Zhidenko}},\ }\bibfield  {title} {\bibinfo {title} {Stability and
  quasinormal modes of the massive scalar field around kerr black holes},\
  }\href@noop {} {\bibfield  {journal} {\bibinfo  {journal} {Physical Review
  D—Particles, Fields, Gravitation, and Cosmology}\ }\textbf {\bibinfo
  {volume} {73}},\ \bibinfo {pages} {124040} (\bibinfo {year}
  {2006})}\BibitemShut {NoStop}%
\bibitem [{\citenamefont {Cardoso}\ and\ \citenamefont
  {Yoshida}(2005)}]{cardoso2005superradiant}%
  \BibitemOpen
  \bibfield  {author} {\bibinfo {author} {\bibfnamefont {V.}~\bibnamefont
  {Cardoso}}\ and\ \bibinfo {author} {\bibfnamefont {S.}~\bibnamefont
  {Yoshida}},\ }\bibfield  {title} {\bibinfo {title} {Superradiant
  instabilities of rotating black branes and strings},\ }\href@noop {}
  {\bibfield  {journal} {\bibinfo  {journal} {Journal of High Energy Physics}\
  }\textbf {\bibinfo {volume} {2005}},\ \bibinfo {pages} {009} (\bibinfo {year}
  {2005})}\BibitemShut {NoStop}%
\bibitem [{\citenamefont {Fukuda}\ and\ \citenamefont
  {Nakayama}(2020)}]{Fukuda:2019ewf}%
  \BibitemOpen
  \bibfield  {author} {\bibinfo {author} {\bibfnamefont {H.}~\bibnamefont
  {Fukuda}}\ and\ \bibinfo {author} {\bibfnamefont {K.}~\bibnamefont
  {Nakayama}},\ }\bibfield  {title} {\bibinfo {title} {{Aspects of Nonlinear
  Effect on Black Hole Superradiance}},\ }\href
  {https://doi.org/10.1007/JHEP01(2020)128} {\bibfield  {journal} {\bibinfo
  {journal} {JHEP}\ }\textbf {\bibinfo {volume} {01}},\ \bibinfo {pages}
  {128}},\ \Eprint {https://arxiv.org/abs/1910.06308} {arXiv:1910.06308
  [hep-ph]} \BibitemShut {NoStop}%
\bibitem [{\citenamefont {Rosa}\ and\ \citenamefont
  {Kephart}(2018)}]{Rosa:2017ury}%
  \BibitemOpen
  \bibfield  {author} {\bibinfo {author} {\bibfnamefont {J.~a.~G.}\
  \bibnamefont {Rosa}}\ and\ \bibinfo {author} {\bibfnamefont {T.~W.}\
  \bibnamefont {Kephart}},\ }\bibfield  {title} {\bibinfo {title} {{Stimulated
  Axion Decay in Superradiant Clouds around Primordial Black Holes}},\ }\href
  {https://doi.org/10.1103/PhysRevLett.120.231102} {\bibfield  {journal}
  {\bibinfo  {journal} {Phys. Rev. Lett.}\ }\textbf {\bibinfo {volume} {120}},\
  \bibinfo {pages} {231102} (\bibinfo {year} {2018})},\ \Eprint
  {https://arxiv.org/abs/1709.06581} {arXiv:1709.06581 [gr-qc]} \BibitemShut
  {NoStop}%
\bibitem [{\citenamefont {Ikeda}\ \emph {et~al.}(2019)\citenamefont {Ikeda},
  \citenamefont {Brito},\ and\ \citenamefont {Cardoso}}]{Ikeda:2018nhb}%
  \BibitemOpen
  \bibfield  {author} {\bibinfo {author} {\bibfnamefont {T.}~\bibnamefont
  {Ikeda}}, \bibinfo {author} {\bibfnamefont {R.}~\bibnamefont {Brito}},\ and\
  \bibinfo {author} {\bibfnamefont {V.}~\bibnamefont {Cardoso}},\ }\bibfield
  {title} {\bibinfo {title} {{Blasts of Light from Axions}},\ }\href
  {https://doi.org/10.1103/PhysRevLett.122.081101} {\bibfield  {journal}
  {\bibinfo  {journal} {Phys. Rev. Lett.}\ }\textbf {\bibinfo {volume} {122}},\
  \bibinfo {pages} {081101} (\bibinfo {year} {2019})},\ \Eprint
  {https://arxiv.org/abs/1811.04950} {arXiv:1811.04950 [gr-qc]} \BibitemShut
  {NoStop}%
\bibitem [{\citenamefont {Mathur}\ \emph {et~al.}(2020)\citenamefont {Mathur},
  \citenamefont {Rajendran},\ and\ \citenamefont {Tanin}}]{Mathur:2020aqv}%
  \BibitemOpen
  \bibfield  {author} {\bibinfo {author} {\bibfnamefont {A.}~\bibnamefont
  {Mathur}}, \bibinfo {author} {\bibfnamefont {S.}~\bibnamefont {Rajendran}},\
  and\ \bibinfo {author} {\bibfnamefont {E.~H.}\ \bibnamefont {Tanin}},\
  }\bibfield  {title} {\bibinfo {title} {{Clockwork mechanism to remove
  superradiance limits}},\ }\href {https://doi.org/10.1103/PhysRevD.102.055015}
  {\bibfield  {journal} {\bibinfo  {journal} {Phys. Rev. D}\ }\textbf {\bibinfo
  {volume} {102}},\ \bibinfo {pages} {055015} (\bibinfo {year} {2020})},\
  \Eprint {https://arxiv.org/abs/2004.12326} {arXiv:2004.12326 [hep-ph]}
  \BibitemShut {NoStop}%
\bibitem [{\citenamefont {Blas}\ and\ \citenamefont
  {Witte}(2020{\natexlab{a}})}]{Blas:2020kaa}%
  \BibitemOpen
  \bibfield  {author} {\bibinfo {author} {\bibfnamefont {D.}~\bibnamefont
  {Blas}}\ and\ \bibinfo {author} {\bibfnamefont {S.~J.}\ \bibnamefont
  {Witte}},\ }\bibfield  {title} {\bibinfo {title} {{Quenching Mechanisms of
  Photon Superradiance}},\ }\href {https://doi.org/10.1103/PhysRevD.102.123018}
  {\bibfield  {journal} {\bibinfo  {journal} {Phys. Rev. D}\ }\textbf {\bibinfo
  {volume} {102}},\ \bibinfo {pages} {123018} (\bibinfo {year}
  {2020}{\natexlab{a}})},\ \Eprint {https://arxiv.org/abs/2009.10075}
  {arXiv:2009.10075 [hep-ph]} \BibitemShut {NoStop}%
\bibitem [{\citenamefont {Blas}\ and\ \citenamefont
  {Witte}(2020{\natexlab{b}})}]{Blas:2020nbs}%
  \BibitemOpen
  \bibfield  {author} {\bibinfo {author} {\bibfnamefont {D.}~\bibnamefont
  {Blas}}\ and\ \bibinfo {author} {\bibfnamefont {S.~J.}\ \bibnamefont
  {Witte}},\ }\bibfield  {title} {\bibinfo {title} {{Imprints of Axion
  Superradiance in the CMB}},\ }\href
  {https://doi.org/10.1103/PhysRevD.102.103018} {\bibfield  {journal} {\bibinfo
   {journal} {Phys. Rev. D}\ }\textbf {\bibinfo {volume} {102}},\ \bibinfo
  {pages} {103018} (\bibinfo {year} {2020}{\natexlab{b}})},\ \Eprint
  {https://arxiv.org/abs/2009.10074} {arXiv:2009.10074 [astro-ph.CO]}
  \BibitemShut {NoStop}%
\bibitem [{\citenamefont {Caputo}\ \emph {et~al.}(2021)\citenamefont {Caputo},
  \citenamefont {Witte}, \citenamefont {Blas},\ and\ \citenamefont
  {Pani}}]{Caputo:2021efm}%
  \BibitemOpen
  \bibfield  {author} {\bibinfo {author} {\bibfnamefont {A.}~\bibnamefont
  {Caputo}}, \bibinfo {author} {\bibfnamefont {S.~J.}\ \bibnamefont {Witte}},
  \bibinfo {author} {\bibfnamefont {D.}~\bibnamefont {Blas}},\ and\ \bibinfo
  {author} {\bibfnamefont {P.}~\bibnamefont {Pani}},\ }\bibfield  {title}
  {\bibinfo {title} {{Electromagnetic signatures of dark photon
  superradiance}},\ }\href {https://doi.org/10.1103/PhysRevD.104.043006}
  {\bibfield  {journal} {\bibinfo  {journal} {Phys. Rev. D}\ }\textbf {\bibinfo
  {volume} {104}},\ \bibinfo {pages} {043006} (\bibinfo {year} {2021})},\
  \Eprint {https://arxiv.org/abs/2102.11280} {arXiv:2102.11280 [hep-ph]}
  \BibitemShut {NoStop}%
\bibitem [{\citenamefont {Siemonsen}\ \emph {et~al.}(2023)\citenamefont
  {Siemonsen}, \citenamefont {Mondino}, \citenamefont {Egana-Ugrinovic},
  \citenamefont {Huang}, \citenamefont {Baryakhtar},\ and\ \citenamefont
  {East}}]{Siemonsen:2022ivj}%
  \BibitemOpen
  \bibfield  {author} {\bibinfo {author} {\bibfnamefont {N.}~\bibnamefont
  {Siemonsen}}, \bibinfo {author} {\bibfnamefont {C.}~\bibnamefont {Mondino}},
  \bibinfo {author} {\bibfnamefont {D.}~\bibnamefont {Egana-Ugrinovic}},
  \bibinfo {author} {\bibfnamefont {J.}~\bibnamefont {Huang}}, \bibinfo
  {author} {\bibfnamefont {M.}~\bibnamefont {Baryakhtar}},\ and\ \bibinfo
  {author} {\bibfnamefont {W.~E.}\ \bibnamefont {East}},\ }\bibfield  {title}
  {\bibinfo {title} {{Dark photon superradiance: Electrodynamics and
  multimessenger signals}},\ }\href
  {https://doi.org/10.1103/PhysRevD.107.075025} {\bibfield  {journal} {\bibinfo
   {journal} {Phys. Rev. D}\ }\textbf {\bibinfo {volume} {107}},\ \bibinfo
  {pages} {075025} (\bibinfo {year} {2023})},\ \Eprint
  {https://arxiv.org/abs/2212.09772} {arXiv:2212.09772 [astro-ph.HE]}
  \BibitemShut {NoStop}%
\bibitem [{\citenamefont {Spieksma}\ \emph {et~al.}(2023)\citenamefont
  {Spieksma}, \citenamefont {Cannizzaro}, \citenamefont {Ikeda}, \citenamefont
  {Cardoso},\ and\ \citenamefont {Chen}}]{Spieksma:2023vwl}%
  \BibitemOpen
  \bibfield  {author} {\bibinfo {author} {\bibfnamefont {T.~F.~M.}\
  \bibnamefont {Spieksma}}, \bibinfo {author} {\bibfnamefont {E.}~\bibnamefont
  {Cannizzaro}}, \bibinfo {author} {\bibfnamefont {T.}~\bibnamefont {Ikeda}},
  \bibinfo {author} {\bibfnamefont {V.}~\bibnamefont {Cardoso}},\ and\ \bibinfo
  {author} {\bibfnamefont {Y.}~\bibnamefont {Chen}},\ }\bibfield  {title}
  {\bibinfo {title} {{Superradiance: Axionic couplings and plasma effects}},\
  }\href {https://doi.org/10.1103/PhysRevD.108.063013} {\bibfield  {journal}
  {\bibinfo  {journal} {Phys. Rev. D}\ }\textbf {\bibinfo {volume} {108}},\
  \bibinfo {pages} {063013} (\bibinfo {year} {2023})},\ \Eprint
  {https://arxiv.org/abs/2306.16447} {arXiv:2306.16447 [gr-qc]} \BibitemShut
  {NoStop}%
\bibitem [{\citenamefont {Ferreira}\ and\ \citenamefont
  {Gil~Muyor}(2024)}]{Ferreira:2024ktd}%
  \BibitemOpen
  \bibfield  {author} {\bibinfo {author} {\bibfnamefont {R.~Z.}\ \bibnamefont
  {Ferreira}}\ and\ \bibinfo {author} {\bibfnamefont {A.}~\bibnamefont
  {Gil~Muyor}},\ }\bibfield  {title} {\bibinfo {title} {{Lightening up
  primordial black holes in the galaxy with the QCD axion: Signals at the LOFAR
  telescope}},\ }\href {https://doi.org/10.1103/PhysRevD.110.083013} {\bibfield
   {journal} {\bibinfo  {journal} {Phys. Rev. D}\ }\textbf {\bibinfo {volume}
  {110}},\ \bibinfo {pages} {083013} (\bibinfo {year} {2024})},\ \Eprint
  {https://arxiv.org/abs/2404.12437} {arXiv:2404.12437 [hep-ph]} \BibitemShut
  {NoStop}%
\bibitem [{\citenamefont {Brito}\ \emph
  {et~al.}(2015{\natexlab{b}})\citenamefont {Brito}, \citenamefont {Cardoso},\
  and\ \citenamefont {Pani}}]{Brito:2014wla}%
  \BibitemOpen
  \bibfield  {author} {\bibinfo {author} {\bibfnamefont {R.}~\bibnamefont
  {Brito}}, \bibinfo {author} {\bibfnamefont {V.}~\bibnamefont {Cardoso}},\
  and\ \bibinfo {author} {\bibfnamefont {P.}~\bibnamefont {Pani}},\ }\bibfield
  {title} {\bibinfo {title} {{Black holes as particle detectors: evolution of
  superradiant instabilities}},\ }\href
  {https://doi.org/10.1088/0264-9381/32/13/134001} {\bibfield  {journal}
  {\bibinfo  {journal} {Class. Quant. Grav.}\ }\textbf {\bibinfo {volume}
  {32}},\ \bibinfo {pages} {134001} (\bibinfo {year} {2015}{\natexlab{b}})},\
  \Eprint {https://arxiv.org/abs/1411.0686} {arXiv:1411.0686 [gr-qc]}
  \BibitemShut {NoStop}%
\bibitem [{\citenamefont {Stegmann}\ \emph {et~al.}(2025)\citenamefont
  {Stegmann}, \citenamefont {Olejak},\ and\ \citenamefont
  {de~Mink}}]{Stegmann:2025cja}%
  \BibitemOpen
  \bibfield  {author} {\bibinfo {author} {\bibfnamefont {J.}~\bibnamefont
  {Stegmann}}, \bibinfo {author} {\bibfnamefont {A.}~\bibnamefont {Olejak}},\
  and\ \bibinfo {author} {\bibfnamefont {S.~E.}\ \bibnamefont {de~Mink}},\
  }\bibfield  {title} {\bibinfo {title} {{Resolving Black Hole Family Issues
  Among the Massive Ancestors of Very High-Spin Gravitational-Wave Events Like
  GW231123}},\ }\href@noop {} {\  (\bibinfo {year} {2025})},\ \Eprint
  {https://arxiv.org/abs/2507.15967} {arXiv:2507.15967 [astro-ph.HE]}
  \BibitemShut {NoStop}%
\bibitem [{\citenamefont {Li}\ \emph {et~al.}(2025)\citenamefont {Li},
  \citenamefont {Tang}, \citenamefont {Xue},\ and\ \citenamefont
  {Fan}}]{Li:2025fnf}%
  \BibitemOpen
  \bibfield  {author} {\bibinfo {author} {\bibfnamefont {Y.-J.}\ \bibnamefont
  {Li}}, \bibinfo {author} {\bibfnamefont {S.-P.}\ \bibnamefont {Tang}},
  \bibinfo {author} {\bibfnamefont {L.-Q.}\ \bibnamefont {Xue}},\ and\ \bibinfo
  {author} {\bibfnamefont {Y.-Z.}\ \bibnamefont {Fan}},\ }\bibfield  {title}
  {\bibinfo {title} {{GW231123: a product of successive mergers from $\sim 10 $
  stellar-mass black holes}},\ }\href@noop {} {\  (\bibinfo {year} {2025})},\
  \Eprint {https://arxiv.org/abs/2507.17551} {arXiv:2507.17551 [astro-ph.HE]}
  \BibitemShut {NoStop}%
\bibitem [{\citenamefont {Rodriguez}\ \emph {et~al.}(2019)\citenamefont
  {Rodriguez}, \citenamefont {Zevin}, \citenamefont {Amaro-Seoane},
  \citenamefont {Chatterjee}, \citenamefont {Kremer}, \citenamefont {Rasio},\
  and\ \citenamefont {Ye}}]{Rodriguez:2019huv}%
  \BibitemOpen
  \bibfield  {author} {\bibinfo {author} {\bibfnamefont {C.~L.}\ \bibnamefont
  {Rodriguez}}, \bibinfo {author} {\bibfnamefont {M.}~\bibnamefont {Zevin}},
  \bibinfo {author} {\bibfnamefont {P.}~\bibnamefont {Amaro-Seoane}}, \bibinfo
  {author} {\bibfnamefont {S.}~\bibnamefont {Chatterjee}}, \bibinfo {author}
  {\bibfnamefont {K.}~\bibnamefont {Kremer}}, \bibinfo {author} {\bibfnamefont
  {F.~A.}\ \bibnamefont {Rasio}},\ and\ \bibinfo {author} {\bibfnamefont
  {C.~S.}\ \bibnamefont {Ye}},\ }\bibfield  {title} {\bibinfo {title} {{Black
  holes: The next generation{\textemdash}repeated mergers in dense star
  clusters and their gravitational-wave properties}},\ }\href
  {https://doi.org/10.1103/PhysRevD.100.043027} {\bibfield  {journal} {\bibinfo
   {journal} {Phys. Rev. D}\ }\textbf {\bibinfo {volume} {100}},\ \bibinfo
  {pages} {043027} (\bibinfo {year} {2019})},\ \Eprint
  {https://arxiv.org/abs/1906.10260} {arXiv:1906.10260 [astro-ph.HE]}
  \BibitemShut {NoStop}%
\bibitem [{\citenamefont {Antonini}\ and\ \citenamefont
  {Rasio}(2016)}]{Antonini:2016gqe}%
  \BibitemOpen
  \bibfield  {author} {\bibinfo {author} {\bibfnamefont {F.}~\bibnamefont
  {Antonini}}\ and\ \bibinfo {author} {\bibfnamefont {F.~A.}\ \bibnamefont
  {Rasio}},\ }\bibfield  {title} {\bibinfo {title} {{Merging black hole
  binaries in galactic nuclei: implications for advanced-LIGO detections}},\
  }\href {https://doi.org/10.3847/0004-637X/831/2/187} {\bibfield  {journal}
  {\bibinfo  {journal} {Astrophys. J.}\ }\textbf {\bibinfo {volume} {831}},\
  \bibinfo {pages} {187} (\bibinfo {year} {2016})},\ \Eprint
  {https://arxiv.org/abs/1606.04889} {arXiv:1606.04889 [astro-ph.HE]}
  \BibitemShut {NoStop}%
\bibitem [{\citenamefont {Mapelli}\ \emph
  {et~al.}(2021{\natexlab{a}})\citenamefont {Mapelli}, \citenamefont
  {Dall’Amico}, \citenamefont {Bouffanais}, \citenamefont {Giacobbo},
  \citenamefont {Arca~Sedda}, \citenamefont {Artale}, \citenamefont {Ballone},
  \citenamefont {Di~Carlo}, \citenamefont {Iorio}, \citenamefont
  {Santoliquido},\ and\ \citenamefont {Torniamenti}}]{Mapelli:2021syv}%
  \BibitemOpen
  \bibfield  {author} {\bibinfo {author} {\bibfnamefont {M.}~\bibnamefont
  {Mapelli}}, \bibinfo {author} {\bibfnamefont {M.}~\bibnamefont
  {Dall’Amico}}, \bibinfo {author} {\bibfnamefont {Y.}~\bibnamefont
  {Bouffanais}}, \bibinfo {author} {\bibfnamefont {N.}~\bibnamefont
  {Giacobbo}}, \bibinfo {author} {\bibfnamefont {M.}~\bibnamefont
  {Arca~Sedda}}, \bibinfo {author} {\bibfnamefont {M.~C.}\ \bibnamefont
  {Artale}}, \bibinfo {author} {\bibfnamefont {A.}~\bibnamefont {Ballone}},
  \bibinfo {author} {\bibfnamefont {U.~N.}\ \bibnamefont {Di~Carlo}}, \bibinfo
  {author} {\bibfnamefont {G.}~\bibnamefont {Iorio}}, \bibinfo {author}
  {\bibfnamefont {F.}~\bibnamefont {Santoliquido}},\ and\ \bibinfo {author}
  {\bibfnamefont {S.}~\bibnamefont {Torniamenti}},\ }\bibfield  {title}
  {\bibinfo {title} {Hierarchical black hole mergers in young, globular and
  nuclear star clusters: the effect of metallicity, spin and cluster
  properties},\ }\href {https://doi.org/10.1093/mnras/stab1240} {\bibfield
  {journal} {\bibinfo  {journal} {Mon. Not. Roy. Astron. Soc.}\ }\textbf
  {\bibinfo {volume} {505}},\ \bibinfo {pages} {339} (\bibinfo {year}
  {2021}{\natexlab{a}})},\ \Eprint {https://arxiv.org/abs/2103.05016}
  {arXiv:2103.05016 [astro-ph.HE]} \BibitemShut {NoStop}%
\bibitem [{\citenamefont {Antonini}\ \emph {et~al.}(2023)\citenamefont
  {Antonini}, \citenamefont {Gieles}, \citenamefont {Dosopoulou},\ and\
  \citenamefont {Chattopadhyay}}]{Antonini:2022kly}%
  \BibitemOpen
  \bibfield  {author} {\bibinfo {author} {\bibfnamefont {F.}~\bibnamefont
  {Antonini}}, \bibinfo {author} {\bibfnamefont {M.}~\bibnamefont {Gieles}},
  \bibinfo {author} {\bibfnamefont {F.}~\bibnamefont {Dosopoulou}},\ and\
  \bibinfo {author} {\bibfnamefont {D.}~\bibnamefont {Chattopadhyay}},\
  }\bibfield  {title} {\bibinfo {title} {Coalescing black hole binaries from
  globular clusters: mass distributions and comparison to gravitational wave
  data from gwtc-3},\ }\href {https://doi.org/10.1093/mnras/stad983} {\bibfield
   {journal} {\bibinfo  {journal} {Mon. Not. Roy. Astron. Soc.}\ }\textbf
  {\bibinfo {volume} {522}},\ \bibinfo {pages} {466} (\bibinfo {year}
  {2023})},\ \Eprint {https://arxiv.org/abs/2208.01081} {arXiv:2208.01081
  [astro-ph.HE]} \BibitemShut {NoStop}%
\bibitem [{\citenamefont {Chattopadhyay}\ \emph {et~al.}(2023)\citenamefont
  {Chattopadhyay}, \citenamefont {Stegmann}, \citenamefont {Antonini},
  \citenamefont {Barber},\ and\ \citenamefont
  {Romero-Shaw}}]{Chattopadhyay:2023yyk}%
  \BibitemOpen
  \bibfield  {author} {\bibinfo {author} {\bibfnamefont {D.}~\bibnamefont
  {Chattopadhyay}}, \bibinfo {author} {\bibfnamefont {J.}~\bibnamefont
  {Stegmann}}, \bibinfo {author} {\bibfnamefont {F.}~\bibnamefont {Antonini}},
  \bibinfo {author} {\bibfnamefont {J.}~\bibnamefont {Barber}},\ and\ \bibinfo
  {author} {\bibfnamefont {I.~M.}\ \bibnamefont {Romero-Shaw}},\ }\bibfield
  {title} {\bibinfo {title} {Double black hole mergers in nuclear star
  clusters: eccentricities, spins, masses, and the growth of massive seeds},\
  }\href {https://doi.org/10.1093/mnras/stad3600} {\bibfield  {journal}
  {\bibinfo  {journal} {Mon. Not. Roy. Astron. Soc.}\ }\textbf {\bibinfo
  {volume} {526}},\ \bibinfo {pages} {4908} (\bibinfo {year} {2023})},\ \Eprint
  {https://arxiv.org/abs/2308.10884} {arXiv:2308.10884 [astro-ph.HE]}
  \BibitemShut {NoStop}%
\bibitem [{\citenamefont {Mahapatra}\ \emph {et~al.}(2025)\citenamefont
  {Mahapatra}, \citenamefont {Chattopadhyay}, \citenamefont {Gupta},
  \citenamefont {Favata}, \citenamefont {Sathyaprakash},\ and\ \citenamefont
  {Arun}}]{Mahapatra:2022orr}%
  \BibitemOpen
  \bibfield  {author} {\bibinfo {author} {\bibfnamefont {P.}~\bibnamefont
  {Mahapatra}}, \bibinfo {author} {\bibfnamefont {D.}~\bibnamefont
  {Chattopadhyay}}, \bibinfo {author} {\bibfnamefont {A.}~\bibnamefont
  {Gupta}}, \bibinfo {author} {\bibfnamefont {M.}~\bibnamefont {Favata}},
  \bibinfo {author} {\bibfnamefont {B.~S.}\ \bibnamefont {Sathyaprakash}},\
  and\ \bibinfo {author} {\bibfnamefont {K.~G.}\ \bibnamefont {Arun}},\
  }\bibfield  {title} {\bibinfo {title} {Predictions of a simple parametric
  model of hierarchical black hole mergers},\ }\href
  {https://doi.org/10.1103/PhysRevD.111.023013} {\bibfield  {journal} {\bibinfo
   {journal} {Phys. Rev. D}\ }\textbf {\bibinfo {volume} {111}},\ \bibinfo
  {pages} {023013} (\bibinfo {year} {2025})},\ \Eprint
  {https://arxiv.org/abs/2209.05766} {arXiv:2209.05766 [astro-ph.HE]}
  \BibitemShut {NoStop}%
\bibitem [{\citenamefont {Yang}\ \emph {et~al.}(2019)\citenamefont {Yang},
  \citenamefont {Bartos}, \citenamefont {Gayathri}, \citenamefont {Ford},
  \citenamefont {Haiman}, \citenamefont {Klimenko}, \citenamefont {Kocsis},
  \citenamefont {Márka}, \citenamefont {Márka}, \citenamefont {McKernan},\
  and\ \citenamefont {O'Shaughnessy}}]{Yang:2019cbr}%
  \BibitemOpen
  \bibfield  {author} {\bibinfo {author} {\bibfnamefont {Y.}~\bibnamefont
  {Yang}}, \bibinfo {author} {\bibfnamefont {I.}~\bibnamefont {Bartos}},
  \bibinfo {author} {\bibfnamefont {V.}~\bibnamefont {Gayathri}}, \bibinfo
  {author} {\bibfnamefont {K.~E.~S.}\ \bibnamefont {Ford}}, \bibinfo {author}
  {\bibfnamefont {Z.}~\bibnamefont {Haiman}}, \bibinfo {author} {\bibfnamefont
  {S.}~\bibnamefont {Klimenko}}, \bibinfo {author} {\bibfnamefont
  {B.}~\bibnamefont {Kocsis}}, \bibinfo {author} {\bibfnamefont
  {S.}~\bibnamefont {Márka}}, \bibinfo {author} {\bibfnamefont
  {Z.}~\bibnamefont {Márka}}, \bibinfo {author} {\bibfnamefont
  {B.}~\bibnamefont {McKernan}},\ and\ \bibinfo {author} {\bibfnamefont
  {R.}~\bibnamefont {O'Shaughnessy}},\ }\bibfield  {title} {\bibinfo {title}
  {Hierarchical black hole mergers in active galactic nuclei},\ }\href
  {https://doi.org/10.1103/PhysRevLett.123.181101} {\bibfield  {journal}
  {\bibinfo  {journal} {Phys. Rev. Lett.}\ }\textbf {\bibinfo {volume} {123}},\
  \bibinfo {pages} {181101} (\bibinfo {year} {2019})},\ \Eprint
  {https://arxiv.org/abs/1906.09281} {arXiv:1906.09281 [astro-ph.HE]}
  \BibitemShut {NoStop}%
\bibitem [{\citenamefont {Arca~Sedda}\ \emph {et~al.}(2020)\citenamefont
  {Arca~Sedda}, \citenamefont {Mapelli}, \citenamefont {Spera}, \citenamefont
  {Benacquista},\ and\ \citenamefont {Giacobbo}}]{ArcaSedda:2020hbe}%
  \BibitemOpen
  \bibfield  {author} {\bibinfo {author} {\bibfnamefont {M.}~\bibnamefont
  {Arca~Sedda}}, \bibinfo {author} {\bibfnamefont {M.}~\bibnamefont {Mapelli}},
  \bibinfo {author} {\bibfnamefont {M.}~\bibnamefont {Spera}}, \bibinfo
  {author} {\bibfnamefont {M.}~\bibnamefont {Benacquista}},\ and\ \bibinfo
  {author} {\bibfnamefont {N.}~\bibnamefont {Giacobbo}},\ }\bibfield  {title}
  {\bibinfo {title} {Fingerprints of binary black hole formation channels
  encoded in the mass and spin of merger remnants},\ }\href
  {https://doi.org/10.3847/1538-4357/ab8863} {\bibfield  {journal} {\bibinfo
  {journal} {Astrophys. J.}\ }\textbf {\bibinfo {volume} {894}},\ \bibinfo
  {pages} {133} (\bibinfo {year} {2020})},\ \Eprint
  {https://arxiv.org/abs/2003.07409} {arXiv:2003.07409 [astro-ph.GA]}
  \BibitemShut {NoStop}%
\bibitem [{\citenamefont {Vaccaro}\ \emph
  {et~al.}(2024{\natexlab{a}})\citenamefont {Vaccaro}, \citenamefont {Mapelli},
  \citenamefont {Périgois}, \citenamefont {Barone}, \citenamefont {Artale},
  \citenamefont {Dall’Amico}, \citenamefont {Iorio},\ and\ \citenamefont
  {Torniamenti}}]{Vaccaro:2023gas}%
  \BibitemOpen
  \bibfield  {author} {\bibinfo {author} {\bibfnamefont {M.~P.}\ \bibnamefont
  {Vaccaro}}, \bibinfo {author} {\bibfnamefont {M.}~\bibnamefont {Mapelli}},
  \bibinfo {author} {\bibfnamefont {C.}~\bibnamefont {Périgois}}, \bibinfo
  {author} {\bibfnamefont {D.}~\bibnamefont {Barone}}, \bibinfo {author}
  {\bibfnamefont {M.~C.}\ \bibnamefont {Artale}}, \bibinfo {author}
  {\bibfnamefont {M.}~\bibnamefont {Dall’Amico}}, \bibinfo {author}
  {\bibfnamefont {G.}~\bibnamefont {Iorio}},\ and\ \bibinfo {author}
  {\bibfnamefont {S.}~\bibnamefont {Torniamenti}},\ }\bibfield  {title}
  {\bibinfo {title} {Impact of gas hardening on the population properties of
  hierarchical black hole mergers in active galactic nucleus disks},\ }\href
  {https://doi.org/10.1051/0004-6361/202346072} {\bibfield  {journal} {\bibinfo
   {journal} {Astron. Astrophys.}\ }\textbf {\bibinfo {volume} {685}},\
  \bibinfo {pages} {A51} (\bibinfo {year} {2024}{\natexlab{a}})},\ \Eprint
  {https://arxiv.org/abs/2311.18548} {arXiv:2311.18548 [astro-ph.HE]}
  \BibitemShut {NoStop}%
\bibitem [{\citenamefont {Gilbaum}\ \emph {et~al.}(2025)\citenamefont
  {Gilbaum}, \citenamefont {Grishin}, \citenamefont {Stone},\ and\
  \citenamefont {Mandel}}]{Gilbaum:2024xxq}%
  \BibitemOpen
  \bibfield  {author} {\bibinfo {author} {\bibfnamefont {S.}~\bibnamefont
  {Gilbaum}}, \bibinfo {author} {\bibfnamefont {E.}~\bibnamefont {Grishin}},
  \bibinfo {author} {\bibfnamefont {N.~C.}\ \bibnamefont {Stone}},\ and\
  \bibinfo {author} {\bibfnamefont {I.}~\bibnamefont {Mandel}},\ }\bibfield
  {title} {\bibinfo {title} {{How to Escape from a Trap: Outcomes of Repeated
  Black Hole Mergers in Active Galactic Nuclei}},\ }\href
  {https://doi.org/10.3847/2041-8213/adb7dc} {\bibfield  {journal} {\bibinfo
  {journal} {Astrophys. J. Lett.}\ }\textbf {\bibinfo {volume} {982}},\
  \bibinfo {pages} {L13} (\bibinfo {year} {2025})},\ \Eprint
  {https://arxiv.org/abs/2410.19904} {arXiv:2410.19904 [astro-ph.HE]}
  \BibitemShut {NoStop}%
\bibitem [{\citenamefont {Rodriguez}\ \emph {et~al.}(2020)\citenamefont
  {Rodriguez}, \citenamefont {Kremer}, \citenamefont {Grudić}, \citenamefont
  {Hafen}, \citenamefont {Chatterjee}, \citenamefont {Fragione}, \citenamefont
  {Lamberts}, \citenamefont {Martinez}, \citenamefont {Rasio}, \citenamefont
  {Weatherford},\ and\ \citenamefont {Ye}}]{Rodriguez:2020viw}%
  \BibitemOpen
  \bibfield  {author} {\bibinfo {author} {\bibfnamefont {C.~L.}\ \bibnamefont
  {Rodriguez}}, \bibinfo {author} {\bibfnamefont {K.}~\bibnamefont {Kremer}},
  \bibinfo {author} {\bibfnamefont {M.~Y.}\ \bibnamefont {Grudić}}, \bibinfo
  {author} {\bibfnamefont {Z.}~\bibnamefont {Hafen}}, \bibinfo {author}
  {\bibfnamefont {S.}~\bibnamefont {Chatterjee}}, \bibinfo {author}
  {\bibfnamefont {G.}~\bibnamefont {Fragione}}, \bibinfo {author}
  {\bibfnamefont {A.}~\bibnamefont {Lamberts}}, \bibinfo {author}
  {\bibfnamefont {M.~A.~S.}\ \bibnamefont {Martinez}}, \bibinfo {author}
  {\bibfnamefont {F.~A.}\ \bibnamefont {Rasio}}, \bibinfo {author}
  {\bibfnamefont {N.}~\bibnamefont {Weatherford}},\ and\ \bibinfo {author}
  {\bibfnamefont {C.~S.}\ \bibnamefont {Ye}},\ }\bibfield  {title} {\bibinfo
  {title} {Gw190412 as a third-generation black hole merger from a super star
  cluster},\ }\href {https://doi.org/10.3847/2041-8213/ab96c6} {\bibfield
  {journal} {\bibinfo  {journal} {Astrophys. J. Lett.}\ }\textbf {\bibinfo
  {volume} {896}},\ \bibinfo {pages} {L10} (\bibinfo {year} {2020})},\ \Eprint
  {https://arxiv.org/abs/2005.04239} {arXiv:2005.04239 [astro-ph.HE]}
  \BibitemShut {NoStop}%
\bibitem [{\citenamefont {Araújo-Álvarez}\ \emph {et~al.}(2024)\citenamefont
  {Araújo-Álvarez}, \citenamefont {Wong}, \citenamefont {Liu},\ and\
  \citenamefont {Calderón~Bustillo}}]{AraujoAlvarez:2024xxx}%
  \BibitemOpen
  \bibfield  {author} {\bibinfo {author} {\bibfnamefont {C.}~\bibnamefont
  {Araújo-Álvarez}}, \bibinfo {author} {\bibfnamefont {H.~W.~Y.}\
  \bibnamefont {Wong}}, \bibinfo {author} {\bibfnamefont {A.}~\bibnamefont
  {Liu}},\ and\ \bibinfo {author} {\bibfnamefont {J.}~\bibnamefont
  {Calderón~Bustillo}},\ }\bibfield  {title} {\bibinfo {title} {Kicking time
  back in black hole mergers: Ancestral masses, spins, birth recoils, and
  hierarchical-formation viability of gw190521},\ }\href
  {https://doi.org/10.3847/1538-4357/ad2b88} {\bibfield  {journal} {\bibinfo
  {journal} {Astrophys. J.}\ }\textbf {\bibinfo {volume} {977}},\ \bibinfo
  {pages} {220} (\bibinfo {year} {2024})},\ \Eprint
  {https://arxiv.org/abs/2404.00720} {arXiv:2404.00720 [astro-ph.HE]}
  \BibitemShut {NoStop}%
\bibitem [{\citenamefont {Mapelli}\ \emph
  {et~al.}(2021{\natexlab{b}})\citenamefont {Mapelli}, \citenamefont
  {Santoliquido}, \citenamefont {Bouffanais}, \citenamefont {Sedda},
  \citenamefont {Artale},\ and\ \citenamefont {Ballone}}]{Mapelli:2020xeq}%
  \BibitemOpen
  \bibfield  {author} {\bibinfo {author} {\bibfnamefont {M.}~\bibnamefont
  {Mapelli}}, \bibinfo {author} {\bibfnamefont {F.}~\bibnamefont
  {Santoliquido}}, \bibinfo {author} {\bibfnamefont {Y.}~\bibnamefont
  {Bouffanais}}, \bibinfo {author} {\bibfnamefont {M.~A.}\ \bibnamefont
  {Sedda}}, \bibinfo {author} {\bibfnamefont {M.~C.}\ \bibnamefont {Artale}},\
  and\ \bibinfo {author} {\bibfnamefont {A.}~\bibnamefont {Ballone}},\
  }\bibfield  {title} {\bibinfo {title} {{Mass and Rate of Hierarchical Black
  Hole Mergers in Young, Globular and Nuclear Star Clusters}},\ }\href
  {https://doi.org/10.3390/sym13091678} {\bibfield  {journal} {\bibinfo
  {journal} {Symmetry}\ }\textbf {\bibinfo {volume} {13}},\ \bibinfo {pages}
  {1678} (\bibinfo {year} {2021}{\natexlab{b}})},\ \Eprint
  {https://arxiv.org/abs/2007.15022} {arXiv:2007.15022 [astro-ph.HE]}
  \BibitemShut {NoStop}%
\bibitem [{\citenamefont {{Fragione}}\ and\ \citenamefont
  {{Silk}}(2020)}]{FragioneSilk}%
  \BibitemOpen
  \bibfield  {author} {\bibinfo {author} {\bibfnamefont {G.}~\bibnamefont
  {{Fragione}}}\ and\ \bibinfo {author} {\bibfnamefont {J.}~\bibnamefont
  {{Silk}}},\ }\bibfield  {title} {\bibinfo {title} {{Repeated mergers and
  ejection of black holes within nuclear star clusters}},\ }\href
  {https://doi.org/10.1093/mnras/staa2629} {\bibfield  {journal} {\bibinfo
  {journal} {Monthly Notices of the Royal Astronomical Society}\ }\textbf
  {\bibinfo {volume} {498}},\ \bibinfo {pages} {4591} (\bibinfo {year}
  {2020})},\ \Eprint {https://arxiv.org/abs/2006.01867} {arXiv:2006.01867
  [astro-ph.GA]} \BibitemShut {NoStop}%
\bibitem [{\citenamefont {{\'A}lvarez}\ \emph {et~al.}(2024)\citenamefont
  {{\'A}lvarez}, \citenamefont {Wong}, \citenamefont {Liu},\ and\ \citenamefont
  {Calder{\'o}n~Bustillo}}]{Alvarez:2024dpd}%
  \BibitemOpen
  \bibfield  {author} {\bibinfo {author} {\bibfnamefont {C.~A.}\ \bibnamefont
  {{\'A}lvarez}}, \bibinfo {author} {\bibfnamefont {H.~W.~Y.}\ \bibnamefont
  {Wong}}, \bibinfo {author} {\bibfnamefont {A.}~\bibnamefont {Liu}},\ and\
  \bibinfo {author} {\bibfnamefont {J.}~\bibnamefont {Calder{\'o}n~Bustillo}},\
  }\bibfield  {title} {\bibinfo {title} {{Kicking Time Back in Black Hole
  Mergers: Ancestral Masses, Spins, Birth Recoils, and Hierarchical-formation
  Viability of GW190521}},\ }\href {https://doi.org/10.3847/1538-4357/ad90a9}
  {\bibfield  {journal} {\bibinfo  {journal} {Astrophys. J.}\ }\textbf
  {\bibinfo {volume} {977}},\ \bibinfo {pages} {220} (\bibinfo {year}
  {2024})},\ \Eprint {https://arxiv.org/abs/2404.00720} {arXiv:2404.00720
  [astro-ph.HE]} \BibitemShut {NoStop}%
\bibitem [{\citenamefont {Antonini}\ \emph {et~al.}(2019)\citenamefont
  {Antonini}, \citenamefont {Gieles},\ and\ \citenamefont
  {Gualandris}}]{Antonini:2018auk}%
  \BibitemOpen
  \bibfield  {author} {\bibinfo {author} {\bibfnamefont {F.}~\bibnamefont
  {Antonini}}, \bibinfo {author} {\bibfnamefont {M.}~\bibnamefont {Gieles}},\
  and\ \bibinfo {author} {\bibfnamefont {A.}~\bibnamefont {Gualandris}},\
  }\bibfield  {title} {\bibinfo {title} {{Black hole growth through
  hierarchical black hole mergers in dense star clusters: implications for
  gravitational wave detections}},\ }\href
  {https://doi.org/10.1093/mnras/stz1149} {\bibfield  {journal} {\bibinfo
  {journal} {Mon. Not. Roy. Astron. Soc.}\ }\textbf {\bibinfo {volume} {486}},\
  \bibinfo {pages} {5008} (\bibinfo {year} {2019})},\ \Eprint
  {https://arxiv.org/abs/1811.03640} {arXiv:1811.03640 [astro-ph.HE]}
  \BibitemShut {NoStop}%
\bibitem [{\citenamefont {Portegies~Zwart}\ and\ \citenamefont
  {McMillan}(2002)}]{PortegiesZwart:2002}%
  \BibitemOpen
  \bibfield  {author} {\bibinfo {author} {\bibfnamefont {S.~F.}\ \bibnamefont
  {Portegies~Zwart}}\ and\ \bibinfo {author} {\bibfnamefont {S.~L.~W.}\
  \bibnamefont {McMillan}},\ }\bibfield  {title} {\bibinfo {title} {The runaway
  growth of intermediate-mass black holes in dense star clusters},\ }\href
  {https://doi.org/10.1086/341798} {\bibfield  {journal} {\bibinfo  {journal}
  {Astrophys. J.}\ }\textbf {\bibinfo {volume} {576}},\ \bibinfo {pages} {899}
  (\bibinfo {year} {2002})},\ \Eprint {https://arxiv.org/abs/astro-ph/0201055}
  {arXiv:astro-ph/0201055 [astro-ph]} \BibitemShut {NoStop}%
\bibitem [{\citenamefont {Lupi}\ \emph {et~al.}(2014)\citenamefont {Lupi},
  \citenamefont {Colpi}, \citenamefont {Devecchi}, \citenamefont {Galanti},\
  and\ \citenamefont {Volonteri}}]{Lupi:2014vza}%
  \BibitemOpen
  \bibfield  {author} {\bibinfo {author} {\bibfnamefont {A.}~\bibnamefont
  {Lupi}}, \bibinfo {author} {\bibfnamefont {M.}~\bibnamefont {Colpi}},
  \bibinfo {author} {\bibfnamefont {B.}~\bibnamefont {Devecchi}}, \bibinfo
  {author} {\bibfnamefont {G.}~\bibnamefont {Galanti}},\ and\ \bibinfo {author}
  {\bibfnamefont {M.}~\bibnamefont {Volonteri}},\ }\bibfield  {title} {\bibinfo
  {title} {{Constraining the high redshift formation of black hole seeds in
  nuclear star clusters with gas inflows}},\ }\href
  {https://doi.org/10.1093/mnras/stu1120} {\bibfield  {journal} {\bibinfo
  {journal} {Mon. Not. Roy. Astron. Soc.}\ }\textbf {\bibinfo {volume} {442}},\
  \bibinfo {pages} {3616} (\bibinfo {year} {2014})},\ \Eprint
  {https://arxiv.org/abs/1406.2325} {arXiv:1406.2325 [astro-ph.GA]}
  \BibitemShut {NoStop}%
\bibitem [{\citenamefont {Rodriguez}\ \emph {et~al.}(2018)\citenamefont
  {Rodriguez}, \citenamefont {Amaro-Seoane}, \citenamefont {Chatterjee},\ and\
  \citenamefont {Rasio}}]{Rodriguez:2017pec}%
  \BibitemOpen
  \bibfield  {author} {\bibinfo {author} {\bibfnamefont {C.~L.}\ \bibnamefont
  {Rodriguez}}, \bibinfo {author} {\bibfnamefont {P.}~\bibnamefont
  {Amaro-Seoane}}, \bibinfo {author} {\bibfnamefont {S.}~\bibnamefont
  {Chatterjee}},\ and\ \bibinfo {author} {\bibfnamefont {F.~A.}\ \bibnamefont
  {Rasio}},\ }\bibfield  {title} {\bibinfo {title} {{Post-Newtonian Dynamics in
  Dense Star Clusters: Highly-Eccentric, Highly-Spinning, and Repeated Binary
  Black Hole Mergers}},\ }\href
  {https://doi.org/10.1103/PhysRevLett.120.151101} {\bibfield  {journal}
  {\bibinfo  {journal} {Phys. Rev. Lett.}\ }\textbf {\bibinfo {volume} {120}},\
  \bibinfo {pages} {151101} (\bibinfo {year} {2018})},\ \Eprint
  {https://arxiv.org/abs/1712.04937} {arXiv:1712.04937 [astro-ph.HE]}
  \BibitemShut {NoStop}%
\bibitem [{\citenamefont {Samsing}\ and\ \citenamefont
  {Hotokezaka}(2021)}]{Samsing:2020qqd}%
  \BibitemOpen
  \bibfield  {author} {\bibinfo {author} {\bibfnamefont {J.}~\bibnamefont
  {Samsing}}\ and\ \bibinfo {author} {\bibfnamefont {K.}~\bibnamefont
  {Hotokezaka}},\ }\bibfield  {title} {\bibinfo {title} {{Populating the Black
  Hole Mass Gaps in Stellar Clusters: General Relations and Upper Limits}},\
  }\href {https://doi.org/10.3847/1538-4357/ac2b27} {\bibfield  {journal}
  {\bibinfo  {journal} {Astrophys. J.}\ }\textbf {\bibinfo {volume} {923}},\
  \bibinfo {pages} {126} (\bibinfo {year} {2021})},\ \Eprint
  {https://arxiv.org/abs/2006.09744} {arXiv:2006.09744 [astro-ph.HE]}
  \BibitemShut {NoStop}%
\bibitem [{\citenamefont {Tagawa}\ \emph {et~al.}(2020)\citenamefont {Tagawa},
  \citenamefont {Haiman},\ and\ \citenamefont {Kocsis}}]{Tagawa:2019osr}%
  \BibitemOpen
  \bibfield  {author} {\bibinfo {author} {\bibfnamefont {H.}~\bibnamefont
  {Tagawa}}, \bibinfo {author} {\bibfnamefont {Z.}~\bibnamefont {Haiman}},\
  and\ \bibinfo {author} {\bibfnamefont {B.}~\bibnamefont {Kocsis}},\
  }\bibfield  {title} {\bibinfo {title} {{Formation and Evolution of Compact
  Object Binaries in AGN Disks}},\ }\href
  {https://doi.org/10.3847/1538-4357/ab9b8c} {\bibfield  {journal} {\bibinfo
  {journal} {Astrophys. J.}\ }\textbf {\bibinfo {volume} {898}},\ \bibinfo
  {pages} {25} (\bibinfo {year} {2020})},\ \Eprint
  {https://arxiv.org/abs/1912.08218} {arXiv:1912.08218 [astro-ph.GA]}
  \BibitemShut {NoStop}%
\bibitem [{\citenamefont {Gr{\"o}bner}\ \emph {et~al.}(2020)\citenamefont
  {Gr{\"o}bner}, \citenamefont {Ishibashi}, \citenamefont {Tiwari},
  \citenamefont {Haney},\ and\ \citenamefont {Jetzer}}]{Grobner:2020drr}%
  \BibitemOpen
  \bibfield  {author} {\bibinfo {author} {\bibfnamefont {M.}~\bibnamefont
  {Gr{\"o}bner}}, \bibinfo {author} {\bibfnamefont {W.}~\bibnamefont
  {Ishibashi}}, \bibinfo {author} {\bibfnamefont {S.}~\bibnamefont {Tiwari}},
  \bibinfo {author} {\bibfnamefont {M.}~\bibnamefont {Haney}},\ and\ \bibinfo
  {author} {\bibfnamefont {P.}~\bibnamefont {Jetzer}},\ }\bibfield  {title}
  {\bibinfo {title} {{Binary black hole mergers in AGN accretion discs:
  gravitational wave rate density estimates}},\ }\href
  {https://doi.org/10.1051/0004-6361/202037681} {\bibfield  {journal} {\bibinfo
   {journal} {Astron. Astrophys.}\ }\textbf {\bibinfo {volume} {638}},\
  \bibinfo {pages} {A119} (\bibinfo {year} {2020})},\ \Eprint
  {https://arxiv.org/abs/2005.03571} {arXiv:2005.03571 [astro-ph.GA]}
  \BibitemShut {NoStop}%
\bibitem [{\citenamefont {Bartos}\ \emph {et~al.}(2017)\citenamefont {Bartos},
  \citenamefont {Kocsis}, \citenamefont {Haiman},\ and\ \citenamefont
  {M{\'a}rka}}]{Bartos:2016dgn}%
  \BibitemOpen
  \bibfield  {author} {\bibinfo {author} {\bibfnamefont {I.}~\bibnamefont
  {Bartos}}, \bibinfo {author} {\bibfnamefont {B.}~\bibnamefont {Kocsis}},
  \bibinfo {author} {\bibfnamefont {Z.}~\bibnamefont {Haiman}},\ and\ \bibinfo
  {author} {\bibfnamefont {S.}~\bibnamefont {M{\'a}rka}},\ }\bibfield  {title}
  {\bibinfo {title} {{Rapid and Bright Stellar-mass Binary Black Hole Mergers
  in Active Galactic Nuclei}},\ }\href
  {https://doi.org/10.3847/1538-4357/835/2/165} {\bibfield  {journal} {\bibinfo
   {journal} {Astrophys. J.}\ }\textbf {\bibinfo {volume} {835}},\ \bibinfo
  {pages} {165} (\bibinfo {year} {2017})},\ \Eprint
  {https://arxiv.org/abs/1602.03831} {arXiv:1602.03831 [astro-ph.HE]}
  \BibitemShut {NoStop}%
\bibitem [{\citenamefont {Leigh}\ \emph {et~al.}(2018)\citenamefont {Leigh},
  \citenamefont {Stone}, \citenamefont {Geller}, \citenamefont {Shara},
  \citenamefont {Mudryk},\ and\ \citenamefont {Fregeau}}]{Leigh:2017zjd}%
  \BibitemOpen
  \bibfield  {author} {\bibinfo {author} {\bibfnamefont {N.~W.~C.}\
  \bibnamefont {Leigh}}, \bibinfo {author} {\bibfnamefont {N.~C.}\ \bibnamefont
  {Stone}}, \bibinfo {author} {\bibfnamefont {A.~M.}\ \bibnamefont {Geller}},
  \bibinfo {author} {\bibfnamefont {M.~M.}\ \bibnamefont {Shara}}, \bibinfo
  {author} {\bibfnamefont {L.}~\bibnamefont {Mudryk}},\ and\ \bibinfo {author}
  {\bibfnamefont {J.~M.}\ \bibnamefont {Fregeau}},\ }\bibfield  {title}
  {\bibinfo {title} {On the rate of black hole binary mergers in galactic
  nuclei due to dynamical hardening},\ }\href
  {https://doi.org/10.1093/mnras/stx3134} {\bibfield  {journal} {\bibinfo
  {journal} {Mon. Not. Roy. Astron. Soc.}\ }\textbf {\bibinfo {volume} {474}},\
  \bibinfo {pages} {5672} (\bibinfo {year} {2018})},\ \Eprint
  {https://arxiv.org/abs/1711.10494} {arXiv:1711.10494 [astro-ph.HE]}
  \BibitemShut {NoStop}%
\bibitem [{\citenamefont {McKernan}\ \emph
  {et~al.}(2018{\natexlab{a}})\citenamefont {McKernan}, \citenamefont {Ford},
  \citenamefont {Bellovary}, \citenamefont {Leigh}, \citenamefont {Haiman},
  \citenamefont {Kocsis}, \citenamefont {Lyra}, \citenamefont {Macsai},
  \citenamefont {O’Dowd}, \citenamefont {S{\"a}ndor},\ and\ \citenamefont
  {Winter}}]{McKernan:2017tvu}%
  \BibitemOpen
  \bibfield  {author} {\bibinfo {author} {\bibfnamefont {B.}~\bibnamefont
  {McKernan}}, \bibinfo {author} {\bibfnamefont {K.~E.~S.}\ \bibnamefont
  {Ford}}, \bibinfo {author} {\bibfnamefont {J.}~\bibnamefont {Bellovary}},
  \bibinfo {author} {\bibfnamefont {N.~W.~C.}\ \bibnamefont {Leigh}}, \bibinfo
  {author} {\bibfnamefont {Z.}~\bibnamefont {Haiman}}, \bibinfo {author}
  {\bibfnamefont {B.}~\bibnamefont {Kocsis}}, \bibinfo {author} {\bibfnamefont
  {W.}~\bibnamefont {Lyra}}, \bibinfo {author} {\bibfnamefont {A.}~\bibnamefont
  {Macsai}}, \bibinfo {author} {\bibfnamefont {M.}~\bibnamefont {O’Dowd}},
  \bibinfo {author} {\bibfnamefont {Z.}~\bibnamefont {S{\"a}ndor}},\ and\
  \bibinfo {author} {\bibfnamefont {L.~M.}\ \bibnamefont {Winter}},\ }\bibfield
   {title} {\bibinfo {title} {Constraining stellar-mass black hole mergers in
  agn disks detectable with ligo},\ }\href
  {https://doi.org/10.3847/1538-4357/aadd52} {\bibfield  {journal} {\bibinfo
  {journal} {Astrophys. J.}\ }\textbf {\bibinfo {volume} {866}},\ \bibinfo
  {pages} {66} (\bibinfo {year} {2018}{\natexlab{a}})},\ \Eprint
  {https://arxiv.org/abs/1702.07818} {arXiv:1702.07818 [astro-ph.HE]}
  \BibitemShut {NoStop}%
\bibitem [{\citenamefont {Secunda}\ \emph {et~al.}(2019)\citenamefont
  {Secunda}, \citenamefont {Bellovary}, \citenamefont {Mac~Low}, \citenamefont
  {Ford}, \citenamefont {McKernan},\ and\ \citenamefont
  {Leigh}}]{Secunda:2018zfg}%
  \BibitemOpen
  \bibfield  {author} {\bibinfo {author} {\bibfnamefont {A.}~\bibnamefont
  {Secunda}}, \bibinfo {author} {\bibfnamefont {J.}~\bibnamefont {Bellovary}},
  \bibinfo {author} {\bibfnamefont {M.-M.}\ \bibnamefont {Mac~Low}}, \bibinfo
  {author} {\bibfnamefont {K.~E.~S.}\ \bibnamefont {Ford}}, \bibinfo {author}
  {\bibfnamefont {B.}~\bibnamefont {McKernan}},\ and\ \bibinfo {author}
  {\bibfnamefont {N.~W.~C.}\ \bibnamefont {Leigh}},\ }\bibfield  {title}
  {\bibinfo {title} {Orbital migration of interacting stellar mass black holes
  in disks around supermassive black holes},\ }\href
  {https://doi.org/10.3847/1538-4357/ab1b3c} {\bibfield  {journal} {\bibinfo
  {journal} {Astrophys. J.}\ }\textbf {\bibinfo {volume} {878}},\ \bibinfo
  {pages} {85} (\bibinfo {year} {2019})},\ \Eprint
  {https://arxiv.org/abs/1807.02859} {arXiv:1807.02859 [astro-ph.HE]}
  \BibitemShut {NoStop}%
\bibitem [{\citenamefont {{Hong}}\ and\ \citenamefont
  {{Lee}}(2015)}]{2015MNRAS.448..754H}%
  \BibitemOpen
  \bibfield  {author} {\bibinfo {author} {\bibfnamefont {J.}~\bibnamefont
  {{Hong}}}\ and\ \bibinfo {author} {\bibfnamefont {H.~M.}\ \bibnamefont
  {{Lee}}},\ }\bibfield  {title} {\bibinfo {title} {{Black hole binaries in
  galactic nuclei and gravitational wave sources}},\ }\href
  {https://doi.org/10.1093/mnras/stv035} {\bibfield  {journal} {\bibinfo
  {journal} {Monthly Notices of the Royal Astronomical Society}\ }\textbf
  {\bibinfo {volume} {448}},\ \bibinfo {pages} {754} (\bibinfo {year}
  {2015})},\ \Eprint {https://arxiv.org/abs/1501.02717} {arXiv:1501.02717
  [astro-ph.GA]} \BibitemShut {NoStop}%
\bibitem [{\citenamefont {O'Leary}\ \emph {et~al.}(2009)\citenamefont
  {O'Leary}, \citenamefont {Kocsis},\ and\ \citenamefont
  {Loeb}}]{OLeary:2008myb}%
  \BibitemOpen
  \bibfield  {author} {\bibinfo {author} {\bibfnamefont {R.~M.}\ \bibnamefont
  {O'Leary}}, \bibinfo {author} {\bibfnamefont {B.}~\bibnamefont {Kocsis}},\
  and\ \bibinfo {author} {\bibfnamefont {A.}~\bibnamefont {Loeb}},\ }\bibfield
  {title} {\bibinfo {title} {{Gravitational waves from scattering of
  stellar-mass black holes in galactic nuclei}},\ }\href
  {https://doi.org/10.1111/j.1365-2966.2009.14653.x} {\bibfield  {journal}
  {\bibinfo  {journal} {Mon. Not. Roy. Astron. Soc.}\ }\textbf {\bibinfo
  {volume} {395}},\ \bibinfo {pages} {2127} (\bibinfo {year} {2009})},\ \Eprint
  {https://arxiv.org/abs/0807.2638} {arXiv:0807.2638 [astro-ph]} \BibitemShut
  {NoStop}%
\bibitem [{\citenamefont {McKernan}\ \emph
  {et~al.}(2018{\natexlab{b}})\citenamefont {McKernan}, \citenamefont
  {Saavik~Ford}, \citenamefont {Bellovary}, \citenamefont {Leigh},
  \citenamefont {Haiman}, \citenamefont {Kocsis}, \citenamefont {Lyra},
  \citenamefont {Mac~Low}, \citenamefont {Metzger}, \citenamefont {O’Dowd},
  \citenamefont {Endlich},\ and\ \citenamefont {Rosen}}]{McKernan_2018}%
  \BibitemOpen
  \bibfield  {author} {\bibinfo {author} {\bibfnamefont {B.}~\bibnamefont
  {McKernan}}, \bibinfo {author} {\bibfnamefont {K.~E.}\ \bibnamefont
  {Saavik~Ford}}, \bibinfo {author} {\bibfnamefont {J.}~\bibnamefont
  {Bellovary}}, \bibinfo {author} {\bibfnamefont {N.~W.~C.}\ \bibnamefont
  {Leigh}}, \bibinfo {author} {\bibfnamefont {Z.}~\bibnamefont {Haiman}},
  \bibinfo {author} {\bibfnamefont {B.}~\bibnamefont {Kocsis}}, \bibinfo
  {author} {\bibfnamefont {W.}~\bibnamefont {Lyra}}, \bibinfo {author}
  {\bibfnamefont {M.-M.}\ \bibnamefont {Mac~Low}}, \bibinfo {author}
  {\bibfnamefont {B.}~\bibnamefont {Metzger}}, \bibinfo {author} {\bibfnamefont
  {M.}~\bibnamefont {O’Dowd}}, \bibinfo {author} {\bibfnamefont
  {S.}~\bibnamefont {Endlich}},\ and\ \bibinfo {author} {\bibfnamefont {D.~J.}\
  \bibnamefont {Rosen}},\ }\bibfield  {title} {\bibinfo {title} {Constraining
  stellar-mass black hole mergers in agn disks detectable with ligo},\ }\href
  {https://doi.org/10.3847/1538-4357/aadae5} {\bibfield  {journal} {\bibinfo
  {journal} {The Astrophysical Journal}\ }\textbf {\bibinfo {volume} {866}},\
  \bibinfo {pages} {66} (\bibinfo {year} {2018}{\natexlab{b}})}\BibitemShut
  {NoStop}%
\bibitem [{\citenamefont {McKernan}\ \emph {et~al.}(2012)\citenamefont
  {McKernan}, \citenamefont {Ford}, \citenamefont {Lyra},\ and\ \citenamefont
  {Perets}}]{McKernan:2012}%
  \BibitemOpen
  \bibfield  {author} {\bibinfo {author} {\bibfnamefont {B.}~\bibnamefont
  {McKernan}}, \bibinfo {author} {\bibfnamefont {K.~E.~S.}\ \bibnamefont
  {Ford}}, \bibinfo {author} {\bibfnamefont {W.}~\bibnamefont {Lyra}},\ and\
  \bibinfo {author} {\bibfnamefont {H.~B.}\ \bibnamefont {Perets}},\ }\bibfield
   {title} {\bibinfo {title} {Intermediate mass black holes in agn discs - i.
  production and growth},\ }\href
  {https://doi.org/10.1111/j.1365-2966.2012.21560.x} {\bibfield  {journal}
  {\bibinfo  {journal} {Mon. Not. R. Astron. Soc.}\ }\textbf {\bibinfo {volume}
  {425}},\ \bibinfo {pages} {460} (\bibinfo {year} {2012})},\ \Eprint
  {https://arxiv.org/abs/1206.2309} {arXiv:1206.2309 [astro-ph.HE]}
  \BibitemShut {NoStop}%
\bibitem [{\citenamefont {McKernan}\ \emph {et~al.}(2014)\citenamefont
  {McKernan}, \citenamefont {Ford}, \citenamefont {Kocsis}, \citenamefont
  {Lyra},\ and\ \citenamefont {Winter}}]{McKernan:2014}%
  \BibitemOpen
  \bibfield  {author} {\bibinfo {author} {\bibfnamefont {B.}~\bibnamefont
  {McKernan}}, \bibinfo {author} {\bibfnamefont {K.~E.~S.}\ \bibnamefont
  {Ford}}, \bibinfo {author} {\bibfnamefont {B.}~\bibnamefont {Kocsis}},
  \bibinfo {author} {\bibfnamefont {W.}~\bibnamefont {Lyra}},\ and\ \bibinfo
  {author} {\bibfnamefont {L.~M.}\ \bibnamefont {Winter}},\ }\bibfield  {title}
  {\bibinfo {title} {Intermediate-mass black holes in agn discs - ii. model
  predictions and observational constraints},\ }\href
  {https://doi.org/10.1093/mnras/stu553} {\bibfield  {journal} {\bibinfo
  {journal} {Mon. Not. R. Astron. Soc.}\ }\textbf {\bibinfo {volume} {441}},\
  \bibinfo {pages} {900} (\bibinfo {year} {2014})},\ \Eprint
  {https://arxiv.org/abs/1403.6433} {arXiv:1403.6433 [astro-ph.HE]}
  \BibitemShut {NoStop}%
\bibitem [{\citenamefont {McKernan}\ \emph {et~al.}(2020)\citenamefont
  {McKernan}, \citenamefont {Ford}, \citenamefont {O'Shaughnessy},\ and\
  \citenamefont {Wysocki}}]{McKernan:2020}%
  \BibitemOpen
  \bibfield  {author} {\bibinfo {author} {\bibfnamefont {B.}~\bibnamefont
  {McKernan}}, \bibinfo {author} {\bibfnamefont {K.~E.~S.}\ \bibnamefont
  {Ford}}, \bibinfo {author} {\bibfnamefont {R.}~\bibnamefont
  {O'Shaughnessy}},\ and\ \bibinfo {author} {\bibfnamefont {D.}~\bibnamefont
  {Wysocki}},\ }\bibfield  {title} {\bibinfo {title} {Monte carlo simulations
  of black hole mergers in agn discs: Low chi\_eff mergers and predictions for
  ligo},\ }\href {https://doi.org/10.1093/mnras/staa731} {\bibfield  {journal}
  {\bibinfo  {journal} {Mon. Not. R. Astron. Soc.}\ }\textbf {\bibinfo {volume}
  {494}},\ \bibinfo {pages} {1203} (\bibinfo {year} {2020})},\ \Eprint
  {https://arxiv.org/abs/1907.04356} {arXiv:1907.04356 [astro-ph.HE]}
  \BibitemShut {NoStop}%
\bibitem [{\citenamefont {Weatherford}\ \emph {et~al.}(2023)\citenamefont
  {Weatherford}, \citenamefont {Kıroğlu}, \citenamefont {Fragione},
  \citenamefont {Chatterjee}, \citenamefont {Kremer},\ and\ \citenamefont
  {Rasio}}]{Weatherford_2023}%
  \BibitemOpen
  \bibfield  {author} {\bibinfo {author} {\bibfnamefont {N.~C.}\ \bibnamefont
  {Weatherford}}, \bibinfo {author} {\bibfnamefont {F.}~\bibnamefont
  {Kıroğlu}}, \bibinfo {author} {\bibfnamefont {G.}~\bibnamefont {Fragione}},
  \bibinfo {author} {\bibfnamefont {S.}~\bibnamefont {Chatterjee}}, \bibinfo
  {author} {\bibfnamefont {K.}~\bibnamefont {Kremer}},\ and\ \bibinfo {author}
  {\bibfnamefont {F.~A.}\ \bibnamefont {Rasio}},\ }\bibfield  {title} {\bibinfo
  {title} {Stellar escape from globular clusters. i. escape mechanisms and
  properties at ejection},\ }\href {https://doi.org/10.3847/1538-4357/acbcc1}
  {\bibfield  {journal} {\bibinfo  {journal} {The Astrophysical Journal}\
  }\textbf {\bibinfo {volume} {946}},\ \bibinfo {pages} {104} (\bibinfo {year}
  {2023})}\BibitemShut {NoStop}%
\bibitem [{\citenamefont {Fragione}\ and\ \citenamefont
  {Kocsis}(2018)}]{Fragione:2018vty}%
  \BibitemOpen
  \bibfield  {author} {\bibinfo {author} {\bibfnamefont {G.}~\bibnamefont
  {Fragione}}\ and\ \bibinfo {author} {\bibfnamefont {B.}~\bibnamefont
  {Kocsis}},\ }\bibfield  {title} {\bibinfo {title} {{Black hole mergers from
  an evolving population of globular clusters}},\ }\href
  {https://doi.org/10.1103/PhysRevLett.121.161103} {\bibfield  {journal}
  {\bibinfo  {journal} {Phys. Rev. Lett.}\ }\textbf {\bibinfo {volume} {121}},\
  \bibinfo {pages} {161103} (\bibinfo {year} {2018})},\ \Eprint
  {https://arxiv.org/abs/1806.02351} {arXiv:1806.02351 [astro-ph.GA]}
  \BibitemShut {NoStop}%
\bibitem [{\citenamefont {Oh}\ and\ \citenamefont {Kroupa}(2016)}]{Oh_2016}%
  \BibitemOpen
  \bibfield  {author} {\bibinfo {author} {\bibfnamefont {S.}~\bibnamefont
  {Oh}}\ and\ \bibinfo {author} {\bibfnamefont {P.}~\bibnamefont {Kroupa}},\
  }\bibfield  {title} {\bibinfo {title} {Dynamical ejections of massive stars
  from young star clusters under diverse initial conditions},\ }\href
  {https://doi.org/10.1051/0004-6361/201628233} {\bibfield  {journal} {\bibinfo
   {journal} {Astronomy \& Astrophysics}\ }\textbf {\bibinfo {volume} {590}},\
  \bibinfo {pages} {A107} (\bibinfo {year} {2016})}\BibitemShut {NoStop}%
\bibitem [{\citenamefont {Mouri}\ and\ \citenamefont
  {Taniguchi}(2002)}]{Mouri_2002}%
  \BibitemOpen
  \bibfield  {author} {\bibinfo {author} {\bibfnamefont {H.}~\bibnamefont
  {Mouri}}\ and\ \bibinfo {author} {\bibfnamefont {Y.}~\bibnamefont
  {Taniguchi}},\ }\bibfield  {title} {\bibinfo {title} {Runaway merging of
  black holes: Analytical constraint on the timescale},\ }\href
  {https://doi.org/10.1086/339472} {\bibfield  {journal} {\bibinfo  {journal}
  {The Astrophysical Journal}\ }\textbf {\bibinfo {volume} {566}},\ \bibinfo
  {pages} {L17–L20} (\bibinfo {year} {2002})}\BibitemShut {NoStop}%
\bibitem [{\citenamefont {{Goodman}}\ and\ \citenamefont
  {{Hut}}(1993)}]{Goodman1993}%
  \BibitemOpen
  \bibfield  {author} {\bibinfo {author} {\bibfnamefont {J.}~\bibnamefont
  {{Goodman}}}\ and\ \bibinfo {author} {\bibfnamefont {P.}~\bibnamefont
  {{Hut}}},\ }\bibfield  {title} {\bibinfo {title} {{Binary--Single-Star
  Scattering. V. Steady State Binary Distribution in a Homogeneous Static
  Background of Single Stars}},\ }\href {https://doi.org/10.1086/172200}
  {\bibfield  {journal} {\bibinfo  {journal} {\apj}\ }\textbf {\bibinfo
  {volume} {403}},\ \bibinfo {pages} {271} (\bibinfo {year}
  {1993})}\BibitemShut {NoStop}%
\bibitem [{\citenamefont {Pina}\ and\ \citenamefont {Gieles}(2023)}]{pina2023}%
  \BibitemOpen
  \bibfield  {author} {\bibinfo {author} {\bibfnamefont {D.~M.}\ \bibnamefont
  {Pina}}\ and\ \bibinfo {author} {\bibfnamefont {M.}~\bibnamefont {Gieles}},\
  }\href {https://arxiv.org/abs/2308.10318} {\bibinfo {title} {Demographics of
  three-body binary black holes in star clusters: implications for
  gravitational waves}} (\bibinfo {year} {2023}),\ \Eprint
  {https://arxiv.org/abs/2308.10318} {arXiv:2308.10318 [astro-ph.GA]}
  \BibitemShut {NoStop}%
\bibitem [{\citenamefont {Peters}\ and\ \citenamefont
  {Mathews}(1963)}]{Peters:1963ux}%
  \BibitemOpen
  \bibfield  {author} {\bibinfo {author} {\bibfnamefont {P.~C.}\ \bibnamefont
  {Peters}}\ and\ \bibinfo {author} {\bibfnamefont {J.}~\bibnamefont
  {Mathews}},\ }\bibfield  {title} {\bibinfo {title} {{Gravitational radiation
  from point masses in a Keplerian orbit}},\ }\href
  {https://doi.org/10.1103/PhysRev.131.435} {\bibfield  {journal} {\bibinfo
  {journal} {Phys. Rev.}\ }\textbf {\bibinfo {volume} {131}},\ \bibinfo {pages}
  {435} (\bibinfo {year} {1963})}\BibitemShut {NoStop}%
\bibitem [{\citenamefont {Peters}(1964)}]{Peters:1964zz}%
  \BibitemOpen
  \bibfield  {author} {\bibinfo {author} {\bibfnamefont {P.~C.}\ \bibnamefont
  {Peters}},\ }\bibfield  {title} {\bibinfo {title} {{Gravitational Radiation
  and the Motion of Two Point Masses}},\ }\href
  {https://doi.org/10.1103/PhysRev.136.B1224} {\bibfield  {journal} {\bibinfo
  {journal} {Phys. Rev.}\ }\textbf {\bibinfo {volume} {136}},\ \bibinfo {pages}
  {B1224} (\bibinfo {year} {1964})}\BibitemShut {NoStop}%
\bibitem [{\citenamefont {Bellovary}\ \emph {et~al.}(2016)\citenamefont
  {Bellovary}, \citenamefont {Mac~Low}, \citenamefont {McKernan},\ and\
  \citenamefont {Ford}}]{Bellovary:2015ifg}%
  \BibitemOpen
  \bibfield  {author} {\bibinfo {author} {\bibfnamefont {J.~M.}\ \bibnamefont
  {Bellovary}}, \bibinfo {author} {\bibfnamefont {M.-M.}\ \bibnamefont
  {Mac~Low}}, \bibinfo {author} {\bibfnamefont {B.}~\bibnamefont {McKernan}},\
  and\ \bibinfo {author} {\bibfnamefont {K.~E.~S.}\ \bibnamefont {Ford}},\
  }\bibfield  {title} {\bibinfo {title} {{Migration Traps in Disks Around
  Supermassive Black Holes}},\ }\href
  {https://doi.org/10.3847/2041-8205/819/2/L17} {\bibfield  {journal} {\bibinfo
   {journal} {Astrophys. J. Lett.}\ }\textbf {\bibinfo {volume} {819}},\
  \bibinfo {pages} {L17} (\bibinfo {year} {2016})},\ \Eprint
  {https://arxiv.org/abs/1511.00005} {arXiv:1511.00005 [astro-ph.GA]}
  \BibitemShut {NoStop}%
\bibitem [{\citenamefont {{Silverman}}\ \emph {et~al.}(2005)\citenamefont
  {{Silverman}}, \citenamefont {{Green}}, \citenamefont {{Barkhouse}},
  \citenamefont {{Cameron}}, \citenamefont {{Foltz}}, \citenamefont
  {{Jannuzi}}, \citenamefont {{Kim}}, \citenamefont {{Kim}}, \citenamefont
  {{Mossman}}, \citenamefont {{Tananbaum}}, \citenamefont {{Wilkes}},
  \citenamefont {{Smith}}, \citenamefont {{Smith}},\ and\ \citenamefont
  {{Smith}}}]{2005ApJ...624..630S}%
  \BibitemOpen
  \bibfield  {author} {\bibinfo {author} {\bibfnamefont {J.~D.}\ \bibnamefont
  {{Silverman}}}, \bibinfo {author} {\bibfnamefont {P.~J.}\ \bibnamefont
  {{Green}}}, \bibinfo {author} {\bibfnamefont {W.~A.}\ \bibnamefont
  {{Barkhouse}}}, \bibinfo {author} {\bibfnamefont {R.~A.}\ \bibnamefont
  {{Cameron}}}, \bibinfo {author} {\bibfnamefont {C.}~\bibnamefont {{Foltz}}},
  \bibinfo {author} {\bibfnamefont {B.~T.}\ \bibnamefont {{Jannuzi}}}, \bibinfo
  {author} {\bibfnamefont {D.~W.}\ \bibnamefont {{Kim}}}, \bibinfo {author}
  {\bibfnamefont {M.}~\bibnamefont {{Kim}}}, \bibinfo {author} {\bibfnamefont
  {A.}~\bibnamefont {{Mossman}}}, \bibinfo {author} {\bibfnamefont
  {H.}~\bibnamefont {{Tananbaum}}}, \bibinfo {author} {\bibfnamefont {B.~J.}\
  \bibnamefont {{Wilkes}}}, \bibinfo {author} {\bibfnamefont {M.~G.}\
  \bibnamefont {{Smith}}}, \bibinfo {author} {\bibfnamefont {R.~C.}\
  \bibnamefont {{Smith}}},\ and\ \bibinfo {author} {\bibfnamefont {P.~S.}\
  \bibnamefont {{Smith}}},\ }\bibfield  {title} {\bibinfo {title} {{Comoving
  Space Density of X-Ray-selected Active Galactic Nuclei}},\ }\href
  {https://doi.org/10.1086/429361} {\bibfield  {journal} {\bibinfo  {journal}
  {\apj}\ }\textbf {\bibinfo {volume} {624}},\ \bibinfo {pages} {630} (\bibinfo
  {year} {2005})},\ \Eprint {https://arxiv.org/abs/astro-ph/0406330}
  {arXiv:astro-ph/0406330 [astro-ph]} \BibitemShut {NoStop}%
\bibitem [{\citenamefont {Secunda}\ \emph {et~al.}(2020)\citenamefont
  {Secunda}, \citenamefont {Bellovary}, \citenamefont {Mac~Low}, \citenamefont
  {Ford}, \citenamefont {McKernan}, \citenamefont {Leigh}, \citenamefont
  {Lyra}, \citenamefont {Sandor},\ and\ \citenamefont
  {Adorno}}]{Secunda:2020mhd}%
  \BibitemOpen
  \bibfield  {author} {\bibinfo {author} {\bibfnamefont {A.}~\bibnamefont
  {Secunda}}, \bibinfo {author} {\bibfnamefont {J.}~\bibnamefont {Bellovary}},
  \bibinfo {author} {\bibfnamefont {M.-M.}\ \bibnamefont {Mac~Low}}, \bibinfo
  {author} {\bibfnamefont {K.~E.~S.}\ \bibnamefont {Ford}}, \bibinfo {author}
  {\bibfnamefont {B.}~\bibnamefont {McKernan}}, \bibinfo {author}
  {\bibfnamefont {N.~W.~C.}\ \bibnamefont {Leigh}}, \bibinfo {author}
  {\bibfnamefont {W.}~\bibnamefont {Lyra}}, \bibinfo {author} {\bibfnamefont
  {Z.}~\bibnamefont {Sandor}},\ and\ \bibinfo {author} {\bibfnamefont {J.~I.}\
  \bibnamefont {Adorno}},\ }\bibfield  {title} {\bibinfo {title} {{Orbital
  Migration of Interacting Stellar Mass Black Holes in Disks around
  Supermassive Black Holes II. Spins and Incoming Objects}},\ }\href
  {https://doi.org/10.3847/1538-4357/abbc1d} {\bibfield  {journal} {\bibinfo
  {journal} {Astrophys. J.}\ }\textbf {\bibinfo {volume} {903}},\ \bibinfo
  {pages} {133} (\bibinfo {year} {2020})},\ \Eprint
  {https://arxiv.org/abs/2004.11936} {arXiv:2004.11936 [astro-ph.HE]}
  \BibitemShut {NoStop}%
\bibitem [{\citenamefont {Neumayer}\ \emph {et~al.}(2020)\citenamefont
  {Neumayer}, \citenamefont {Seth},\ and\ \citenamefont
  {Boeker}}]{Neumayer:2020gno}%
  \BibitemOpen
  \bibfield  {author} {\bibinfo {author} {\bibfnamefont {N.}~\bibnamefont
  {Neumayer}}, \bibinfo {author} {\bibfnamefont {A.}~\bibnamefont {Seth}},\
  and\ \bibinfo {author} {\bibfnamefont {T.}~\bibnamefont {Boeker}},\
  }\bibfield  {title} {\bibinfo {title} {{Nuclear star clusters}},\ }\href
  {https://doi.org/10.1007/s00159-020-00125-0} {\bibfield  {journal} {\bibinfo
  {journal} {Astron. Astrophys. Rev.}\ }\textbf {\bibinfo {volume} {28}},\
  \bibinfo {pages} {4} (\bibinfo {year} {2020})},\ \Eprint
  {https://arxiv.org/abs/2001.03626} {arXiv:2001.03626 [astro-ph.GA]}
  \BibitemShut {NoStop}%
\bibitem [{\citenamefont {{S{\'a}nchez-Janssen}}\ \emph
  {et~al.}(2019)\citenamefont {{S{\'a}nchez-Janssen}}, \citenamefont
  {{C{\^o}t{\'e}}}, \citenamefont {{Ferrarese}}, \citenamefont {{Peng}},
  \citenamefont {{Roediger}}, \citenamefont {{Blakeslee}}, \citenamefont
  {{Emsellem}}, \citenamefont {{Puzia}}, \citenamefont {{Spengler}},
  \citenamefont {{Taylor}}, \citenamefont {{{\'A}lamo-Mart{\'\i}nez}},
  \citenamefont {{Boselli}}, \citenamefont {{Cantiello}}, \citenamefont
  {{Cuillandre}}, \citenamefont {{Duc}}, \citenamefont {{Durrell}},
  \citenamefont {{Gwyn}}, \citenamefont {{MacArthur}}, \citenamefont
  {{Lan{\c{c}}on}}, \citenamefont {{Lim}}, \citenamefont {{Liu}}, \citenamefont
  {{Mei}}, \citenamefont {{Miller}}, \citenamefont {{Mu{\~n}oz}}, \citenamefont
  {{Mihos}}, \citenamefont {{Paudel}}, \citenamefont {{Powalka}},\ and\
  \citenamefont {{Toloba}}}]{NSCfraction}%
  \BibitemOpen
  \bibfield  {author} {\bibinfo {author} {\bibfnamefont {R.}~\bibnamefont
  {{S{\'a}nchez-Janssen}}}, \bibinfo {author} {\bibfnamefont {P.}~\bibnamefont
  {{C{\^o}t{\'e}}}}, \bibinfo {author} {\bibfnamefont {L.}~\bibnamefont
  {{Ferrarese}}}, \bibinfo {author} {\bibfnamefont {E.~W.}\ \bibnamefont
  {{Peng}}}, \bibinfo {author} {\bibfnamefont {J.}~\bibnamefont {{Roediger}}},
  \bibinfo {author} {\bibfnamefont {J.~P.}\ \bibnamefont {{Blakeslee}}},
  \bibinfo {author} {\bibfnamefont {E.}~\bibnamefont {{Emsellem}}}, \bibinfo
  {author} {\bibfnamefont {T.~H.}\ \bibnamefont {{Puzia}}}, \bibinfo {author}
  {\bibfnamefont {C.}~\bibnamefont {{Spengler}}}, \bibinfo {author}
  {\bibfnamefont {J.}~\bibnamefont {{Taylor}}}, \bibinfo {author}
  {\bibfnamefont {K.~A.}\ \bibnamefont {{{\'A}lamo-Mart{\'\i}nez}}}, \bibinfo
  {author} {\bibfnamefont {A.}~\bibnamefont {{Boselli}}}, \bibinfo {author}
  {\bibfnamefont {M.}~\bibnamefont {{Cantiello}}}, \bibinfo {author}
  {\bibfnamefont {J.-C.}\ \bibnamefont {{Cuillandre}}}, \bibinfo {author}
  {\bibfnamefont {P.-A.}\ \bibnamefont {{Duc}}}, \bibinfo {author}
  {\bibfnamefont {P.}~\bibnamefont {{Durrell}}}, \bibinfo {author}
  {\bibfnamefont {S.}~\bibnamefont {{Gwyn}}}, \bibinfo {author} {\bibfnamefont
  {L.~A.}\ \bibnamefont {{MacArthur}}}, \bibinfo {author} {\bibfnamefont
  {A.}~\bibnamefont {{Lan{\c{c}}on}}}, \bibinfo {author} {\bibfnamefont
  {S.}~\bibnamefont {{Lim}}}, \bibinfo {author} {\bibfnamefont
  {C.}~\bibnamefont {{Liu}}}, \bibinfo {author} {\bibfnamefont
  {S.}~\bibnamefont {{Mei}}}, \bibinfo {author} {\bibfnamefont
  {B.}~\bibnamefont {{Miller}}}, \bibinfo {author} {\bibfnamefont
  {R.}~\bibnamefont {{Mu{\~n}oz}}}, \bibinfo {author} {\bibfnamefont {J.~C.}\
  \bibnamefont {{Mihos}}}, \bibinfo {author} {\bibfnamefont {S.}~\bibnamefont
  {{Paudel}}}, \bibinfo {author} {\bibfnamefont {M.}~\bibnamefont
  {{Powalka}}},\ and\ \bibinfo {author} {\bibfnamefont {E.}~\bibnamefont
  {{Toloba}}},\ }\bibfield  {title} {\bibinfo {title} {{The Next Generation
  Virgo Cluster Survey. XXIII. Fundamentals of Nuclear Star Clusters over Seven
  Decades in Galaxy Mass}},\ }\href {https://doi.org/10.3847/1538-4357/aaf4fd}
  {\bibfield  {journal} {\bibinfo  {journal} {\apj}\ }\textbf {\bibinfo
  {volume} {878}},\ \bibinfo {eid} {18} (\bibinfo {year} {2019})},\ \Eprint
  {https://arxiv.org/abs/1812.01019} {arXiv:1812.01019 [astro-ph.GA]}
  \BibitemShut {NoStop}%
\bibitem [{\citenamefont {{Wright}}\ \emph {et~al.}(2017)\citenamefont
  {{Wright}}, \citenamefont {{Robotham}}, \citenamefont {{Driver}},
  \citenamefont {{Alpaslan}}, \citenamefont {{Andrews}}, \citenamefont
  {{Baldry}}, \citenamefont {{Bland-Hawthorn}}, \citenamefont {{Brough}},
  \citenamefont {{Brown}}, \citenamefont {{Colless}}, \citenamefont {{da
  Cunha}}, \citenamefont {{Davies}}, \citenamefont {{Graham}}, \citenamefont
  {{Holwerda}}, \citenamefont {{Hopkins}}, \citenamefont {{Kafle}},
  \citenamefont {{Kelvin}}, \citenamefont {{Loveday}}, \citenamefont
  {{Maddox}}, \citenamefont {{Meyer}}, \citenamefont {{Moffett}}, \citenamefont
  {{Norberg}}, \citenamefont {{Phillipps}}, \citenamefont {{Rowlands}},
  \citenamefont {{Taylor}}, \citenamefont {{Wang}},\ and\ \citenamefont
  {{Wilkins}}}]{2017MNRAS.470..283W}%
  \BibitemOpen
  \bibfield  {author} {\bibinfo {author} {\bibfnamefont {A.~H.}\ \bibnamefont
  {{Wright}}}, \bibinfo {author} {\bibfnamefont {A.~S.~G.}\ \bibnamefont
  {{Robotham}}}, \bibinfo {author} {\bibfnamefont {S.~P.}\ \bibnamefont
  {{Driver}}}, \bibinfo {author} {\bibfnamefont {M.}~\bibnamefont
  {{Alpaslan}}}, \bibinfo {author} {\bibfnamefont {S.~K.}\ \bibnamefont
  {{Andrews}}}, \bibinfo {author} {\bibfnamefont {I.~K.}\ \bibnamefont
  {{Baldry}}}, \bibinfo {author} {\bibfnamefont {J.}~\bibnamefont
  {{Bland-Hawthorn}}}, \bibinfo {author} {\bibfnamefont {S.}~\bibnamefont
  {{Brough}}}, \bibinfo {author} {\bibfnamefont {M.~J.~I.}\ \bibnamefont
  {{Brown}}}, \bibinfo {author} {\bibfnamefont {M.}~\bibnamefont {{Colless}}},
  \bibinfo {author} {\bibfnamefont {E.}~\bibnamefont {{da Cunha}}}, \bibinfo
  {author} {\bibfnamefont {L.~J.~M.}\ \bibnamefont {{Davies}}}, \bibinfo
  {author} {\bibfnamefont {A.~W.}\ \bibnamefont {{Graham}}}, \bibinfo {author}
  {\bibfnamefont {B.~W.}\ \bibnamefont {{Holwerda}}}, \bibinfo {author}
  {\bibfnamefont {A.~M.}\ \bibnamefont {{Hopkins}}}, \bibinfo {author}
  {\bibfnamefont {P.~R.}\ \bibnamefont {{Kafle}}}, \bibinfo {author}
  {\bibfnamefont {L.~S.}\ \bibnamefont {{Kelvin}}}, \bibinfo {author}
  {\bibfnamefont {J.}~\bibnamefont {{Loveday}}}, \bibinfo {author}
  {\bibfnamefont {S.~J.}\ \bibnamefont {{Maddox}}}, \bibinfo {author}
  {\bibfnamefont {M.~J.}\ \bibnamefont {{Meyer}}}, \bibinfo {author}
  {\bibfnamefont {A.~J.}\ \bibnamefont {{Moffett}}}, \bibinfo {author}
  {\bibfnamefont {P.}~\bibnamefont {{Norberg}}}, \bibinfo {author}
  {\bibfnamefont {S.}~\bibnamefont {{Phillipps}}}, \bibinfo {author}
  {\bibfnamefont {K.}~\bibnamefont {{Rowlands}}}, \bibinfo {author}
  {\bibfnamefont {E.~N.}\ \bibnamefont {{Taylor}}}, \bibinfo {author}
  {\bibfnamefont {L.}~\bibnamefont {{Wang}}},\ and\ \bibinfo {author}
  {\bibfnamefont {S.~M.}\ \bibnamefont {{Wilkins}}},\ }\bibfield  {title}
  {\bibinfo {title} {{Galaxy And Mass Assembly (GAMA): the galaxy stellar mass
  function to z = 0.1 from the r-band selected equatorial regions}},\ }\href
  {https://doi.org/10.1093/mnras/stx1149} {\bibfield  {journal} {\bibinfo
  {journal} {Monthly Notices of the Royal Astronomical Society}\ }\textbf
  {\bibinfo {volume} {470}},\ \bibinfo {pages} {283} (\bibinfo {year}
  {2017})},\ \Eprint {https://arxiv.org/abs/1705.04074} {arXiv:1705.04074
  [astro-ph.GA]} \BibitemShut {NoStop}%
\bibitem [{\citenamefont {Weigel}\ \emph {et~al.}(2016)\citenamefont {Weigel},
  \citenamefont {Schawinski},\ and\ \citenamefont {Bruderer}}]{Weigel_2016}%
  \BibitemOpen
  \bibfield  {author} {\bibinfo {author} {\bibfnamefont {A.~K.}\ \bibnamefont
  {Weigel}}, \bibinfo {author} {\bibfnamefont {K.}~\bibnamefont {Schawinski}},\
  and\ \bibinfo {author} {\bibfnamefont {C.}~\bibnamefont {Bruderer}},\
  }\bibfield  {title} {\bibinfo {title} {Stellar mass functions: methods,
  systematics and results for the local universe},\ }\href
  {https://doi.org/10.1093/mnras/stw756} {\bibfield  {journal} {\bibinfo
  {journal} {Monthly Notices of the Royal Astronomical Society}\ }\textbf
  {\bibinfo {volume} {459}},\ \bibinfo {pages} {2150–2187} (\bibinfo {year}
  {2016})}\BibitemShut {NoStop}%
\bibitem [{\citenamefont {{Baldry}}\ \emph {et~al.}(2012)\citenamefont
  {{Baldry}}, \citenamefont {{Driver}}, \citenamefont {{Loveday}},
  \citenamefont {{Taylor}}, \citenamefont {{Kelvin}}, \citenamefont {{Liske}},
  \citenamefont {{Norberg}}, \citenamefont {{Robotham}}, \citenamefont
  {{Brough}}, \citenamefont {{Hopkins}}, \citenamefont {{Bamford}},
  \citenamefont {{Peacock}}, \citenamefont {{Bland-Hawthorn}}, \citenamefont
  {{Conselice}}, \citenamefont {{Croom}}, \citenamefont {{Jones}},
  \citenamefont {{Parkinson}}, \citenamefont {{Popescu}}, \citenamefont
  {{Prescott}}, \citenamefont {{Sharp}},\ and\ \citenamefont
  {{Tuffs}}}]{2012MNRAS.421..621B}%
  \BibitemOpen
  \bibfield  {author} {\bibinfo {author} {\bibfnamefont {I.~K.}\ \bibnamefont
  {{Baldry}}}, \bibinfo {author} {\bibfnamefont {S.~P.}\ \bibnamefont
  {{Driver}}}, \bibinfo {author} {\bibfnamefont {J.}~\bibnamefont {{Loveday}}},
  \bibinfo {author} {\bibfnamefont {E.~N.}\ \bibnamefont {{Taylor}}}, \bibinfo
  {author} {\bibfnamefont {L.~S.}\ \bibnamefont {{Kelvin}}}, \bibinfo {author}
  {\bibfnamefont {J.}~\bibnamefont {{Liske}}}, \bibinfo {author} {\bibfnamefont
  {P.}~\bibnamefont {{Norberg}}}, \bibinfo {author} {\bibfnamefont {A.~S.~G.}\
  \bibnamefont {{Robotham}}}, \bibinfo {author} {\bibfnamefont
  {S.}~\bibnamefont {{Brough}}}, \bibinfo {author} {\bibfnamefont {A.~M.}\
  \bibnamefont {{Hopkins}}}, \bibinfo {author} {\bibfnamefont {S.~P.}\
  \bibnamefont {{Bamford}}}, \bibinfo {author} {\bibfnamefont {J.~A.}\
  \bibnamefont {{Peacock}}}, \bibinfo {author} {\bibfnamefont {J.}~\bibnamefont
  {{Bland-Hawthorn}}}, \bibinfo {author} {\bibfnamefont {C.~J.}\ \bibnamefont
  {{Conselice}}}, \bibinfo {author} {\bibfnamefont {S.~M.}\ \bibnamefont
  {{Croom}}}, \bibinfo {author} {\bibfnamefont {D.~H.}\ \bibnamefont
  {{Jones}}}, \bibinfo {author} {\bibfnamefont {H.~R.}\ \bibnamefont
  {{Parkinson}}}, \bibinfo {author} {\bibfnamefont {C.~C.}\ \bibnamefont
  {{Popescu}}}, \bibinfo {author} {\bibfnamefont {M.}~\bibnamefont
  {{Prescott}}}, \bibinfo {author} {\bibfnamefont {R.~G.}\ \bibnamefont
  {{Sharp}}},\ and\ \bibinfo {author} {\bibfnamefont {R.~J.}\ \bibnamefont
  {{Tuffs}}},\ }\bibfield  {title} {\bibinfo {title} {{Galaxy And Mass Assembly
  (GAMA): the galaxy stellar mass function at z < 0.06}},\ }\href
  {https://doi.org/10.1111/j.1365-2966.2012.20340.x} {\bibfield  {journal}
  {\bibinfo  {journal} {Monthly Notices of the Royal Astronomical Society}\
  }\textbf {\bibinfo {volume} {421}},\ \bibinfo {pages} {621} (\bibinfo {year}
  {2012})},\ \Eprint {https://arxiv.org/abs/1111.5707} {arXiv:1111.5707
  [astro-ph.CO]} \BibitemShut {NoStop}%
\bibitem [{\citenamefont {{Kroupa}}(2001)}]{2001MNRAS.322..231K}%
  \BibitemOpen
  \bibfield  {author} {\bibinfo {author} {\bibfnamefont {P.}~\bibnamefont
  {{Kroupa}}},\ }\bibfield  {title} {\bibinfo {title} {{On the variation of the
  initial mass function}},\ }\href
  {https://doi.org/10.1046/j.1365-8711.2001.04022.x} {\bibfield  {journal}
  {\bibinfo  {journal} {Monthly Notices of the Royal Astronomical Society}\
  }\textbf {\bibinfo {volume} {322}},\ \bibinfo {pages} {231} (\bibinfo {year}
  {2001})},\ \Eprint {https://arxiv.org/abs/astro-ph/0009005}
  {arXiv:astro-ph/0009005 [astro-ph]} \BibitemShut {NoStop}%
\bibitem [{\citenamefont {Mapelli}(2016)}]{Mapelli:2016vca}%
  \BibitemOpen
  \bibfield  {author} {\bibinfo {author} {\bibfnamefont {M.}~\bibnamefont
  {Mapelli}},\ }\bibfield  {title} {\bibinfo {title} {{Massive black hole
  binaries from runaway collisions: the impact of metallicity}},\ }\href
  {https://doi.org/10.1093/mnras/stw869} {\bibfield  {journal} {\bibinfo
  {journal} {Mon. Not. Roy. Astron. Soc.}\ }\textbf {\bibinfo {volume} {459}},\
  \bibinfo {pages} {3432} (\bibinfo {year} {2016})},\ \Eprint
  {https://arxiv.org/abs/1604.03559} {arXiv:1604.03559 [astro-ph.GA]}
  \BibitemShut {NoStop}%
\bibitem [{\citenamefont {Fryer}\ \emph {et~al.}(2012)\citenamefont {Fryer},
  \citenamefont {Belczynski}, \citenamefont {Wiktorowicz}, \citenamefont
  {Dominik}, \citenamefont {Kalogera},\ and\ \citenamefont
  {Holz}}]{Fryer_2012}%
  \BibitemOpen
  \bibfield  {author} {\bibinfo {author} {\bibfnamefont {C.~L.}\ \bibnamefont
  {Fryer}}, \bibinfo {author} {\bibfnamefont {K.}~\bibnamefont {Belczynski}},
  \bibinfo {author} {\bibfnamefont {G.}~\bibnamefont {Wiktorowicz}}, \bibinfo
  {author} {\bibfnamefont {M.}~\bibnamefont {Dominik}}, \bibinfo {author}
  {\bibfnamefont {V.}~\bibnamefont {Kalogera}},\ and\ \bibinfo {author}
  {\bibfnamefont {D.~E.}\ \bibnamefont {Holz}},\ }\bibfield  {title} {\bibinfo
  {title} {Compact remnant mass function: Dependence on the explosion mechanism
  and metallicity},\ }\href {https://doi.org/10.1088/0004-637x/749/1/91}
  {\bibfield  {journal} {\bibinfo  {journal} {The Astrophysical Journal}\
  }\textbf {\bibinfo {volume} {749}},\ \bibinfo {pages} {91} (\bibinfo {year}
  {2012})}\BibitemShut {NoStop}%
\bibitem [{\citenamefont {{Morscher}}\ \emph {et~al.}(2015)\citenamefont
  {{Morscher}}, \citenamefont {{Pattabiraman}}, \citenamefont {{Rodriguez}},
  \citenamefont {{Rasio}},\ and\ \citenamefont {{Umbreit}}}]{Morscher2015}%
  \BibitemOpen
  \bibfield  {author} {\bibinfo {author} {\bibfnamefont {M.}~\bibnamefont
  {{Morscher}}}, \bibinfo {author} {\bibfnamefont {B.}~\bibnamefont
  {{Pattabiraman}}}, \bibinfo {author} {\bibfnamefont {C.}~\bibnamefont
  {{Rodriguez}}}, \bibinfo {author} {\bibfnamefont {F.~A.}\ \bibnamefont
  {{Rasio}}},\ and\ \bibinfo {author} {\bibfnamefont {S.}~\bibnamefont
  {{Umbreit}}},\ }\bibfield  {title} {\bibinfo {title} {{The Dynamical
  Evolution of Stellar Black Holes in Globular Clusters}},\ }\href
  {https://doi.org/10.1088/0004-637X/800/1/9} {\bibfield  {journal} {\bibinfo
  {journal} {\apj}\ }\textbf {\bibinfo {volume} {800}},\ \bibinfo {eid} {9}
  (\bibinfo {year} {2015})},\ \Eprint {https://arxiv.org/abs/1409.0866}
  {arXiv:1409.0866 [astro-ph.GA]} \BibitemShut {NoStop}%
\bibitem [{\citenamefont {Gerosa}\ and\ \citenamefont
  {Berti}(2017)}]{Gerosa:2017kvu}%
  \BibitemOpen
  \bibfield  {author} {\bibinfo {author} {\bibfnamefont {D.}~\bibnamefont
  {Gerosa}}\ and\ \bibinfo {author} {\bibfnamefont {E.}~\bibnamefont {Berti}},\
  }\bibfield  {title} {\bibinfo {title} {{Are merging black holes born from
  stellar collapse or previous mergers?}},\ }\href
  {https://doi.org/10.1103/PhysRevD.95.124046} {\bibfield  {journal} {\bibinfo
  {journal} {Phys. Rev. D}\ }\textbf {\bibinfo {volume} {95}},\ \bibinfo
  {pages} {124046} (\bibinfo {year} {2017})},\ \Eprint
  {https://arxiv.org/abs/1703.06223} {arXiv:1703.06223 [gr-qc]} \BibitemShut
  {NoStop}%
\bibitem [{\citenamefont {Vaccaro}\ \emph
  {et~al.}(2024{\natexlab{b}})\citenamefont {Vaccaro}, \citenamefont {Mapelli},
  \citenamefont {P{\'e}rigois}, \citenamefont {Barone}, \citenamefont {Artale},
  \citenamefont {Dall'Amico}, \citenamefont {Iorio},\ and\ \citenamefont
  {Torniamenti}}]{Vaccaro:2023cwr}%
  \BibitemOpen
  \bibfield  {author} {\bibinfo {author} {\bibfnamefont {M.~P.}\ \bibnamefont
  {Vaccaro}}, \bibinfo {author} {\bibfnamefont {M.}~\bibnamefont {Mapelli}},
  \bibinfo {author} {\bibfnamefont {C.}~\bibnamefont {P{\'e}rigois}}, \bibinfo
  {author} {\bibfnamefont {D.}~\bibnamefont {Barone}}, \bibinfo {author}
  {\bibfnamefont {M.~C.}\ \bibnamefont {Artale}}, \bibinfo {author}
  {\bibfnamefont {M.}~\bibnamefont {Dall'Amico}}, \bibinfo {author}
  {\bibfnamefont {G.}~\bibnamefont {Iorio}},\ and\ \bibinfo {author}
  {\bibfnamefont {S.}~\bibnamefont {Torniamenti}},\ }\bibfield  {title}
  {\bibinfo {title} {{Impact of gas hardening on the population properties of
  hierarchical black hole mergers in active galactic nucleus disks}},\ }\href
  {https://doi.org/10.1051/0004-6361/202348509} {\bibfield  {journal} {\bibinfo
   {journal} {Astron. Astrophys.}\ }\textbf {\bibinfo {volume} {685}},\
  \bibinfo {pages} {A51} (\bibinfo {year} {2024}{\natexlab{b}})},\ \Eprint
  {https://arxiv.org/abs/2311.18548} {arXiv:2311.18548 [astro-ph.HE]}
  \BibitemShut {NoStop}%
\bibitem [{\citenamefont {Borchers}\ \emph {et~al.}(2025)\citenamefont
  {Borchers}, \citenamefont {Ye},\ and\ \citenamefont
  {Fishbach}}]{Borchers:2025sid}%
  \BibitemOpen
  \bibfield  {author} {\bibinfo {author} {\bibfnamefont {A.}~\bibnamefont
  {Borchers}}, \bibinfo {author} {\bibfnamefont {C.~S.}\ \bibnamefont {Ye}},\
  and\ \bibinfo {author} {\bibfnamefont {M.}~\bibnamefont {Fishbach}},\
  }\bibfield  {title} {\bibinfo {title} {{Gravitational-wave kicks impact spins
  of black holes from hierarchical mergers}},\ }\href@noop {} {\  (\bibinfo
  {year} {2025})},\ \Eprint {https://arxiv.org/abs/2503.21278}
  {arXiv:2503.21278 [astro-ph.HE]} \BibitemShut {NoStop}%
\bibitem [{\citenamefont {Abbott}\ \emph {et~al.}(2016)\citenamefont {Abbott}
  \emph {et~al.}}]{KAGRA:2013rdx}%
  \BibitemOpen
  \bibfield  {author} {\bibinfo {author} {\bibfnamefont {B.~P.}\ \bibnamefont
  {Abbott}} \emph {et~al.} (\bibinfo {collaboration} {KAGRA, LIGO Scientific,
  Virgo}),\ }\bibfield  {title} {\bibinfo {title} {{Prospects for observing and
  localizing gravitational-wave transients with Advanced LIGO, Advanced Virgo
  and KAGRA}},\ }\href {https://doi.org/10.1007/s41114-020-00026-9} {\bibfield
  {journal} {\bibinfo  {journal} {Living Rev. Rel.}\ }\textbf {\bibinfo
  {volume} {19}},\ \bibinfo {pages} {1} (\bibinfo {year} {2016})},\ \Eprint
  {https://arxiv.org/abs/1304.0670} {arXiv:1304.0670 [gr-qc]} \BibitemShut
  {NoStop}%
\bibitem [{\citenamefont {Barsotti}\ \emph {et~al.}(2018)\citenamefont
  {Barsotti}, \citenamefont {Gras}, \citenamefont {Evans},\ and\ \citenamefont
  {Fritschel}}]{Barsotti:18}%
  \BibitemOpen
  \bibfield  {author} {\bibinfo {author} {\bibfnamefont {L.}~\bibnamefont
  {Barsotti}}, \bibinfo {author} {\bibfnamefont {S.}~\bibnamefont {Gras}},
  \bibinfo {author} {\bibfnamefont {M.}~\bibnamefont {Evans}},\ and\ \bibinfo
  {author} {\bibfnamefont {P.}~\bibnamefont {Fritschel}},\ }\href
  {https://dcc.ligo.org/LIGO-T1800044/public} {\emph {\bibinfo {title} {The
  updated Advanced LIGO design curve}}},\ \bibinfo {organization} {LIGO}
  (\bibinfo {year} {2018}),\ \bibinfo {note} {{L}IGO Document
  T1800044}\BibitemShut {NoStop}%
\bibitem [{\citenamefont {Witte}(2024)}]{WitteGit}%
  \BibitemOpen
  \bibfield  {author} {\bibinfo {author} {\bibfnamefont {S.~J.}\ \bibnamefont
  {Witte}},\ }\href@noop {} {\bibinfo {title} {Axion\_sr\_tde}},\ \bibinfo
  {howpublished} {\url{https://github.com/SamWitte/Axion_SR_TDE}} (\bibinfo
  {year} {2024})\BibitemShut {NoStop}%
\bibitem [{\citenamefont {Aswathi}\ \emph {et~al.}(2025)\citenamefont
  {Aswathi}, \citenamefont {East}, \citenamefont {Siemonsen}, \citenamefont
  {Sun},\ and\ \citenamefont {Jones}}]{Aswathi:2025nxa}%
  \BibitemOpen
  \bibfield  {author} {\bibinfo {author} {\bibfnamefont {P.~S.}\ \bibnamefont
  {Aswathi}}, \bibinfo {author} {\bibfnamefont {W.~E.}\ \bibnamefont {East}},
  \bibinfo {author} {\bibfnamefont {N.}~\bibnamefont {Siemonsen}}, \bibinfo
  {author} {\bibfnamefont {L.}~\bibnamefont {Sun}},\ and\ \bibinfo {author}
  {\bibfnamefont {D.}~\bibnamefont {Jones}},\ }\bibfield  {title} {\bibinfo
  {title} {{Ultralight boson constraints from gravitational wave observations
  of spinning binary black holes}},\ }\href@noop {} {\  (\bibinfo {year}
  {2025})},\ \Eprint {https://arxiv.org/abs/2507.20979} {arXiv:2507.20979
  [gr-qc]} \BibitemShut {NoStop}%
\end{thebibliography}%

\appendix

\clearpage

\onecolumngrid
\begin{center}
  \textbf{\large Appendix for Superradiance Constraints from GW231123}\\[.2cm]
  \vspace{0.05in}
  {Andrea Caputo, Gabriele Franciolini and Samuel J. Witte} \\[1cm]
\end{center}

\twocolumngrid

In the following, we show the results of our analysis adopting different posterior samples, likelihoods, and merger time-scales. We also discuss more exotic scenarios for GW231123, provide a comprehensive outline of the statistical procedure adopted, discuss the impact of binary systems on the evolution of superradiant clouds, and derive constraints on non-interacting spin-1 bosons.

\section{Additional Results}
\label{secApp:otherresults}

\begin{figure}
    \centering
    \includegraphics[width=\linewidth]{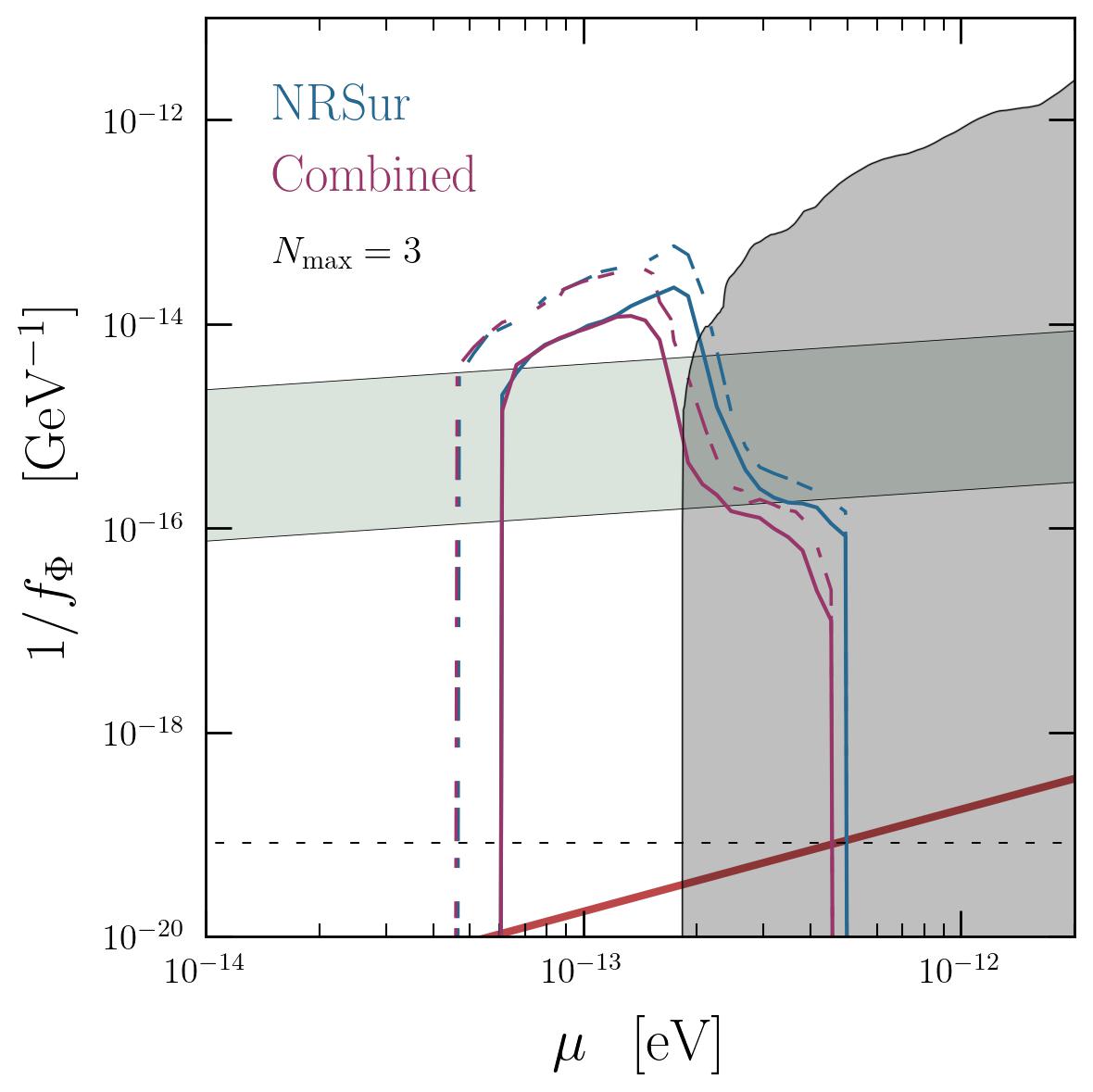}
    \caption{Comparison of limits obtained using {\tt{NRSur}} and {\tt{Combined}} posterior samples from LVK analysis. Results are shown for $\tau_{\rm max} = 10^6$ and $10^7$ years. }
    \label{fig:analysis}
\end{figure}

For the sake of completeness, we show how our limits vary when adopting different assumptions in the underlying analysis. The fiducial analysis shown in the main text has been performed using the {\tt{NRSur}} posterior samples from the LVK collaboration \cite{ligo_scientific_collaboration_2025_16004263}, taking $\tau_{\rm max}=10^5, 10^6,$ and $10^7$ years. In Fig.~\ref{fig:analysis} we compare the impact of instead adopting posterior samples from the `combined' analysis (which combines samples from five BH waveform models, of which one is the {\tt{NRSur}} model); this is done using maximum timescales of $\tau_{\rm max} = 10^6$ and $10^7$ years. We observe small differences in the resulting limits depending on the posterior samples used in the analysis, but primarily for masses $\mu \gtrsim 2 \times 10^{-13}$ eV.

\begin{figure*}
    \centering
    \includegraphics[width=0.49\linewidth]{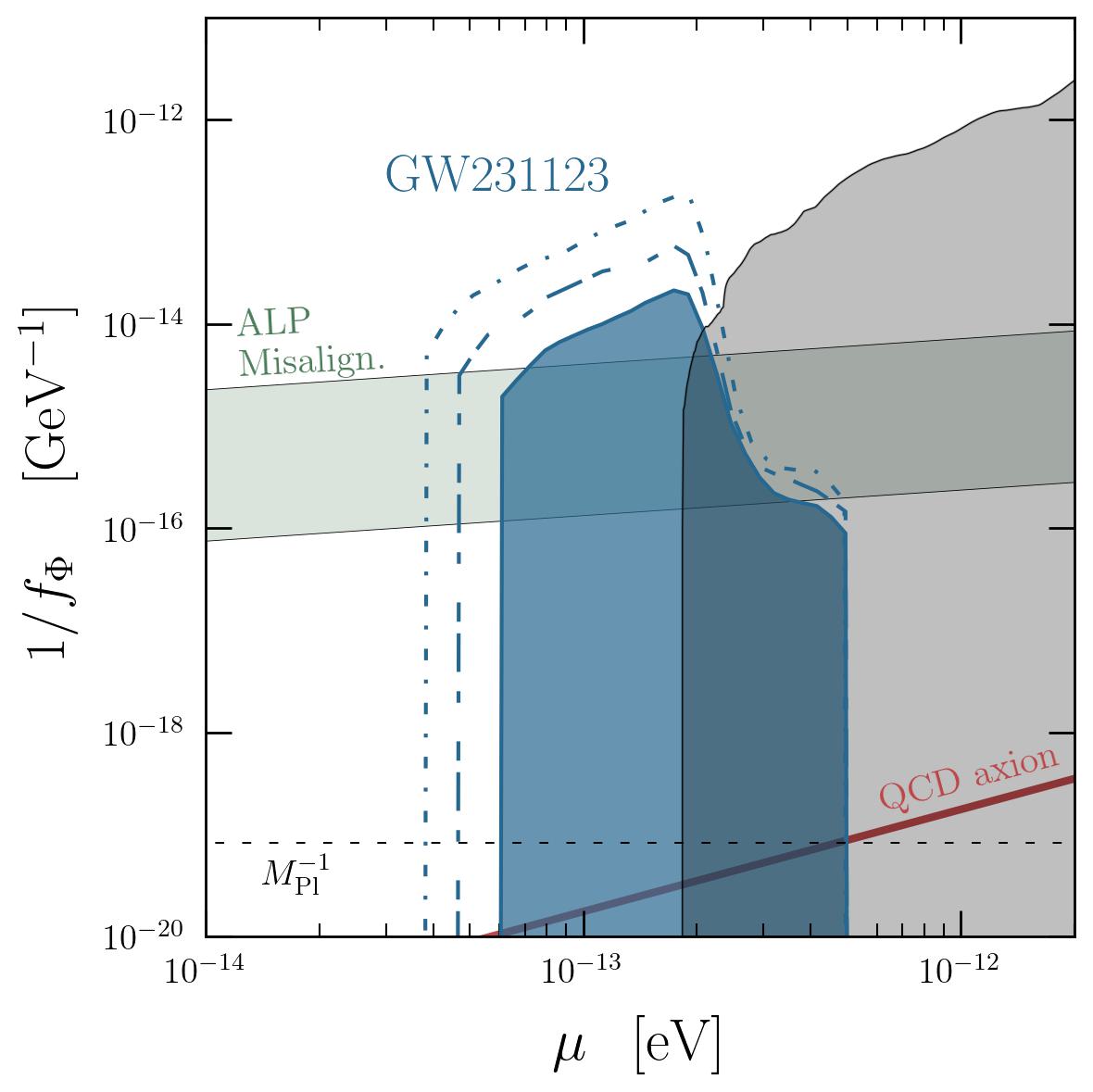}
    \includegraphics[width=0.49\linewidth]{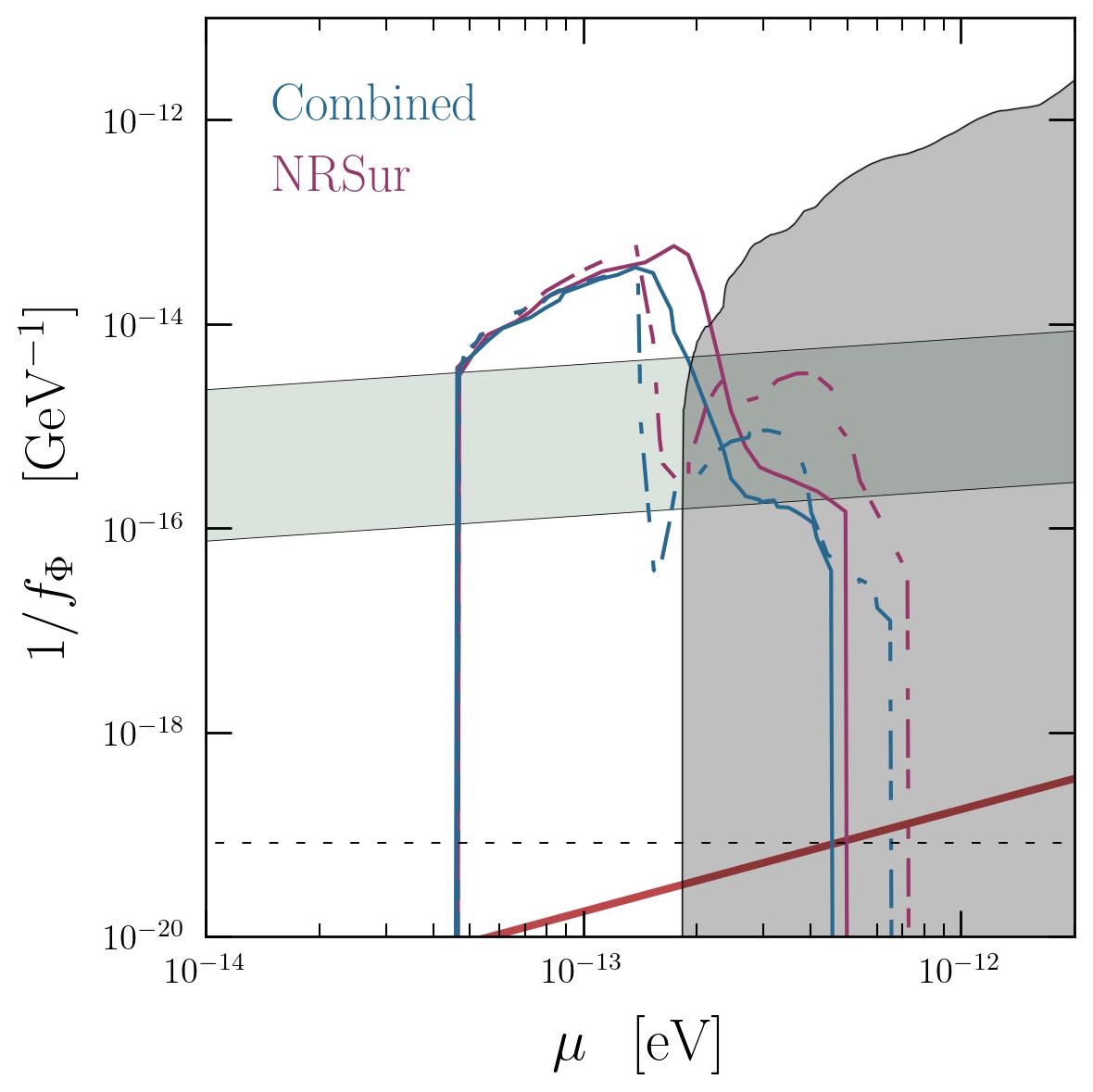}
    \caption{Left: Comparison of limits obtained using {\tt{NRSur}}, $N_{\rm max } = 3$, and maximum timescales $\tau_{\rm max} = 10^5, 10^6,$ and $10^7 $ years. Right: Comparison of limits obtained using {\tt{NRSur}} (purple) and {\tt{Combined}} (blue) posterior samples from LVK analysis, fixing $\tau_{\rm max} = 10^6$, and taking both $N_{\rm max} = 3$ (solid) and $N_{\rm max} = 5$ (dashed). }
    \label{fig:nmax_comp}
\end{figure*}

We also compare the impact of running our analysis with either $N_{\rm max} = 5$, or $N_{\rm max} = 3$. The equivalent of Fig.~\ref{fig:main}, but fixing $N_{\rm max} = 3$, is shown in the left panel of Fig.~\ref{fig:nmax_comp}, while in the right panel of Fig.~\ref{fig:nmax_comp} we fix $\tau_{\rm max} = 10^6$ years, and we show the results for both waveform analyses with $N_{\rm max} = 3, 5$.

\begin{figure*}
    \centering
    \includegraphics[width=0.49\linewidth]{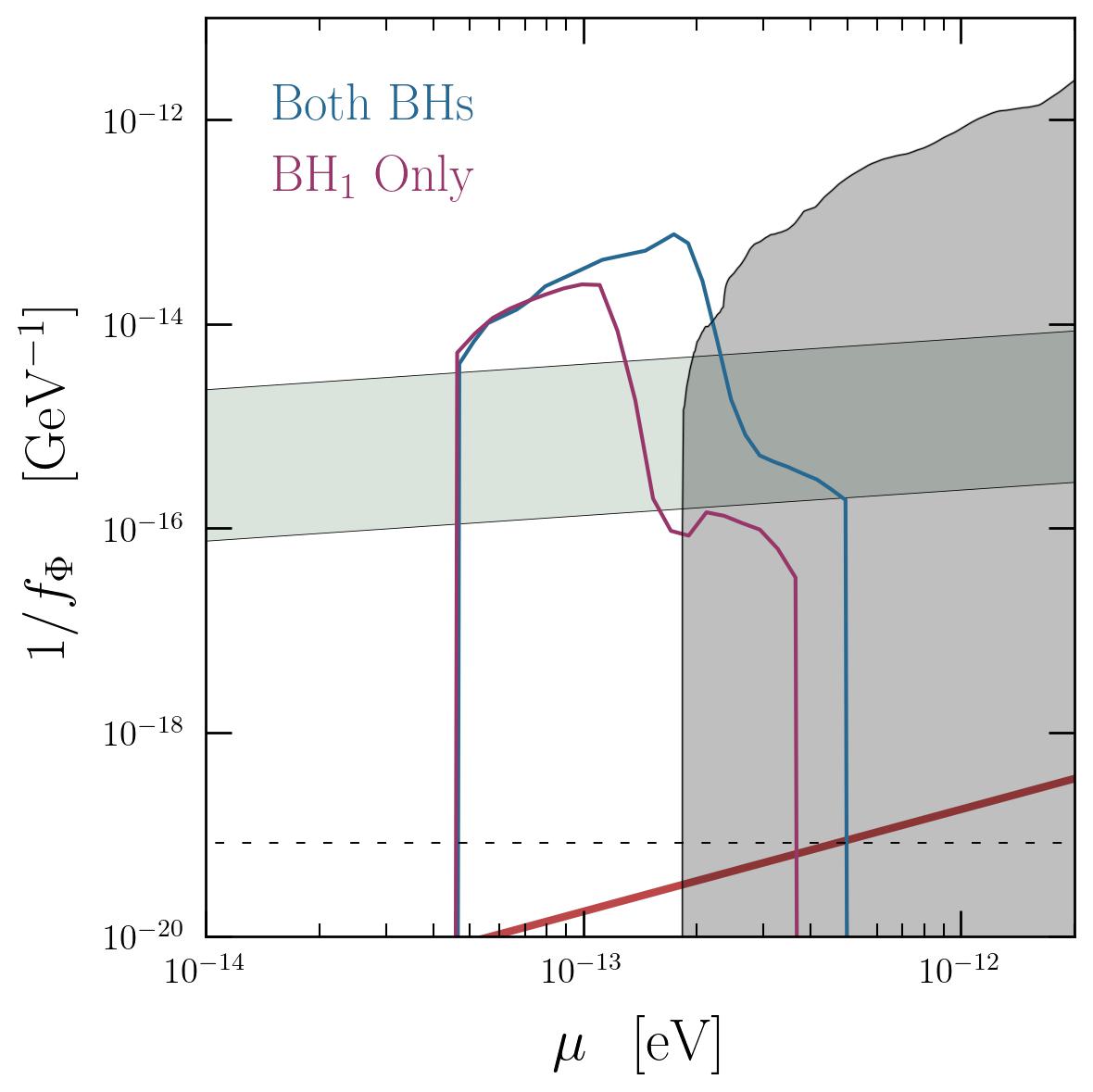}
    \includegraphics[width=0.49\linewidth]{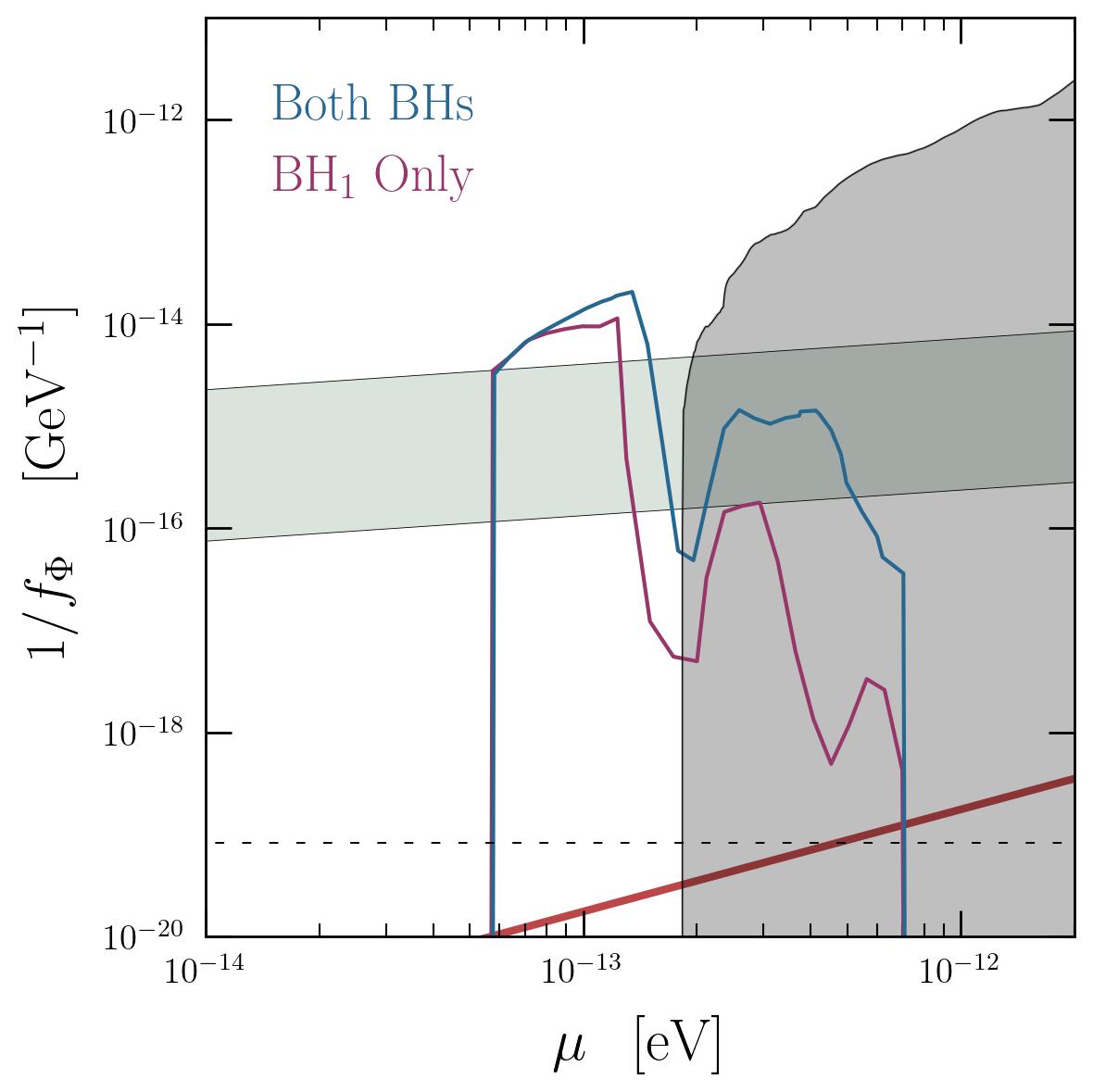}
    \caption{Comparison of limits obtained using only the heavy BH (purple), or using both BHs (blue). Here, we take the {\tt{NRSur}} posterior samples, $N_{\rm max} = 3$ (left) and $N_{\rm max} = 5$ (right), and $\tau_{\rm max} = 10^6$ (left) and $10^5$ years (right). }
    \label{fig:oneBH}
\end{figure*}

Next, we look at the relative importance of including both black holes in the analysis. We plot in Fig.~\ref{fig:oneBH} the limits derived using the correlated mass-spin information on both BHs, versus using only the heavier, more prominently measured, BH. This is done for $N_{\rm max} = 3$ and $\tau_{\rm max} = 10^6$ years (left), and for $N_{\rm max = 5}$ and $\tau_{\rm max} = 10^5$ years (right). For masses $\mu \sim 2 \times 10^{-13}$ eV, the $\left|211\right>$ state is not fully efficient at spinning down the heavy BH; however, samples which are consistent with $\tilde{a}_1 \sim 0.75$ inherently imply that $\tilde{a}_2$ is large, and thus the contribution from the second BH to the likelihood becomes significant.

\begin{figure}
    \centering
    \includegraphics[width=\linewidth]{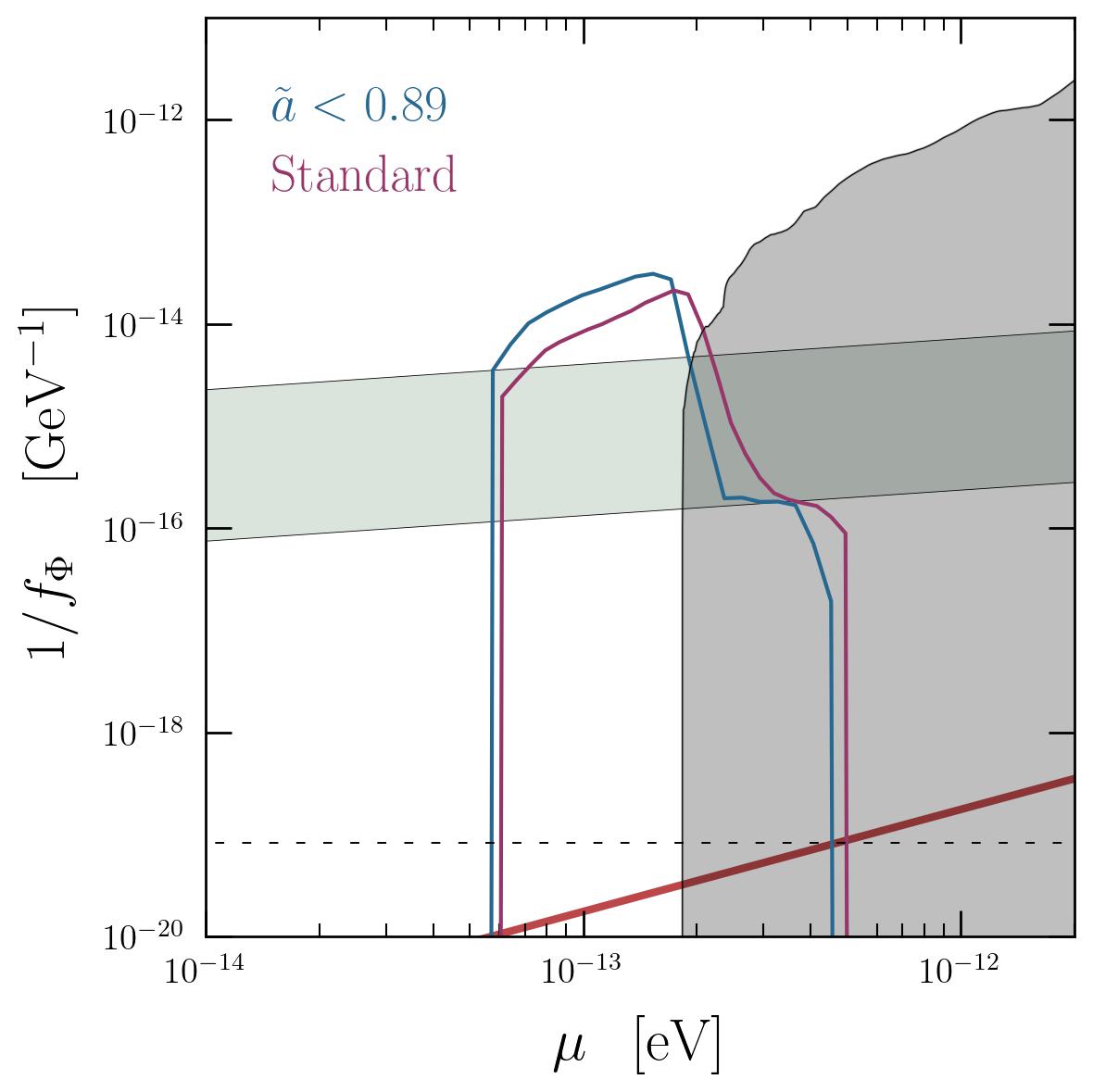}
    \caption{Comparison of limits obtained for $N_{\rm max} = 3$ and $\tau_{\rm max} = 10^5$ years, with those obtained by truncating the prior for spins $\tilde{a} \gtrsim 0.89$. Both runs are made using 
     {\tt{NRSur}} posterior samples. }
    \label{fig:spincut}
\end{figure}

\begin{figure}
    \centering
    \includegraphics[width=\linewidth]{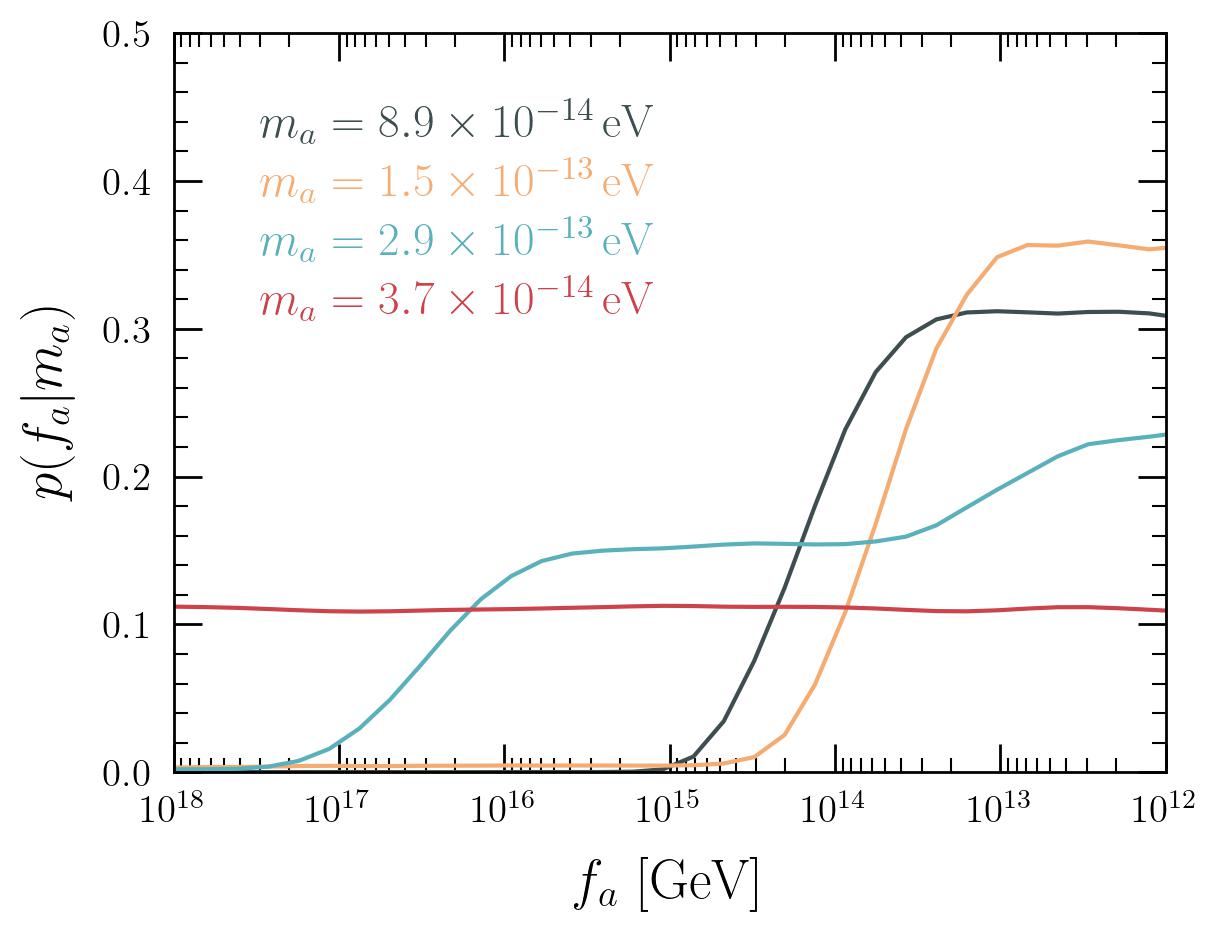}
    \caption{Four illustrations of the one-dimensional posterior distributions on $f_a$ for various fixed axion masses. Example corresponds to the low-spin limit shown in Fig.~\ref{fig:spincut}. }
    \label{fig:oned_posterior}
\end{figure}

Finally, we look at the sensitivity of our derived limit to the high-spin component of the prior. Since the {\tt{NRSur}} likelihood is not calibrated for spins $\tilde{a} \gtrsim 0.89$, we truncate the prior to remove spins above this threshold. The result of performing this analysis, using $N_{\rm max } = 3$ and $\tau_{\rm max} = 10^5$ years, is shown in Fig.~\ref{fig:spincut}. Here, one sees that the constraints are actually strengthened at low masses, and weakened at high masses. This result is as expected. For low mass axions, the $\left| 211\right> - \left|322\right>$ states are in an equilibrium with suppressed occupation numbers, and thus larger initial spins require longer timescales in order to reach the lower edge of the spin posterior. For high masses, the $\left|322\right>$ state grows unimpeded, and thus one recovers the non-interacting result in which most of the time spent growing the cloud, rather than spinning down the black hole, and thus higher spins (which correspond to faster growth timescales) lead to stronger constraints. This result indicates that our derived sensitivity errors on the side of conservative, even though there are large uncertainties in the {\tt{NRSur}} likelihood at $\tilde{a} \gtrsim 0.89$. In order to provide an illustrative example, we also show a few of the one-dimensional posterior distributions on $f_a$ in Fig.~\ref{fig:oned_posterior}.

Note that Ref.~\cite{Aswathi:2025nxa} appeared online at the same time as this work, and also used GW231123 to place constraints on new bosons from superradiance. The constraints on $f_\Phi$ derived in ~\cite{Aswathi:2025nxa} are somewhat weaker than those derived here. Part of this discrepancy comes from the fact that they only used the more massive BH in their analysis -- Fig.~\ref{fig:oneBH} suggests that this likely leads to a sizable difference for axion masses $\mu \gtrsim 10^{-12}$ eV. In addition, we note that the analysis of ~\cite{Aswathi:2025nxa} does not explicitly time-evolve the system, but rather imposes a constraint by comparing posterior spin samples with the maximum allowed spin of a fully evolved state. For a non-interacting boson, the system evolves exponentially fast, and thus this introduces only a small error; for self-interacting axions, however, this approach can produce overly conservative limits, often by a factor of a few (see~\cite{Witte:2024drg} for a discussion). Finally, we note that our fiducial analysis includes $N = 4,5$ states, allowing us to extend our analysis to large axion masses.

\section{Alternative Scenarios}
\label{secApp:Exotic}

In this section, we explore more exotic hypotheses for the origin of the GW231123 event and assess the robustness of the constraints presented in Fig.~\ref{fig:main}. We critically examine alternative formation channels beyond the most plausible hierarchical scenario and evaluate the extent to which our conclusions remain valid under these less conventional assumptions. In particular, we consider population III (Pop III) and primordial BH binaries.\footnote{Ref.~\cite{Cuceu:2025fzi} explored the possibility of GW231123 being emitted from cusps or kinks on a cosmic string, finding this explanation is disfavored compared to BBH.}

\subsection{Pop III binaries}

Remnant BHs in the high-mass range of the mass gap may originate from the evolution of very massive stars formed in environments with low metallicity, where stellar winds are significantly suppressed~\cite{Spera:2017fyx}. Such stars are expected to form from the collapse of pristine gas clouds at high redshift, when the interstellar medium was still poorly enriched with metals. Here, we focus on the first generation of stars, also known as Pop III~\cite{10.1093/mnras/207.3.585,Bromm:2003vv,Belczynski:2004gu,Tanikawa:2021qqi}.

Binaries of Pop III stars can evolve into binary BH systems in which at least one component lies in the high-mass portion of the gap, as supported by population synthesis studies (e.g.,~\cite{Hijikawa:2021hrf}). The resulting mass distribution of the primary and secondary BHs depends on the initial stellar masses and the complex evolution of the binary system.

What is crucial for our analysis is that these binaries form at relatively high redshift, shortly after the onset of structure formation, and are characterized by long coalescence timescales, which can extend up to several Gyr~\cite{Belczynski:2016ieo,Liu:2020lmi}—typically longer than the Salpeter timescale or the timescales adopted in Sec.~\ref{sec:timescale}. It is therefore important to assess whether individual BHs in such binaries could undergo superradiant spin-down within the conservative timescale assumed in Fig.~\ref{fig:main}, before the influence of their companion becomes significant. We elaborate on this possibility in App.~\ref{secApp:binary}.

\subsection{Primordial BH binaries}

The large masses observed in GW231123, potentially lying within the pair-instability mass gap, may also be consistent with a primordial origin~\cite{Zeldovich:1967lct,Hawking:1971ei,Carr:1974nx,Carr:1975qj,Bird:2016dcv,Clesse:2016vqa,Sasaki:2016jop,Eroshenko:2016hmn,Wang:2016ana,Clesse:2020ghq,Hall:2020daa,Franciolini:2022tfm,Escriva:2022bwe} (see e.g.~\cite{Byrnes:2025tji,LISACosmologyWorkingGroup:2023njw} for recent reviews). In this context, the inferred event rate remains compatible with current observational constraints, although accretion may be necessary to remain consistent with the strong bounds on the PBH abundance around $\gtrsim 10^2 M_\odot$~\cite{DeLuca:2020fpg,DeLuca:2020sae} (see also~\cite{Yuan:2025avq}).

The high spins inferred from the signal may appear in tension with conventional stellar-origin scenarios, which typically predict low or moderate spins for binary components~\cite{Mirbabayi:2019uph,DeLuca:2019buf,Harada:2020pzb}. However, this interpretation neglects the role of accretion over cosmological timescales, which can significantly spin up BHs—particularly in high-mass systems with $m \gg M_\odot$ such as GW231123~\cite{DeLuca:2020bjf,DeLuca:2020qqa}. A similar total mass–spin correlation is also found in hierarchical scenarios~\cite{Franciolini:2022iaa,Franciolini:2021xbq}. Importantly, spin-up induced by prolonged accretion would naturally affect the mass ratio, favoring near-equal-mass binaries, in line with the measured properties of this event.

If the binary is indeed primordial, it must have formed before matter–radiation equality, as primordial channels dominate the merger rate~\cite{Sasaki:2016jop,Ali-Haimoud:2017rtz,Raidal:2018bbj,Vaskonen:2019jpv,Franciolini:2022ewd,Raidal:2024bmm}. Moreover, any significant spin-up must have occurred before the end of the reionization epoch (typically $z \sim 6$--$10$), after which accretion becomes inefficient~\cite{DeLuca:2020bjf,Hasinger:2020ptw}. The cosmological time elapsed between that epoch and $z = 0.5$, the approximate redshift of the merger, is $\approx 8.5\,{\rm Gyr}$, providing a wide temporal window for superradiance to occur. Throughout this evolution, the binary would have formed at large separations, allowing superradiant instabilities to proceed as in the isolated BH case until the orbital separation decreased to critical values (discussed in the next appendix), at which point interactions with the companion become non-negligible. This primordial scenario can thus be considered on equal footing with the alternative formation channels discussed in this work, potentially featuring even larger $\tau_{\rm merger}$.

\section{The Impact of the Binary}
\label{secApp:binary}

Binary systems can have a non-negligible impact on the evolution of superradiant clouds, and the spin down of black holes, even when the binary separation $R$ is large (\ie when $R \gg r_c \sim M/\alpha^2$, with $r_c$ being the characteristic radius of cloud)~\cite{Arvanitaki:2014wva,Baumann:2018vus,Baryakhtar:2020gao,Baumann:2022pkl,Tong:2022bbl,Fan:2023jjj,Tomaselli:2023ysb,Zhu:2024bqs,Takahashi:2024fyq,Boskovic:2024fga,Tomaselli:2024bdd, Tomaselli:2025jfo}. This is because the gravitational perturbation can induce a small shift in the, already small, imaginary component of the energy, inducing energy transfer to dissipate bound states and shutting down the superradiant growth in a given level. In this regard, one must ask whether the binary can significantly alter the analysis and conclusions drawn in the main text.

Let us first emphasize that the binary cannot play any significant role in both the AGN and NSC scenarios. In effect, this is because the binary forms at small radial separation. The time from binary formation to merger is always significantly less than the Salpeter timescale, meaning that the BHs must have had high spins prior to binary formation itself. In that sense, our superradiant evolution is testing the dynamical evolution of these systems in `isolation'.

What is not immediately clear, however, is whether one can neglect the binary dynamics in the more exotic scenario in which GW231123 is assumed to be of Pop III origin. Understanding which regions of parameter space may be altered in this scenario is the focus of this section.

\begin{figure}
    \centering
    \includegraphics[width=\linewidth]{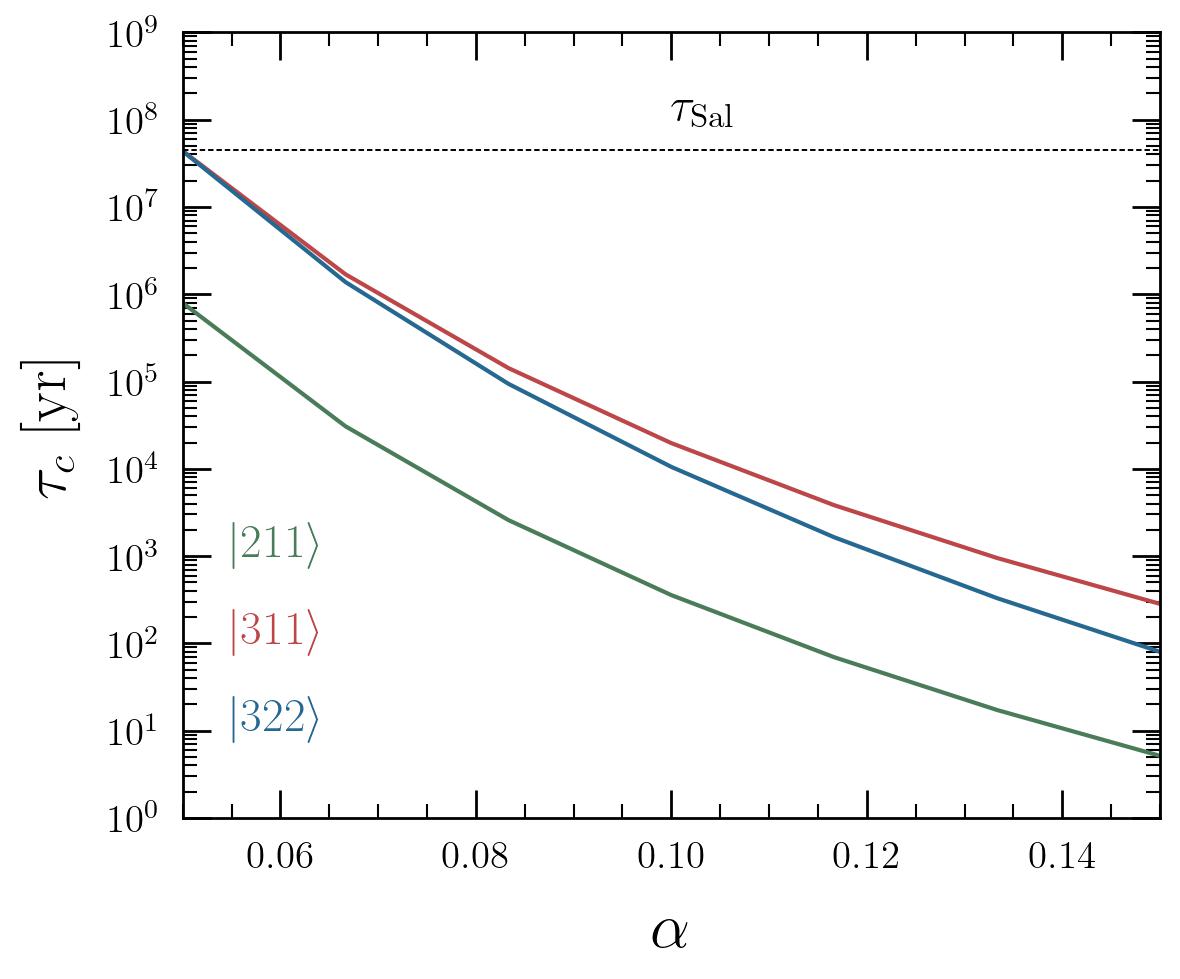}
    \caption{Characteristic timescale prior to merger at which point we expect a sizable correction to the imaginary component of the $\left|n \ell m \right>$ state, shown as a function of $\alpha$. For definiteness, we consider the best-fit GW231123 parameters.
    Shown for comparison is the Salpeter timescale $\tau_{\rm Sal}$. }
    \label{fig:binary}
\end{figure}

The shift in the imaginary component of the eigenvalue $\Gamma_{n\ell m} + \delta \Gamma_{n\ell m}$ from a gravitational perturbation $M_*$, see \eg~\cite{Tomaselli:2023ysb},
\begin{align}
    V_*(t, \vec{r}) = - \sum_{\ell_* = 0}^\infty \sum_{m_* = -\ell_*}^{\ell^*} \frac{4\pi \alpha}{2\ell_* + 1} \frac{M_*}{M} Y_{\ell_* m_*} Y^*_{\ell_* m_*} F(r) \, 
\end{align}
due to the mixing with a state $\left|n^\prime \ell^\prime m^\prime\right>$ is roughly given by~\cite{Arvanitaki:2014wva,Tomaselli:2025jfo}
\begin{eqnarray}\label{eq:gamma_shift}
    \delta \Gamma_{n\ell m} \sim \frac{\Gamma_{(n^\prime \ell^\prime m^\prime)} - \Gamma_{n\ell m}  }{(\omega_{r,n \ell m} - \omega_{r, n^\prime \ell^\prime m^\prime})^2 + (\Gamma_{n^\prime \ell^\prime m^\prime} - \Gamma_{n\ell m})^2} \nonumber \\[10pt] \times \, \left|\left<n^\prime \ell ^\prime m^\prime |V_*| n \ell m\right> \right|^2 \, \hspace{1cm} ,
\end{eqnarray}
where we have assumed we are not on resonance (allowing us to drop the correction to the energy difference coming from the binary), and we have defined
\begin{align}
    F(r) = \begin{cases}
        \frac{r^{\ell_*}}{R_*^{\ell_*+1}}\Theta(R_* - r) + \frac{R_*^{\ell_*}}{r_*^{\ell_*+1}}\Theta(r - R_*) \hspace{.4cm} {\rm for} \, l_* \neq 1 \\
        \left(\frac{R_*}{r^2} - \frac{r}{R_*^2}\right)\Theta(r - R_*)  \hspace{2.3cm } {\rm for} \, l_* = 1 
    \end{cases}
\end{align}
with $R_*$ the binary separation.
The leading order transition at large radii is the hyperfine transition, mixing $\left| 211 \right> $ with $\left| 21-1 \right> $, and $\left| 322 \right> $ with $\left| 320 \right> $. 

We compute the radial distance at which Eq.~\ref{eq:gamma_shift} becomes comparable to $\Gamma_{n\ell m}$, and translate that distance into a characteristic timescale $\tau_c$ using Peter's equation -- the result in shown in Fig.~\ref{fig:binary} for the leading states as a function of $\alpha$. In this calculation, we adopt the central values of the mass and spin -- reasonable variations of these quantities do not significantly alter the conclusions. If this timescale $\tau_c \gg \tau_{\rm Sal}$ (shown in black), then one cannot safely preclude the possibility that all of the BH spins were obtained in the last epoch right before merger (being driven by Eddington limited accretion), and after the binary perturbation and suppressed superradiance. We can see that this is not the case over the parameter space of interest (the lower limit in Fig.~\ref{fig:main} roughly corresponds to $\alpha \sim 0.05$), and thus our analysis remains conservative even should Pop III stars be responsible for the merger. 

In writing Eq.~\ref{eq:gamma_shift}, we have implicitly neglected resonant excitations/de-excitations induced by the binary. 
These resonances, occurring when $\omega_{n\ell m} - \omega_{n^\prime \ell^\prime m^\prime} \sim g \Omega$ (with $g$ an integer),  tend to occur at much larger radii, see \eg\cite{Tomaselli:2024bdd}. In order for the Pop III BH to have merged, it must have formed at a distance of no greater than $\sim 2.4 \times 10^5 r / (G \times M)$. If we are interested in testing the hypothesis that maximal accretion created the BH spins at very late times (less than a Salpeter timescale), then we can once again use Peter's formula to restrict our attention to resonances that occur between $1.1 \times 10^5 \lesssim  r / (G \times M) \lesssim 2.4\times10^5$. Given the window for resonances is extremely narrow, and the timescale between resonances extremely large, one can safely neglect the resonant contribution. 

\section{Spin-1 Fields}\label{app:spin1}
In the case of spin-1 fields, the superradiance rate can be greatly enhanced, making spin measurements potentially more powerful than in the case of a scalar. Here, we provide a rough estimate of where these constraints lie. 

First, it is important to note that if the spin-1 mass is generated from a Higgs mechanism, then constraints are significantly weakened because the large field values reached during the superradiant evolution work to restore the symmetry, back-reacting on the mass and turning off the instability -- see ~\cite{Fukuda:2019ewf}. Evading this constraint typically requires extremely small gauge couplings and a very large Higgs mass, making this scenario extremely fine-tuned.  For this reason, we focus on the case in which the mass arises from the Stueckelberg mechanism. It is also worth noting that if the spin-1 field has a large kinetic mixing with the Standard Model,  additional effects may serve to quench the growth of the superradiant cloud~\cite{Caputo:2021efm,Siemonsen:2022ivj}. For the sake of simplicity, in what follows we will assume the kinetic mixing is sufficiently small that the spin-1 field can be treated as non-interacting.

In order to derive constraints on the non-interacting spin-1 boson, we use the relativistic superradiance rates of the fastest growing state derived in~\cite{cardoso2018constraining}. In the case of the axion, we had fixed the mass and sampled over $f_\Phi$; here, there is no interaction parameter, and thus we adopt a log-flat prior on the mass between $[10^{-14.5}, 10^{-12}]$ eV, which is the range roughly identified by analytic estimates as being of interest for GW231123 (note that since the posterior is unbounded, adopting overly wide priors will lead to overestimated constraints -- for this reason we take them to be as narrow as possible). We take the posterior samples, use a KDE to reconstruct a one-dimensional posterior distribution, and determine the mass range for which the probability distribution crosses the $10\%$ threshold. Taking $\tau_{\rm max} = 10^5$ years, this corresponds to an excluded mass range of $m_{\gamma^\prime} \in [1.35 \times 10^{-14}, 2.60 \times 10^{-13}]$ eV.

\section{Statistical Procedure}

In this section, we formalize the discussion of the statistical procedure used to derive the limits on $f_a$.

Let us begin by defining our model space as $\vec{X} = (f_\Phi, M_1, q, \tilde{a}_1, \tilde{a}_2)$, where following the choice of LVK, we will use the mass ratio $q \equiv M_2/M_1$, instead of $M_2$. Our data $\mathcal{D}$, however, is directly compared against $(M_1, q, \tilde{a}_{1,0}, \tilde{a}_{2,0})$, where $\tilde{a}_{i,0}$ represents the spin of the $i^{\rm th}$-BH at merger, which is distinct from the spin $\tilde{a}_i$ which is defined at $t_0 - \tau_{\rm max}$. With this mind, we can write the posterior as
\begin{eqnarray}\label{eq:post_us}
   & p (f_\Phi, M_1, q, \tilde{a}_1, \tilde{a}_2 | \mathcal{D})  =  \, \nonumber \\[6pt]
   &   \frac{\pi(f_\Phi, \tilde{a}_1, \tilde{a}_2, M_1, q) \mathcal{L}(\mathcal{D} | f_\Phi, \tilde{a}_1, \tilde{a}_2, M_1, q)}{p(\mathcal{D})} 
\end{eqnarray}
where $\mathcal{D}$ is the data, $\pi$ is the prior distribution, and $\mathcal{L}$ the likelihood. Our goal is to use the LVK posterior as an effective input to use in our likelihood, so let us note that the LVK posterior can be expressed as 
\begin{eqnarray}
   & p_{\rm LVK}(M_1, q, \tilde{a}_{1,0}, \tilde{a}_{2,0} | \mathcal{D})  =  \, \nonumber \\[6pt]
   &   \frac{\pi_{\rm LVK}(M_1, q, \tilde{a}_{1,0}, \tilde{a}_{2,0}) \mathcal{L}_{\rm LVK}(\mathcal{D} |M_1, q, \tilde{a}_{1,0}, \tilde{a}_{2,0})}{p(\mathcal{D})}  \, .
\end{eqnarray}
Using the fact that $a_{i,0}(a_i, M_i, f_\Phi)$, we can write
\begin{eqnarray}
 & \mathcal{L}(\mathcal{D} | f_\Phi, \tilde{a}_1, \tilde{a}_2, M_1, q) =   \, \nonumber \\[6pt] & \mathcal{L}_{\rm LVK}(\mathcal{D} |M_1, q, \tilde{a}_{1,0}(a_1,M_1,f_\Phi), \tilde{a}_{2,0}(a_2,M_2,f_\Phi)) = \, \nonumber \\[6pt] & p_{\rm LVK}(M_1, q, \tilde{a}_{1,0}(a_1,M_1,f_\Phi), \tilde{a}_{2,0}(a_2,M_2,f_\Phi) | \mathcal{D}) \times \nonumber \\[6pt] & \frac{p(\mathcal{D})}{\pi(M_1, q, \tilde{a}_{1,0}, \tilde{a}_{2,0})}  =  \, \nonumber \\[6pt] &p_{\rm LVK}(M_1, q, \tilde{a}_{1,0}(a_1,M_1,f_\Phi), \tilde{a}_{2,0}(a_2,M_2,f_\Phi) | \mathcal{D}) \times \nonumber \\[6pt] &  \frac{p(\mathcal{D})}{\pi(M_1) \pi(q)  \pi(\tilde{a}_{1,0})  \pi(\tilde{a}_{2,0})}
\end{eqnarray}
where in the last line we have used the fact that the LVK priors on each variable are independent. This simplification allows Eqn.~\ref{eq:post_us} to be written as 
\begin{eqnarray}\label{eq:firststop_post}
& p (f_\Phi, M_1, q, \tilde{a}_1, \tilde{a}_2 | \mathcal{D})  =  \, \nonumber \\[6pt]
   &   \frac{\pi(f_\Phi, \tilde{a}_1, \tilde{a}_2, M_1, q)}{\pi(M_1) \pi(q)  \pi(\tilde{a}_{1,0})  \pi(\tilde{a}_{2,0})} \times \, \nonumber \\[6pt] &  p_{\rm LVK}(M_1, q, \tilde{a}_{1,0}(a_1,M_1,f_\Phi), \tilde{a}_{2,0}(a_2,M_2,f_\Phi) | \mathcal{D}) 
\end{eqnarray}
At this point, one can note that the LVK priors are flat, implying they do not enter in the optimization procedure and can be dropped. At this point, one approach would be to adopt flat priors $(\tilde{a}_1, \tilde{a}_2, M_1, q)$, and a log-flat prior on $f_\Phi$, evolve each black hole to determine $\tilde{a}_{1,0}$ and $\tilde{a}_{2,0}$, and use a 4D KDE to evaluate $p_{\rm LVK}(M)$. We choose instead to go one step further, and re-write the LVK posterior as
\begin{eqnarray}
    & p_{\rm LVK}(M_1, q, \tilde{a}_{1,0}, \tilde{a}_{2,0} | \mathcal{D}) = \, \nonumber \\[6pt] & \frac{p(a_{1,0}, a_{2,0}, M_1, q, \mathcal{D})}{p(\mathcal{D})} = \, \nonumber \\[6pt] & p(a_{1,0}, a_{2,0}| M_1, q, \mathcal{D}) p(M_1, q|\mathcal{D}) 
\end{eqnarray}
where for simplicity we have suppressed the explicit functional dependence of $\tilde{a}_{i,0}$. Using the above expression, the posterior can be expressed as
\begin{eqnarray}
& p (f_\Phi, M_1, q, \tilde{a}_1, \tilde{a}_2 | \mathcal{D})  \propto  \, \nonumber \\[6pt]
&  \pi(f_\Phi, \tilde{a}_1, \tilde{a}_2) p(a_{1,0}, a_{2,0}| M_1, q, \mathcal{D}) p(M_1, q|\mathcal{D}) \, .
\end{eqnarray}
Here, one can now sample $(f_\Phi, \tilde{a}_1, \tilde{a}_2)$ from the prior distributions (defined below), sample $(M_1,q)$ from the marginalized posterior LVK distribution, evolve each black hole to obtain $a_{1,0}$ and $a_{2,0}$, and then used a 2D KDE to evaluate the LVK posterior condition on $M_1$ and $q$. 

In general, it would be natural to adopt a log-flat distribution on $f_\Phi$
\begin{eqnarray}
    \pi(f_\Phi) = \frac{\theta(f_\Phi - f_{\Phi, {\rm min}})\theta(f_{\Phi, {\rm max}} - f_\Phi )}{f_\Phi \log(f_{\Phi, {\rm max}} / f_{\Phi, {\rm min}})} \, ,
\end{eqnarray}
and flat priors on $\tilde{a}_i$
\begin{eqnarray}\label{eq:flatspin}
    \pi(\tilde{a}_i) = \frac{\theta(\tilde{a}_{\rm max} - \tilde{a}) \theta(\tilde{a})}{\tilde{a}_{\rm max}} \, ,
\end{eqnarray}
where $\tilde{a}_{\rm max} = 0.998$, and we take $f_{\Phi, {\rm max}} = 10^{20}$ GeV and $f_{\Phi, {\rm min}} = 10^{10}$ GeV. This procedure however leads to a highly inefficient sampling procedure, as there exists a large prior volume which can never be compatible with the data (which comes from the fact that superradiance only decreases the spin, and thus small initial spins will always be disfavored). Given the numerical overhead introduced in the forward modeling (which can be very burdensome for some models), we choose to adopt data-informed priors, which a procedure that often goes under the name of Empirical Bayes method~\cite{raftery2001statistics}. Specifically, we  adopt a half-flat-half-gaussian priors on the spins (where we allow for a conditional mass dependence),
given by:

\begin{eqnarray}
    \pi(\tilde{a}_i| M_1, M_2) = \begin{cases}
        \frac{1}{2(\tilde{a}_{\rm max} - \zeta_i)} \hspace{.65cm} \tilde{a}_i > \zeta_i \\[8pt]
        \frac{e^{-(\tilde{a}_i - \zeta_i)^2 / 2 \beta_i^2}}{N_{\tilde{a}_i} \sqrt{2\pi \beta_i^2}} \hspace{.3cm} \tilde{a}_i \leq \zeta_i \, ,
    \end{cases}
\end{eqnarray}
and with $N_{\tilde{a}_i}$ being the appropriate normalization (given that spins have natural truncation scale), and the implicit conditional dependence on $M_1, M_2$ have entered via the definitions of
\begin{eqnarray}
    \zeta_i(M_1, M_2) &\equiv& \int d\tilde{a}_{i,0} \tilde{a}_{i,0} p(\tilde{a}_{i,0} | M_1, M_2) \\
    \beta_i(M_1, M_2) &\equiv&  \int d\tilde{a}_{i,0} (\tilde{a}_{i,0} -\left<\tilde{a}_{i,0}\right>)^2  p(\tilde{a}_{i,0} | M_1, M_2) \, .
\end{eqnarray}
The prior $\pi(M_1,M_2)$ is obtained from a 2D KDE of the LVK posterior samples. Note that in the above construction, $M_1$ and $q$ can be sampled from the LVK posterior, and once samples are obtained, the spin priors can be directly computed. Finally, limits on $f_\Phi$ are obtained by computing the marginalized posterior.

\begin{figure}
    \centering
    \includegraphics[width=\linewidth]{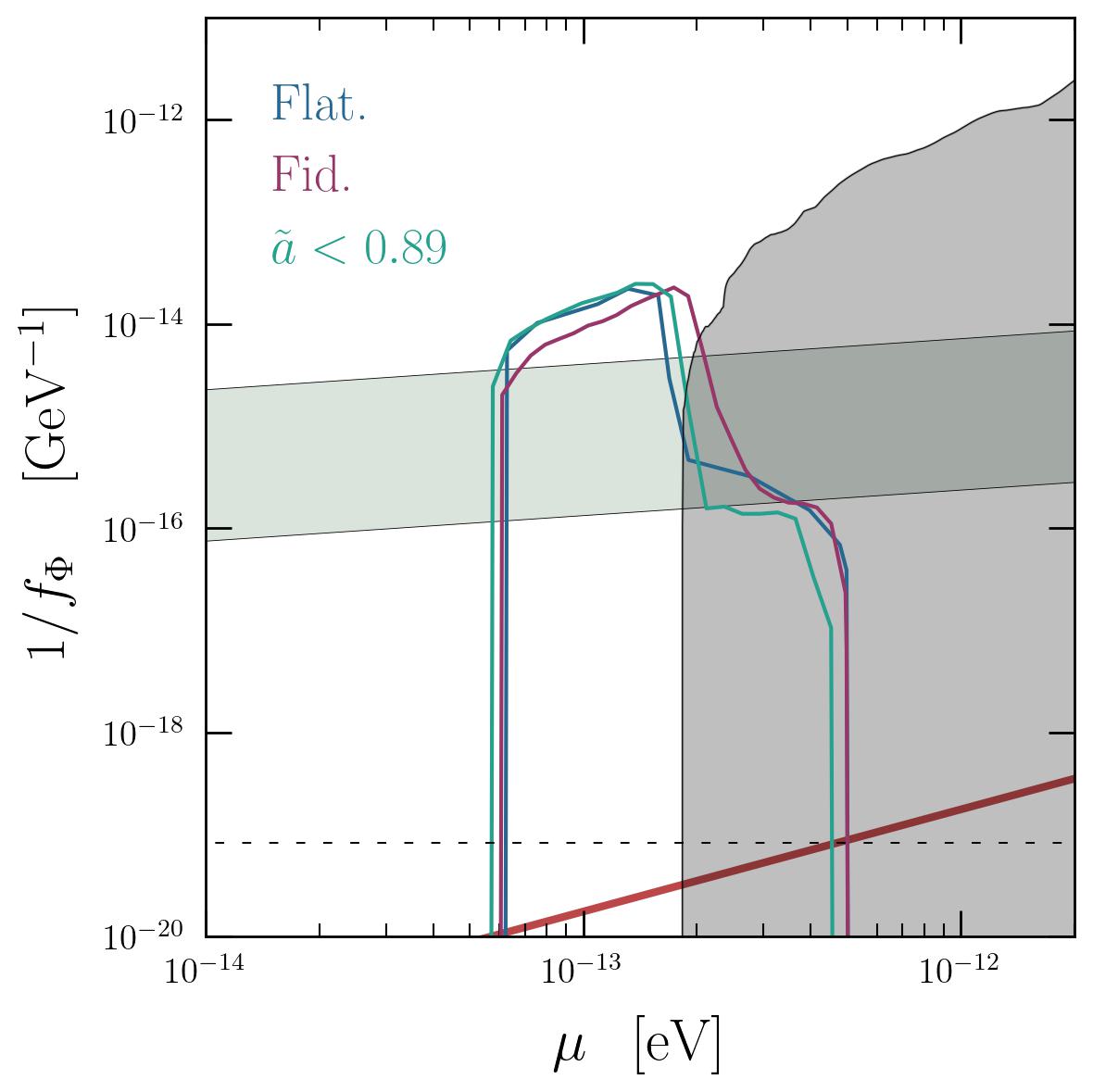}
    \caption{Comparison of fiducial $N_{\rm max}= 3$ limit computed using the {\tt NRSur} likelihood (purple), the analysis performed truncating the spin prior $\tilde{a} < 0.89$ (light blue), and an analysis performed using flat priors on the spins and $(M_1,q)$ and using a 4D KDE for the likelihood (dark blue). } 
    \label{fig:prioranal}
\end{figure}

In order to understand the effect of adopting these data-informed priors on the spin, we perform one independent analysis in which we adopt flat priors on both spins consistent with Eq.~\ref{eq:flatspin}. We further adopt flat priors on $M_1$ and $q$ consistent with those of the LVK analysis, and compute the likelihood using the 4D KDE procedure described above (just below Eq.~\ref{eq:firststop_post}). Since this procedure is far more computationally expensive, we perform this analysis for $N_{\rm max } = 3$ and $\tau_{\rm max} = 10^5$ years -- the result is compared against our fiducial limit, and against our analysis where the spin priors are truncated at $\tilde{a} < 0.89$, in Fig.~\ref{fig:prioranal}. Here, one can see that the flat spin priors tend toward the truncated spin analysis at low masses (which are stronger than our fiducial limit in this region of parameter space), and are weakened near boundary of the $m=1$ spin down; this feature is expected, since the $m=1$ spin-down threshold is strongly sensitive to the spin.

\end{document}